\documentclass[twocolumn, astrosymb]{aastex631}

\usepackage{mathtools}

\newcommand{\white}{\textcolor{white}}

\begin{document}

\title{The JWST-SUSPENSE Ultradeep Spectroscopic Program: \\ Survey Overview and Star-Formation Histories of Quiescent Galaxies at $\boldsymbol{1<z<3}$}

\author[0000-0001-7540-1544]{Martje Slob}
\affiliation{Leiden Observatory, Leiden University, P.O. Box 9513, 2300 RA Leiden, The Netherlands}

\author[0000-0002-7613-9872]{Mariska Kriek}
\affiliation{Leiden Observatory, Leiden University, P.O. Box 9513, 2300 RA Leiden, The Netherlands}

\author[0000-0002-9861-4515]{Aliza G. Beverage}
\affiliation{Department of Astronomy, University of California, Berkeley, CA 94720, USA}

\author[0000-0002-1714-1905]{Katherine A. Suess}
\altaffiliation{NHFP Hubble Fellow}
\affiliation{Kavli Institute for Particle Astrophysics and Cosmology and Department of Physics, Stanford University, Stanford, CA 94305, USA}

\author[0000-0001-6813-875X]{Guillermo Barro}
\affiliation{Department of Physics, University of the Pacific, Stockton, CA 90340 USA}

\author[0000-0003-2680-005X]{Gabriel Brammer}
\affiliation{Cosmic Dawn Center (DAWN), Denmark}
\affiliation{Niels Bohr Institute, University of Copenhagen, Jagtvej 128, DK2200 Copenhagen N, Denmark}

\author[0000-0001-5063-8254]{Rachel Bezanson}
\affiliation{Department of Physics and Astronomy and PITT PACC, University of Pittsburgh, Pittsburgh, PA 15260, USA}

\author[0000-0003-2251-9164]{Chloe M. Cheng}
\affiliation{Leiden Observatory, Leiden University, P.O. Box 9513, 2300 RA Leiden, The Netherlands}

\author[0000-0002-1590-8551]{Charlie Conroy}
\affiliation{Center for Astrophysics \textbar\ Harvard \& Smithsonian, Cambridge, MA, 02138, USA}

\author[0000-0002-2380-9801]{Anna de Graaff}
\affiliation{Max-Planck-Institut f\"ur Astronomie, K\"onigstuhl 17, D-69117, Heidelberg, Germany}

\author[0000-0003-4264-3381]{Natascha M. F\"orster Schreiber}
\affiliation{Max-Planck-Institut f\"ur extraterrestrische Physik, Giessenbachstrasse 1, D-85748 Garching, Germany}

\author[0000-0002-8871-3026]{Marijn Franx}
\affiliation{Leiden Observatory, Leiden University, P.O. Box 9513, 2300 RA Leiden, The Netherlands}

\author[0000-0002-5337-5856]{Brian Lorenz}
\affiliation{Department of Astronomy, University of California, Berkeley, CA 94720, USA}

\author[0000-0001-5175-939X]{Pavel E. Mancera Pi\~na}
\affiliation{Leiden Observatory, Leiden University, P.O. Box 9513, 2300 RA Leiden, The Netherlands}

\author[0000-0001-9002-3502]{Danilo Marchesini}
\affiliation{Department of Physics \& Astronomy, Tufts University, MA 02155, USA}

\author[0000-0002-9330-9108]{Adam Muzzin}
\affiliation{Department of Physics and Astronomy, York University, 4700 Keele Street, Toronto, Ontario, ON MJ3 1P3, Canada}

\author[0000-0001-7769-8660]{Andrew B. Newman}
\affiliation{Observatories of the Carnegie Institution for Science, 813 Santa Barbara St., Pasadena, CA 91101, USA}

\author[0000-0002-0108-4176]{Sedona H. Price}
\affiliation{Department of Physics and Astronomy and PITT PACC, University of Pittsburgh, Pittsburgh, PA 15260, USA}

\author[0000-0003-3509-4855]{Alice E. Shapley}
\affiliation{Department of Physics \& Astronomy, University of California, Los Angeles, CA 90095, USA}

\author[0000-0001-7768-5309]{Mauro Stefanon}
\affiliation{Departament d'Astronomia i Astrof\`isica, Universitat de Val\`encia, C. Dr. Moliner 50, E-46100 Burjassot, Val\`encia,  Spain}
\affiliation{Unidad Asociada CSIC "Grupo de Astrof\'isica Extragal\'actica y Cosmolog\'ia" (Instituto de F\'isica de Cantabria - Universitat de Val\`encia)}

\author[0000-0002-8282-9888]{Pieter van Dokkum}
\affiliation{Astronomy Department, Yale University, 52 Hillhouse Ave,
New Haven, CT 06511, USA}

\author[0000-0002-6442-6030]{Daniel R. Weisz}
\affiliation{Department of Astronomy, University of California, Berkeley, CA 94720, USA}
\correspondingauthor{Martje Slob}
\email{slob@strw.leidenuniv.nl}

\submitjournal{ApJ}
\received{April 19, 2024}

\begin{abstract}
We present an overview and first results from the Spectroscopic Ultradeep Survey Probing Extragalactic Near-infrared Stellar Emission (SUSPENSE), executed with NIRSpec on JWST. The primary goal of the SUSPENSE program is to characterize the stellar, chemical, and kinematic properties of massive quiescent galaxies at cosmic noon. In a single deep NIRSpec/MSA configuration, we target 20 distant quiescent galaxy candidates ($z=1-3$, $H_{AB}\le23$), as well as 53 star-forming galaxies at $z=1-4$. With 16~hr of integration and the G140M-F100LP dispersion-filter combination, we observe numerous Balmer and metal absorption lines for all quiescent candidates. We derive stellar masses (log$M_*/M_{\odot}\sim10.2-11.5$) and detailed star-formation histories (SFHs) and show that all 20 candidate quiescent galaxies indeed have quenched stellar populations. These galaxies show a variety of mass-weighted ages ($0.8-3.3$~Gyr) and star formation timescales ($\sim0.5-4$~Gyr), and four out of 20 galaxies were already quiescent by $z=3$. On average, the $z>1.75$ $[z<1.75]$ galaxies formed 50\% of their stellar mass before $z=4$ $[z=3]$. Furthermore, the typical SFHs of galaxies in these two redshift bins ($z_{\text{mean}}=2.2~[1.3]$) indicate that galaxies at higher redshift formed earlier and over shorter star-formation timescales compared to lower redshifts. Although this evolution is naturally explained by the growth of the quiescent galaxy population over cosmic time, number density calculations imply that mergers and/or late-time star formation also contribute to the evolution. In future work, we will further unravel the early formation, quenching, and late-time evolution of these galaxies by extending this work with studies on their chemical abundances, resolved stellar populations and kinematics.
\end{abstract}


\section{Introduction}\label{sec:intro}

One of the most remarkable discoveries in extra-galactic astronomy from the past two decades is the existence of a population of quiescent galaxies at $z\sim2$ and beyond \citep[e.g.,][]{MFranx2003,ACimatti2004,EDaddi2005,MKriek2006,CStraatman2014,KGlazebrook2017,CSchreiber2018,MTanaka2019, BForrest2020a,BForrest2020b, JEsdaile2021,FValentino2023,JAntwi-Danso2023,ACarnall2023a,ACarnall2023b, KGlazebrook2023, TNanayakkara2024}. The star-formation rates (SFRs) of these galaxies are strongly suppressed and their inferred star formation histories (SFHs) indicate that they formed their stars in vigorous bursts, followed by an efficient quenching process \citep[e.g.,][]{MKriek2016,ACarnall2019,MJafariyazani2020,BForrest2020a,BForrest2020b,FValentino2020,ABeverage2024}. Galaxy formation models struggle to create this galaxy population so early in the universe's history \citep[e.g.,][]{CSchreiber2018, RCecchi2019, AHartley2023, KGould2023, EWeller2024}. Hence, the early and fast formation of these galaxies, as well as the cessation of star formation within them, has remained a problem.

Their subsequent evolution is also a puzzle. Distant quiescent galaxies are much more compact and denser than massive early-type galaxies in the present-day universe \citep[e.g.,][]{EDaddi2005,PvanDokkum2008,AvanderWel2014,KSuess2019a,KSuess2021}. Studies also find that older quiescent galaxies have larger sizes \citep{KWhitaker2012,MYano2016,OAlmaini2017,KSuess2020}, though other results suggest the opposite trend with more compact quiescent galaxies having older stellar populations \citep{AGargiulo2017,PFWu2018,MHamadouche2022}. Furthermore, in contrast to the majority of present-day quiescent galaxies, distant quiescent galaxies appear to have disk-like morphologies \citep[][]{AvanderWel2011, YChang2013}, and may be rotationally supported \citep[][]{SToft2017,ANewman2018b,FDEugenio2023}. Two popular scenarios emerged to explain the observed evolution. In the first scenario, distant quiescent galaxies are the cores of present-day massive elliptical galaxies  \citep[e.g.,][]{TNaab2009,RBezanson2009,PvanDokkum2010}, with their outer regions building up gradually over cosmic time through a series of minor mergers. This scenario can also explain how early quiescent disks evolved into the present-day massive elliptical galaxies, as minor mergers can gradually perturb the ordered rotation \citep{FBournaud2007,TNaab2014,CLagos2018}. The minor merger scenario is also supported by the finding that color gradients increase with galaxy age \citep[][]{KSuess2020} and that distant quiescent galaxies have many small companions \citep{ANewman2012,KSuess2023}. In a second popular scenario, the size evolution can be explained by the growth of the quiescent galaxy population, as galaxies that quench at later times are predicted to be larger \citep[i.e., progenitor bias; e.g.,][]{SKhochfar2006,BPoggianti2013,MCarollo2013,ZJi2022}. Both scenarios are thought to play a role in the observed evolution \citep[e.g.,][]{SBelli2017}.

However, more recent observations show that this picture may be too simplistic. First, neither progenitor bias nor minor mergers can explain the difference in [Fe/H] of 0.2-0.3 dex between distant quiescent galaxies and the cores of nearby early-type galaxies \citep{MKriek2016,MKriek2019,ACarnall2022,MGu2022,ZZhuang2023,ABeverage2024}.  Furthermore, while the cores of nearby massive early-type galaxies are found to have bottom-heavy initial mass functions \citep[IMFs,][]{TTreu2010,CConroy2012d}, the IMF in distant quiescent galaxies appears more consistent with a Milky Way \citep[i.e.,][]{PKroupa2001,GChabrier2003} IMF \citep{JMendel2020,BForrest2022,MKriek2023}. Thus, these observations imply that the evolution may be more complicated than initially proposed and major mergers and/or late-time star formation may play a role as well. 

In order to solve these puzzles and understand the early formation, quenching mechanism, and late-time evolution of the distant quiescent galaxy population, we need detailed stellar population, chemical, and kinematic properties of a significant sample of distant quiescent galaxies. Obtaining such measurements has been exceedingly challenging, as they rely on ultra-deep rest-frame optical spectra. Consequently, the above spectroscopic studies are based on only a few very massive and/or lensed galaxies. Furthermore, despite extreme integration times with the most efficient near-infrared spectrographs on large ground-based telescopes, the spectra still have significant uncertainties and suffer from low spatial resolution. Thus, we have reached the limits of current ground-based facilities. 

With the advent of JWST we are finally capable of breaking this impasse. Early results have demonstrated the power of JWST to observe distant quiescent galaxies ($z\gtrsim2.5$) with the low-resolution mode \citep[][]{KGlazebrook2023,DMarchesini2023,TNanayakkara2024,DSetton2024,AdeGraaff2024}. Furthermore, using medium-resolution spectroscopy, \citet{SBelli2023}, \citet{FDEugenio2023}, and \cite{ACarnall2023a} demonstrate the detection of -- primarily Balmer -- absorption lines in galaxies at $z=2.445$, $z=3.064$, and $z=4.658$ respectively. While these studies attest to the unprecedented sensitivity of NIRSpec, they either lack the required spectral resolution or do not have sufficiently old galaxies to observe metal absorption lines. Such metal absorption lines allow us to study stellar populations and chemical abundances in unparalleled detail. Furthermore, the samples are still very small, and the quiescent galaxies found at $z>3$ are likely not representative of the full massive quiescent galaxy population. Thus, the next step is to obtain larger, more representative, medium-resolution samples and target older quiescent galaxies, for which we can measure a range of metal absorption lines. Taking full advantage of the multi-plexing capabilities of NIRSpec \citep{PFerruit2022} for studying quiescent galaxies, however, remains challenging due to the low number densities of such galaxies; consequently, they are rare in spectroscopic surveys executed in the deep legacy fields.

To successfully overcome these challenges and obtain ultradeep, medium-resolution, rest-frame optical, spatially-resolved spectra of a significant sample of distant quiescent galaxies, we are conducting the Cycle 1 SUSPENSE (Spectroscopic Ultradeep Survey Probing Extragalactic Near-infrared Stellar Emission) program. The SUSPENSE observations were collected with the JWST NIRSpec/MSA in January 2024. SUSPENSE observes 20 quiescent galaxy candidates at $1<z<3$ in a single NIRSpec/MSA configuration. The spectra are of unprecedented depth, show a multitude of Balmer and metal absorption lines, and the MSA slits extend over the full spatial range of the galaxies. 

In this paper we present an overview of the survey, the observational strategy, the data reduction and analysis, the sample characteristics, and the SFHs of 20 quiescent galaxies in our sample. The paper is organized as follows. In Section~\ref{sec:obs} we present the SUSPENSE survey design and observing strategy. In Section~\ref{sec:data_red} we discuss the two-dimensional (2D) data processing and the extraction of the one-dimensional (1D) spectra. Section~\ref{sec:analysis} describes the stellar populations fitting, the sample characteristics, and the SFHs of the quiescent galaxy sample. In Section~\ref{sec:implications} we discuss the implications of our findings for the evolution of the quiescent galaxy population. Finally, in Section~\ref{sec:summary} we present a summary.

Throughout this work we assume a $\Lambda$CDM cosmology with $\Omega_{\rm m}= 0.3$, $\Omega_\Lambda=0.7$, and $H_0 =70\rm \, km s^{-1} \,Mpc^{-1}$, and assume the Solar abundances of \citet{MAsplund2009}, with $Z_{\odot}=0.0142$. All magnitudes are given in the AB-magnitude system \citep{JOke1983}. 

\section{Observations}\label{sec:obs}
\subsection{Target Selection and MSA Configuration}\label{sec:survey}

The JWST-Cycle 1 SUSPENSE program (ID 2110) aims to characterize the stellar, chemical, and kinematic properties of the distant quiescent galaxy population. To reach this goal, we require ultradeep rest-frame optical spectra of a significant sample of distant quiescent galaxies. The NIRspec MSA's unparalleled sensitivity and multiplexing capabilities make it ideally suited to this task. However, fully utilizing its multiplexing capabilities depends on identifying a field with a high density of these galaxies.

To this end, we used the large overlapping area of the UltraVISTA 
\citep[][]{HMcCracken2012,AMuzzin2013b}, COSMOS \citep[F814W;][]{NScoville2007} and COSMOS-DASH \citep[F160W][]{LMowla2018,IMomcheva2017,SCutler2022} surveys to optimize our pointing. We first identified all $z_{\text{phot}}>1.1$ quiescent galaxies in the UltraVISTA DR3 catalog \citep{AMuzzin2013a} using the UVJ selection criteria by \citet{AMuzzin2013b}. The redshift criterion ensured that we observe at least two Balmer lines for all targets. We additionally required $H<22.6$, such that the galaxies are sufficiently bright for resolved kinematic and elemental abundance studies. We identified an extraordinary pointing for which we observe $\sim$20 such galaxies in one NIRSpec/MSA configuration.  

We re-derived and updated the coordinates of all galaxies and stars in our pointing by running Source Extractor \citep{EBertin1996} on the COSMOS-DASH F160W image. For the galaxies we used the brightest-pixel positions, while for the stars we used the barycenter positions. For the fainter galaxies that were not detected in F160W (all 20 candidate quiescent galaxy targets are detected), we used the original UltraVISTA coordinates.

We used the APT MSA planning tool to design two nearly identical MSA configurations, offset in dispersion direction by 8 shutters. We assumed the assigned PA of 71.6174\textdegree, three shutter slitlets, and a three nod pattern (though we only use the two outer nod positions, as discussed in the next section). As our quiescent galaxy candidates are extended, we used the ``unconstrained'' (midbar) option for the source centering constraint, implying that our galaxies may be centered behind both the horizontal and vertical MSA bars. While this was the case for five of the quiescent targets, most galaxies were centered in the open shutter area. We also allowed galaxies in areas of the detector that are affected by failed open shutters. 

Within the group of quiescent galaxies the highest priority was given to galaxies at $z\ge 1.5$, followed by $1.3\le z<1.5$ and $1.1\le z < 1.3$. After we identified our optimal configuration, we extended the slitlets by one microshutter, where possible. Next, we added filler targets by hand, prioritizing bright star-forming galaxies at $z\sim1.5$, for which we expect to detect bright stellar continuum emission and all Balmer emission lines, as well as fainter quiescent galaxies ($H>22.6$). Eventually we considered all targets with $z>1.1$. In contrast to the primary targets, we did not require the filler galaxies to have coverage in all nod and dither positions. Finally, after no more targets could be added, we extended more slits, and opened microshutters on empty sky to construct a master background. For our quiescent targets, the slitlets range from 3 to 7 microshutters, with an average of 5. The slit lengths of the others targets varied between 1 and 7 shutters. Due to failed closed shutters in the MSA we could not open the exact same shutters in the two dithered configurations for all targets, which means that for some targets the number of shutters in a slitlet differs between the two dithers. 

\begin{figure*}
  \begin{center}  
  \includegraphics[width=1.\textwidth]{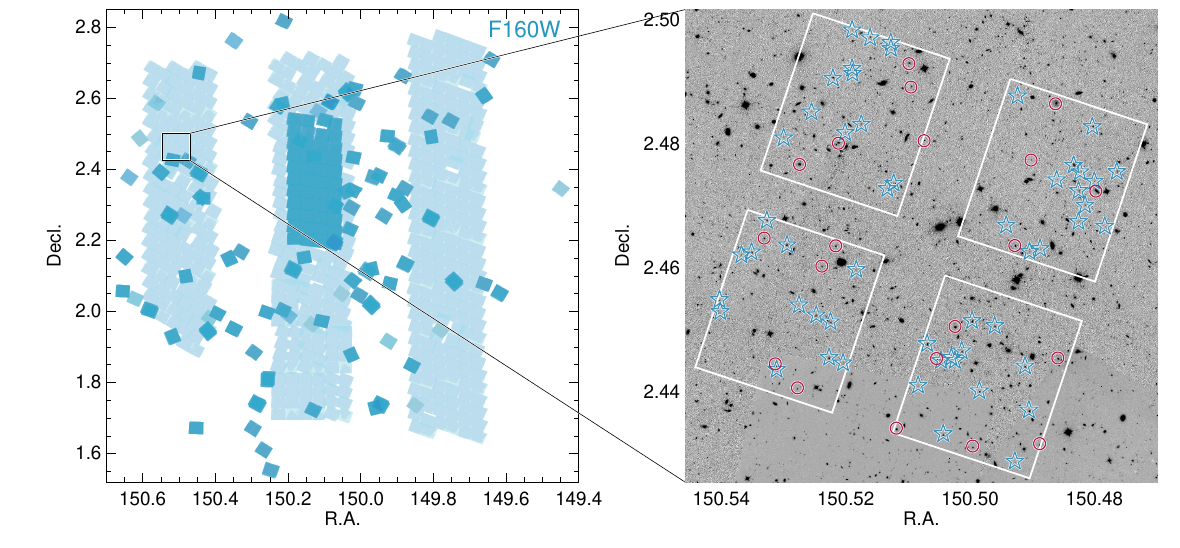}  
  \caption{Footprint of the SUSPENSE observations in the larger COSMOS field. The left panel shows the weight map of all available public HST/F160W imaging, constructed as part of the COSMOS-DASH program \citep{IMomcheva2017}. The darker blue central contiguous area represents the CANDELS survey \citep{AKoekemoer2011,NGrogin2011}, and the three larger and lighter stripes represent the shallower COSMOS-DASH survey, which overlaps with the deep UltraVISTA stripes. In the right panel we zoom in on our pointing, now showing the combined F160W image and the footprint of the NIRSpec/MSA. The red circles and blue stars indicate the quiescent and star-forming targets, respectively.}\label{fig:pointing}
    \end{center}
\end{figure*}

In total we target 73 galaxies, of which 20 are quiescent galaxy candidates. In Figure~\ref{fig:pointing} we show our pointing within the larger COSMOS area, as well as the footprint of the NIRSpec/MSA and the targeted galaxies.

\subsection{Observing Strategy}\label{sec:obs_strat}
Our program was executed on January 2 and 4, 2024, employing the NIRSpec/MSA along with the G140M-F100LP dispersion-filter combination. The wavelength coverage for this setting is 0.97--1.84\,$\mu$m, with a spectral resolution of $R \sim 1000$. This wavelength range corresponds to a rest-frame wavelength coverage of $\sim3700-7000$\,\AA\ for our median redshift ($z\sim1.5$). Though the exact rest-frame wavelength coverage depends on the redshift, for most quiescent galaxies we observed 4600--5400\,\AA, which includes Mg\,I at 5178\,\AA\ as well as several prominent Fe\,I lines. Furthermore, all spectra cover at least one Balmer line.

\begin{figure}
  \begin{center}  
    \includegraphics[width=0.48\textwidth]{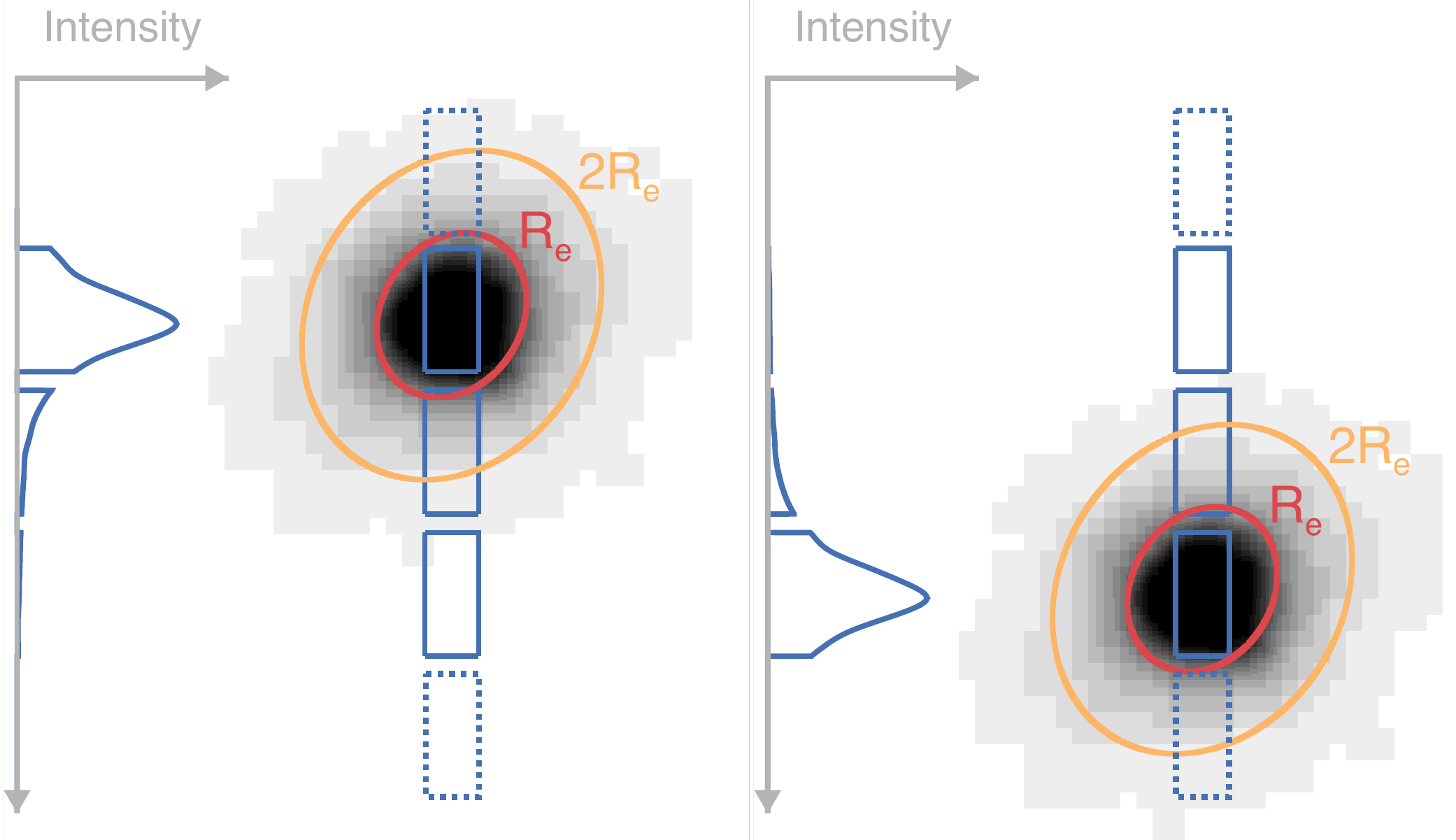}
    \caption{Nod pattern shown for a massive $z\sim2.1$ quiescent galaxy with an $R_e$ of 2.67\,kpc \citep[][]{MKriek2016}. The image is taken with HST/F160W from CANDELS. The red ellipsoid contains half of the light (for the NIRSpec PSF), while the orange ellipsoid indicates the extent of two half-light radii ($R_{\rm e}$). The profiles on the left of the images show the light distribution within the shutters. This figure illustrates that the galaxies extend beyond a single shutter and thus a two-shutter nod is required to minimize self-subtraction of the galaxy's signal in the standard sky-subtraction. \label{fig:nod}}
    \end{center}  
\end{figure}

We used a two-point nod pattern with a cross-dispersion offset of two microshutters, corresponding to an angular offset of 1\farcs 06. This angular offset translates to a distance of $8.6-9.0$\,kpc at the targeted redshift. This offset was strategically chosen based on the typical sizes of quiescent targets at our targeted redshift range; with half-light radii of up to $\sim$3\,kpc \citep[][]{SCutler2022}, a two-micro-shutter offset facilitates the use of the adjacent offset frames as sky, as illustrated in Fig.~\ref{fig:nod}. A cross-dispersion offset of just one microshutter, as used in the majority of NIRSpec/MSA studies of distant galaxies, would have resulted in a partial self-subtraction of the signal during the standard reduction procedure.

We observed our galaxies in two different configurations, offset by 8 micro-shutters in the dispersion direction. This dither, combined with the two-point nod pattern, resulted in four offset positions, needed to adequately mitigate detector and microshutter defects.
Finally, our two offset MSA configurations reduce the detector gap, as the two dithers resulted in spectra which were shifted in the dispersion direction by $\sim$130\,\AA.

Because of microshutter defects, it was not possible to observe exactly the same galaxies in all four positions. 34 galaxies have full coverage, including 18 quiescent targets (see Table~\ref{tab:sample}). Two quiescent galaxies only have partial coverage, either because they were on the edge of the detector (130183) or because they were not prioritized as they are faint (129966). In total, four galaxies have $\sim75$\% coverage, 22 galaxies have $\sim50$\% coverage, and 13 galaxies have $\sim25$\% coverage.

Our observations were split over two identical visits. For both target acquisitions we used the MSATA method and six alignment stars. After the acquisition, we took a confirmation image of 379 seconds of the first configuration. During each visit, we observed both dither configurations. For both configurations we took 10 integrations of 1473 seconds each, with 20 groups per integration and the NRSIRS2 readout pattern. We nodded to a new position after two exposures, and thus for each dither configuration per visit we have 5 nod positions. The total integration time per configuration per visit is 14,730 seconds. Combining both visits and configurations, we have a total on-source integration time of 16.4 hrs. 

\begin{table*}
\centering
\caption{Overview of parameters of the quiescent galaxy candidate sample.}\label{tab:sample}
\begin{tabular}{l c c c c c c r r r r r}
\hline \hline
& \multicolumn{2}{c}{\underline{Coordinates}} & & & \multicolumn{2}{c}{\underline{Rest-frame colors}} & \multicolumn{4}{c}{\underline{\texttt{Prospector} fitting parameters}} & \\
ID & R.A. & Decl. & $H$ & $z_{\rm spec}$ & $U-V$ & $V-J$ & \multicolumn{1}{c}{log $M_{*}^{\ a}$} & \multicolumn{1}{c}{log SFR} & \multicolumn{1}{c}{Age$^{b}$} & \multicolumn{1}{c}{log $Z_{*}$} & \multicolumn{1}{c}{$t_{\rm exp}$} \\
& (hh:mm:ss) & (dd:mm:s) & (mag) & & & & \multicolumn{1}{c}{($M_{\odot}$)} & \multicolumn{1}{c}{($M_{\odot}$\,yr$^{-1}$)} & \multicolumn{1}{c}{(Gyr)} & \multicolumn{1}{c}{($Z_{\odot}$)} & \multicolumn{1}{c}{(hr)} \\
\hline 
127345 & 10:02:02.85 & 2:26:03.4 & 21.3 & 1.168 & 1.85 & 0.97 & 10.72$^{+0.00}_{-0.01}$ & $-2.79^{+0.38}_{-0.61}$ & 2.87$^{+0.02}_{-0.05}$  & -0.06$^{+0.01}_{-0.00}$ & 16.4 \\[2pt] 
130040 & 10:02:06.58 & 2:28:36.2 & 20.4 & 1.170 & 2.03 & 1.33 & 11.21$^{+0.01}_{-0.01}$ & $-1.97^{+0.91}_{-2.46}$ & 2.71$^{+0.04}_{-0.04}$  & -0.30$^{+0.03}_{-0.03}$ & 16.4 \\[2pt] 
128452\textsuperscript{\textdagger}* & 10:02:00.58 & 2:27:02.4 & 20.5 & 1.205 & 1.63 & 0.99 & 10.99$^{+0.01}_{-0.00}$ & $0.50^{+0.01}_{-0.01}$ & 2.41$^{+0.09}_{-0.08}$  & -0.39$^{+0.01}_{-0.01}$ & 16.4 \\[2pt] 
127154 & 10:01:57.32 & 2:25:54.5 & 21.2 & 1.205 & 1.92 & 1.01 & 10.75$^{+0.01}_{-0.01}$ & $-4.07^{+1.56}_{-5.75}$ & 2.35$^{+0.06}_{-0.07}$  & -0.22$^{+0.02}_{-0.02}$ & 16.4 \\[2pt] 
130208 & 10:02:05.08 & 2:28:48.5 & 20.8 & 1.231 & 1.71 & 0.80 & 10.95$^{+0.00}_{-0.00}$ & $-3.32^{+1.15}_{-4.89}$ & 3.26$^{+0.02}_{-0.03}$  & -0.76$^{+0.01}_{-0.01}$ & 16.4 \\[2pt] 
129982 & 10:01:55.14 & 2:28:20.7 & 20.3 & 1.249 & 1.73 & 1.18 & 11.22$^{+0.01}_{-0.01}$ & $0.97^{+0.02}_{-0.02}$ & 2.49$^{+0.04}_{-0.05}$  & -0.38$^{+0.02}_{-0.02}$ & 13.1 \\[2pt] 
127108 & 10:01:59.90 & 2:25:53.1 & 22.5 & 1.335 & 1.63 & 0.84 & 10.24$^{+0.02}_{-0.03}$ & $-1.23^{+0.28}_{-0.97}$ & 2.11$^{+0.10}_{-0.10}$  & -0.34$^{+0.06}_{-0.07}$ & 16.4 \\[2pt] 
129197* & 10:02:07.95 & 2:27:53.5 & 22.3 & 1.474 & 1.88 & 1.07 & 10.52$^{+0.02}_{-0.02}$ & $-2.52^{+1.15}_{-3.71}$ & 1.73$^{+0.11}_{-0.10}$  & 0.17$^{+0.01}_{-0.02}$ & 16.4 \\[2pt] 
130647\textsuperscript{\textdagger} & 10:01:56.70 & 2:29:11.3 & 20.1 & 1.508 & 1.74 & 1.08 & 11.49$^{+0.00}_{-0.00}$ & $0.48^{+0.02}_{-0.02}$ & 3.18$^{+0.01}_{-0.01}$  & -0.15$^{+0.03}_{-0.02}$ & 16.4 \\[2pt] 
130934\textsuperscript{\textdagger}* & 10:02:02.36 & 2:29:34.6 & 21.9 & 1.565 & 1.57 & 1.08 & 10.60$^{+0.05}_{-0.04}$ & $0.97^{+0.08}_{-0.06}$ & 2.17$^{+0.33}_{-0.19}$  & -0.55$^{+0.06}_{-0.09}$ & 16.4 \\[2pt] 
129149 & 10:01:58.27 & 2:27:49.2 & 21.2 & 1.579 & 1.78 & 0.93 & 11.02$^{+0.01}_{-0.01}$ & $-2.76^{+1.17}_{-4.87}$ & 1.82$^{+0.07}_{-0.10}$  & 0.18$^{+0.01}_{-0.01}$ & 16.4 \\[2pt] 
130183\textsuperscript{\textdagger} & 10:02:01.77 & 2:28:49.9 & 22.2 & 1.757 & 1.62 & 1.06 & 10.78$^{+0.04}_{-0.04}$ & $0.77^{+0.21}_{-0.66}$ & 1.09$^{+0.21}_{-0.24}$  & -1.37$^{+0.10}_{-0.02}$ & 6.6 \\[2pt] 
128041 & 10:01:56.61 & 2:26:44.1 & 21.9 & 1.760 & 1.42 & 0.97 & 10.71$^{+0.01}_{-0.01}$ & $-0.38^{+0.21}_{-0.20}$ & 1.79$^{+0.04}_{-0.03}$  & -0.37$^{+0.07}_{-0.06}$ & 16.4 \\[2pt] 
127700 & 10:02:06.66 & 2:26:26.8 & 22.6 & 2.013 & 1.62 & 1.11 & 10.92$^{+0.03}_{-0.05}$ & $0.60^{+0.08}_{-0.11}$ & 1.81$^{+0.15}_{-0.11}$  & -0.35$^{+0.15}_{-0.20}$ & 16.4 \\[2pt] 
129133 & 10:02:05.19 & 2:27:49.0 & 22.1 & 2.139 & 1.73 & 1.10 & 11.09$^{+0.02}_{-0.02}$ & $-0.31^{+0.32}_{-0.52}$ & 1.47$^{+0.06}_{-0.05}$  & -0.46$^{+0.04}_{-0.05}$ & 16.4 \\[2pt] 
127941 & 10:02:07.52 & 2:26:40.7 & 22.5 & 2.141 & 1.55 & 1.14 & 10.80$^{+0.02}_{-0.02}$ & $0.22^{+0.15}_{-0.16}$ & 1.72$^{+0.02}_{-0.02}$  & -0.24$^{+0.03}_{-0.02}$ & 16.4 \\[2pt] 
128036 & 10:02:01.29 & 2:26:43.8 & 22.2 & 2.196 & 1.49 & 0.87 & 10.92$^{+0.04}_{-0.03}$ & $-0.99^{+0.64}_{-2.65}$ & 1.11$^{+0.05}_{-0.05}$  & 0.08$^{+0.04}_{-0.04}$ & 16.4 \\[2pt] 
128913 & 10:02:05.72 & 2:27:37.3 & 22.6 & 2.285 & 1.68 & 0.89 & 10.91$^{+0.03}_{-0.03}$ & $-3.70^{+2.13}_{-9.49}$ & 1.71$^{+0.11}_{-0.42}$  & -0.94$^{+0.82}_{-0.10}$ & 16.4 \\[2pt] 
130725* & 10:02:02.28 & 2:29:21.0 & 22.2 & 2.692 & 1.33 & 0.96 & 11.09$^{+0.05}_{-0.02}$ & $0.10^{+0.41}_{-0.26}$ & 1.17$^{+0.03}_{-0.03}$  & 0.15$^{+0.03}_{-0.06}$ & 16.4 \\[2pt] 
129966 & 10:01:57.65 & 2:28:38.7 & 23.2 & 2.923 & 1.45 & 0.81 & 10.91$^{+0.04}_{-0.04}$ & $0.61^{+0.27}_{-0.39}$ & 0.79$^{+0.07}_{-0.06}$  & 0.01$^{+0.10}_{-0.10}$ & 4.9 \\[2pt] 
\\ 
\hline\hline
\multicolumn{12}{l}{\textsuperscript{\textdagger} Galaxies with a PSB best-fit SFH.} \\
\multicolumn{12}{l}{* Galaxies observed in a failed open shutter area} \\
\multicolumn{12}{l}{$^{a}$ Surviving stellar mass.}\\
\multicolumn{12}{l}{$^{b}$ Mass-weighted ages.}
\end{tabular}
\tablecomments{The derivation of the coordinates is described in Section \ref{sec:survey}, the rest-frame $U-V$ and $V-J$ colors are obtained in Section \ref{sec:samp_chars}, and the measurements of the spectroscopic redshifts and \texttt{Prospector} fitting parameters are described in Section \ref{sec:prosp}.}
\end{table*}

\section{Reduction and data overview}\label{sec:data_red}
\subsection{Data Reduction}
We reduced the data using a modified version of the JWST Science Calibration Pipeline \citep{bushouse_2023_10022973} v1.12.5, and version 1183 of the Calibration Reference Data System (CRDS). Since the two dithers in our observations have slightly different MSA configurations we reduce the dithers separately, and combine the resulting 1D spectra. Below we describe the main data reduction steps and our modifications to the standard JWST Calibration Pipeline.

In Stage 1 of the JWST Calibration Pipeline the master bias frame and dark current were subtracted, and detector artifacts were removed. To remove large cosmic-ray events (snowballs) we set \texttt{expand\_large\_events = True}, \texttt{min\_sat\_area = 15}, and \texttt{min\_jump\_area = 15} in the \texttt{jump} step. The pipeline then flags saturated pixels and removes jumps due to cosmic rays, and obtains the count-rate frames by fitting the slope of each pixel.

After Stage 1 of processing, the count-rate frames still contain $1/f$ correlated vertical read out noise \citep{ESchlawin_2020}. We remove this noise using the correction algorithm from \texttt{grizli} \citep{grizli}.
In Stage 2 of the JWST Calibration Pipeline the background-subtraction for the entire detector of each count-rate frame is performed. Using the pipeline, we construct the background for each frame by taking the average of all frames that were observed during the same visit and dither, but are in the opposite nodding position of our 2-nod pattern.
After background subtraction, the 2D spectra associated with our targets were cut out from the full detector frame, and a flat field, pathloss and barshadow correction was applied to each individual 2D spectrum. Since our sources are extended with respect to the shutter size, we use the CRDS pathloss calibration file corresponding to a uniformly illuminated slit. The fluxes of the corrected 2D spectra were then calibrated to convert the data from count rate units to surface brightness.

In Stage 3, the pipeline combines the frames of all nods and observing days for each target. In order to combine the individual frames, the 2D spectra were first rectified and resampled to a common reference frame.
Next, we identified outlier pixels caused by additional cosmic ray impacts or bad pixels in the individual frames of each target using the outlier detection algorithm from \texttt{MSAEXP}\footnote{\url{https://github.com/gbrammer/msaexp}}, instead of the standard JWST Calibration Pipeline outlier removal algorithm. While masking outliers, we combined the rectified 2D frames for both nods and visits by weighing each pixel by its inverse read noise to construct the final 2D spectrum for each dither. We then used the \citet{KHorne1986} optimal extraction algorithm to get the 1D spectra for each dither position.
We re-scale the 1D spectra to the overlapping photometry (UVISTA $Y$, $J$, and $H$ bands) with a multiplicative factor calculated from the median spectrum and photometry. This re-scaling accounts for the slit loss caused by the fact that our extended targets partly fall outside of the shutter, which is not properly corrected for in the pathloss correction step in the reduction pipeline. 
We only apply this scaling to sources with sufficiently strong continuum emission, including all quiescent galaxies. For the star-forming galaxies for which we only detect emission lines, we will properly account for the absolute flux calibration in future work.

Finally, we combine the 1D spectra for both dither positions. For spectra that are dispersed over both NIRSpec detectors there is a wavelength gap in the spectrum caused by the separation of the two detectors. Our 8-shutter dither in the dispersion direction shifts this detector gap by $\sim$130\,\AA\ between the two dither positions, so that the detector gap is partly filled up in the combined spectrum. The spectrum at the edges of the detector gap is thus constructed from only one of the dithered spectra. We created a mask for each spectrum, indicating the bad pixels and spectral coverage. Finally, we weighed all unmasked pixels in the spectra by their on-source integration time to combine them into the final 1D spectrum.

\subsection{Data Overview}
\begin{figure*}
    \centering
    \vspace{-0.15in}\includegraphics[width = 1.\textwidth]{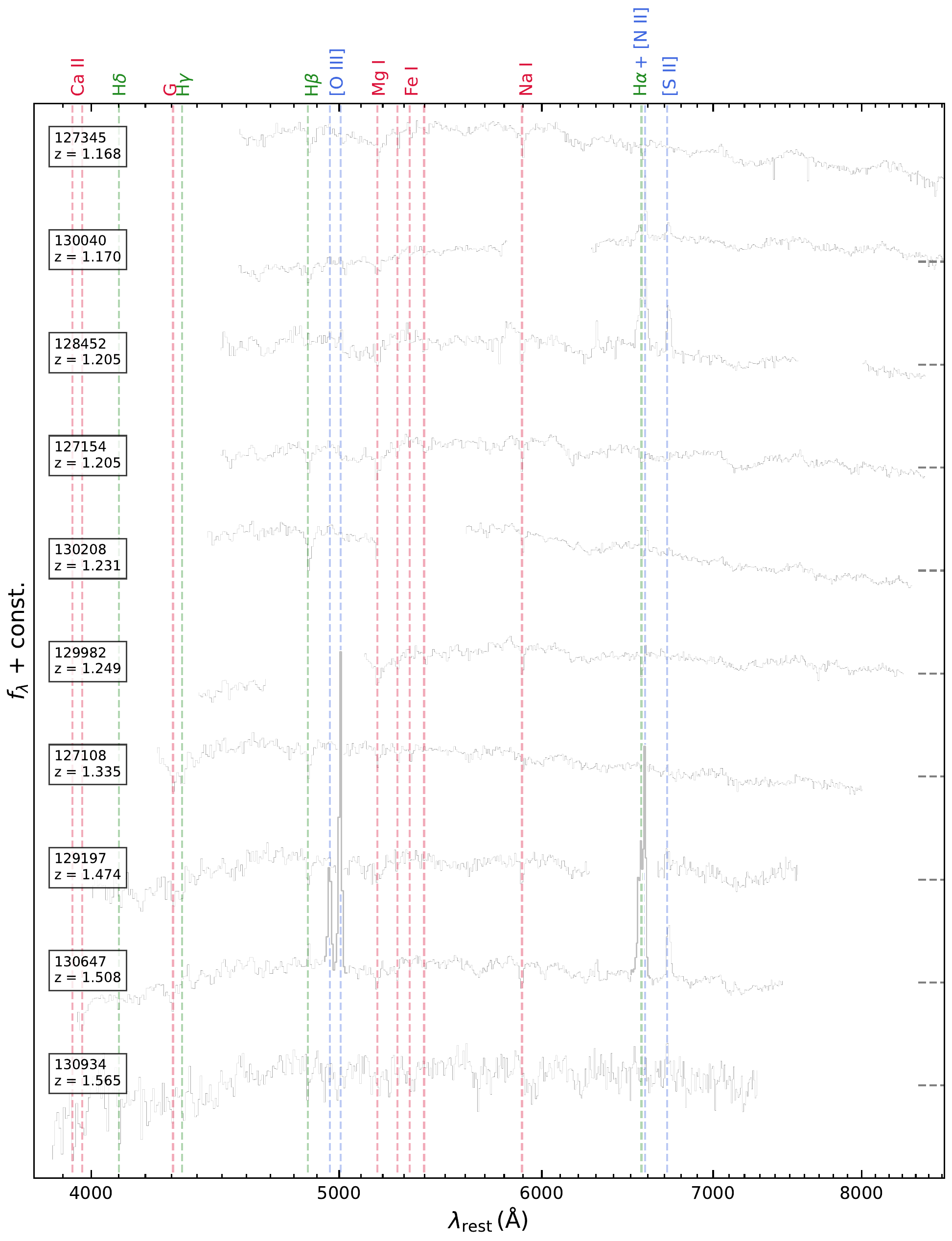}
    \vspace{-0.25in}
    \caption{NIRSpec rest-frame spectra for our sample of 20 candidate quiescent galaxies, sorted by redshift, and offset by a constant in the y-direction. We median bin the spectra over 3 pixels. Dashed blue, red, and green lines mark prominent emission, absorption, and Balmer features, respectively. For clarity purposes, strong emission lines overlapping with other spectra are displayed in gray. The dashed gray lines on the right indicate the zero-points of the spectra.}
    \label{fig:all_q_spec}
\end{figure*}

\begin{figure*}
    \figurenum{3}
    \centering
    \vspace{-0.15in}\includegraphics[width = 1.\textwidth]{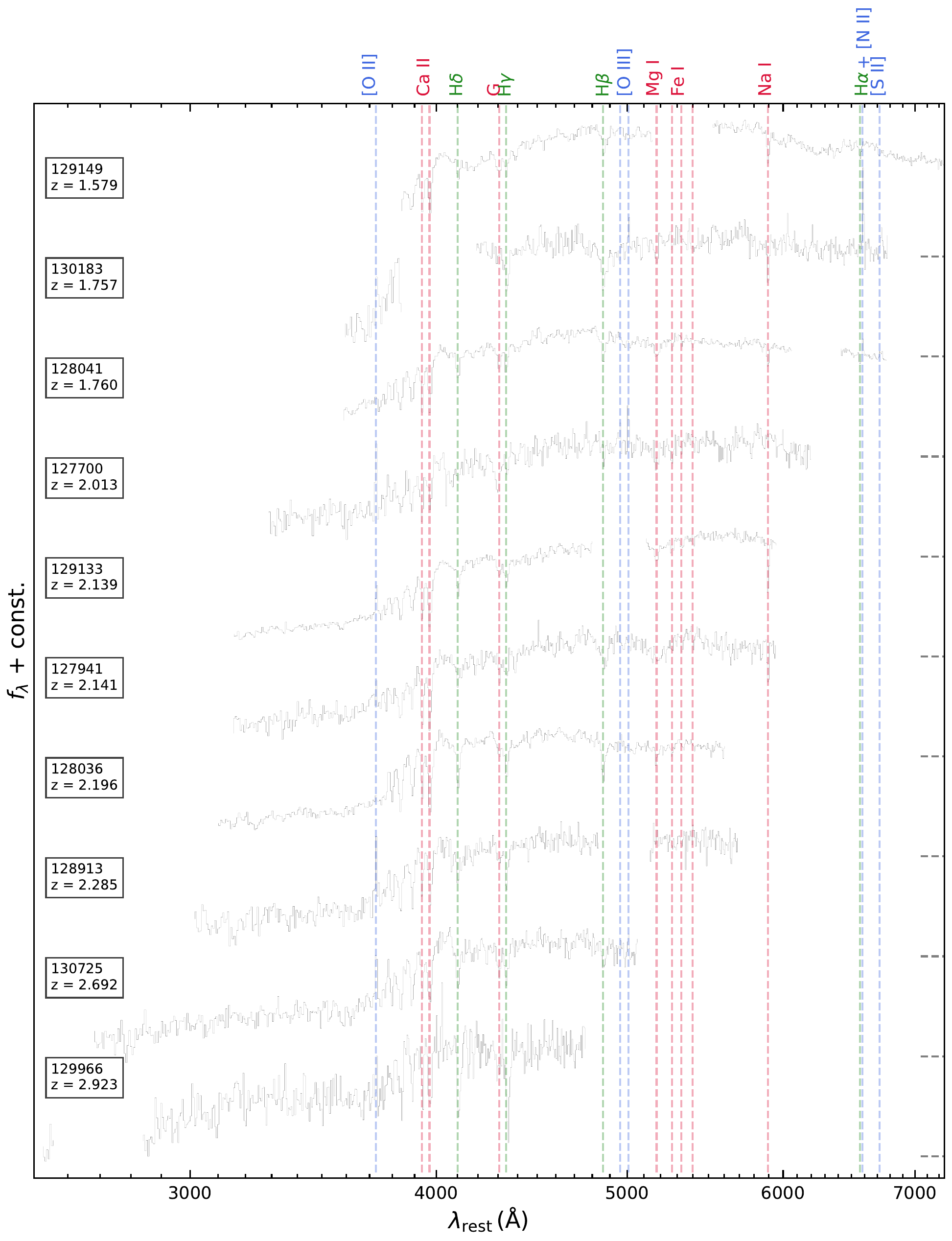}
    \caption{Continued.}
    \label{fig:all_q_spec}
\end{figure*}

\figsetstart
\figsetnum{4}
\figsettitle{Overview of spectra and SEDs}

\figsetgrpstart
\figsetgrpnum{4.1}
\figsetgrptitle{Spectrum of target 129149}
\figsetplot{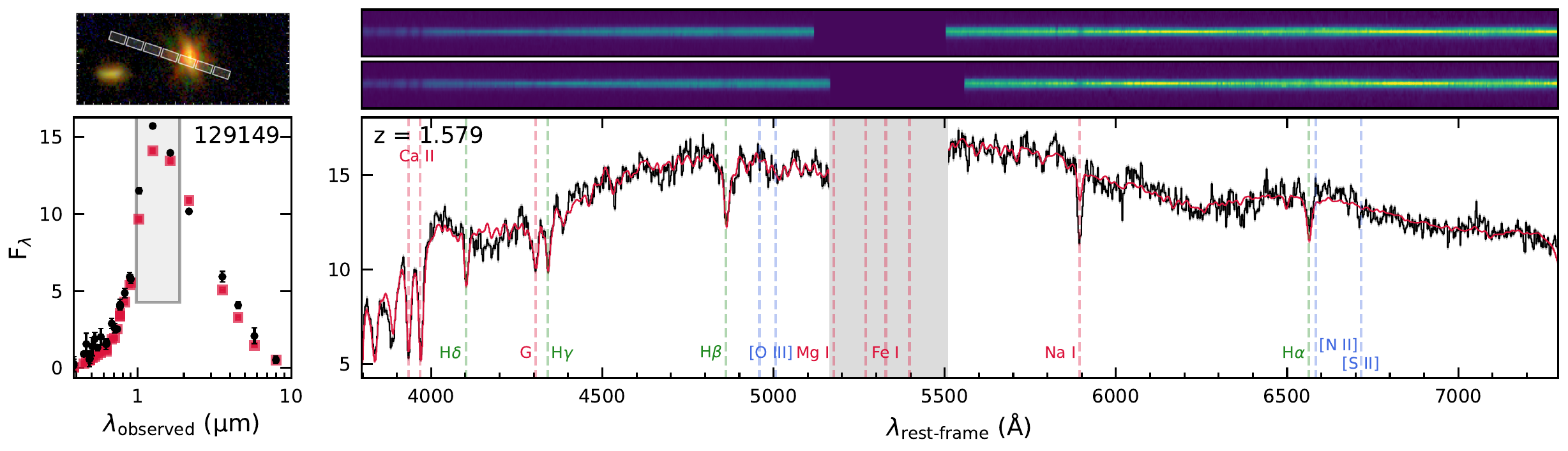}
\figsetgrpnote{Overview of UltraVISTA photometric SEDs (bottom-left), NIRSpec spectra (right), and HST image (top-left) of two example quiescent galaxies in our survey. The 2D spectra for the two observed dithers and the combined 1D spectrum are shown in the top-right and bottom-right rows, respectively. Flux densities ($F_{\lambda}$) are in $10^{-19}$\,erg\,s$^{-1}$\,cm$^{-2}$\,\AA$^{-1}$. The coverage of the NIRSpec spectra are indicated in the SED panels by the gray rectangles. The best-fit \texttt{Prospector} models to the photometry and spectra are shown in red (see Section \ref{sec:prosp}). The images in the top-left panels show the HST-F160W imaging, with an overlay of the NIRSpec MSA slit orientation for one nod position.}
\figsetgrpend

\figsetgrpstart
\figsetgrpnum{4.2}
\figsetgrptitle{Spectrum of target 128036}
\figsetplot{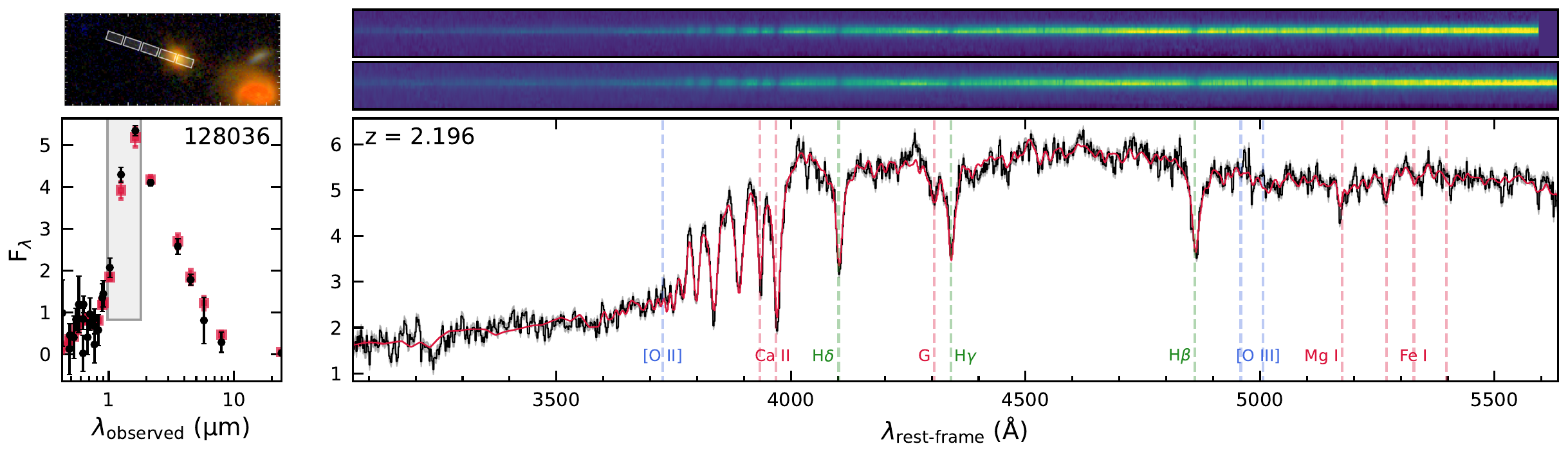}
\figsetgrpnote{Overview of UltraVISTA photometric SEDs (bottom-left), NIRSpec spectra (right), and HST image (top-left) of two example quiescent galaxies in our survey. The 2D spectra for the two observed dithers and the combined 1D spectrum are shown in the top-right and bottom-right rows, respectively. Flux densities ($F_{\lambda}$) are in $10^{-19}$\,erg\,s$^{-1}$\,cm$^{-2}$\,\AA$^{-1}$. The coverage of the NIRSpec spectra are indicated in the SED panels by the gray rectangles. The best-fit \texttt{Prospector} models to the photometry and spectra are shown in red (see Section \ref{sec:prosp}). The images in the top-left panels show the HST-F160W imaging, with an overlay of the NIRSpec MSA slit orientation for one nod position.}
\figsetgrpend

\figsetgrpstart
\figsetgrpnum{4.3}
\figsetgrptitle{Spectrum of target 127345}
\figsetplot{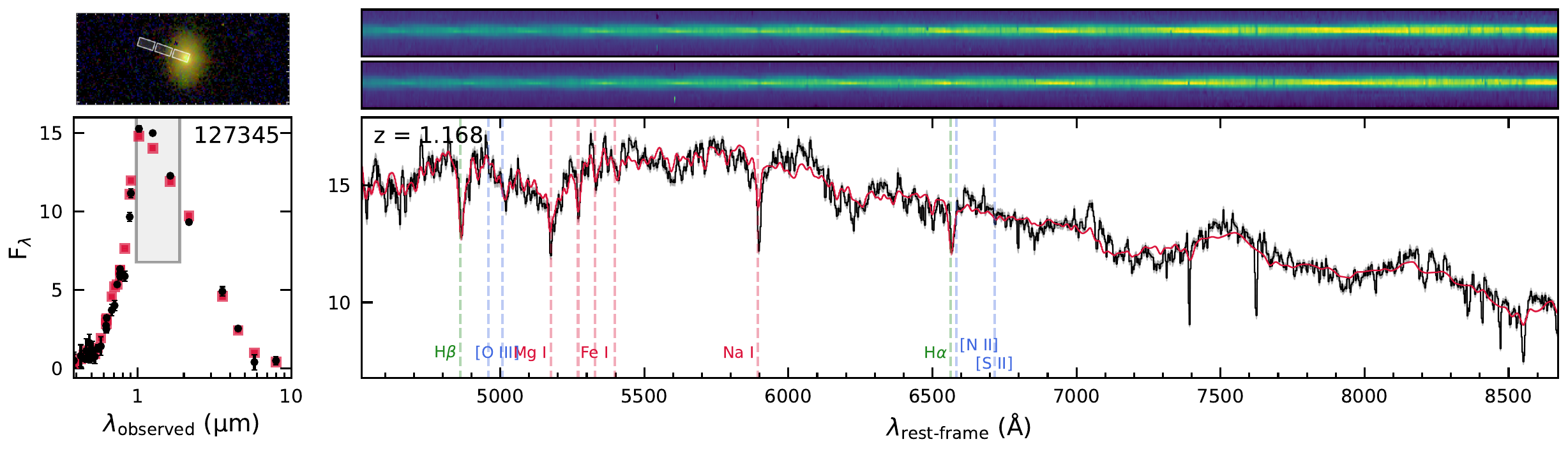}
\figsetgrpnote{Overview of UltraVISTA photometric SEDs (bottom-left), NIRSpec spectra (right), and HST image (top-left) of two example quiescent galaxies in our survey. The 2D spectra for the two observed dithers and the combined 1D spectrum are shown in the top-right and bottom-right rows, respectively. Flux densities ($F_{\lambda}$) are in $10^{-19}$\,erg\,s$^{-1}$\,cm$^{-2}$\,\AA$^{-1}$. The coverage of the NIRSpec spectra are indicated in the SED panels by the gray rectangles. The best-fit \texttt{Prospector} models to the photometry and spectra are shown in red (see Section \ref{sec:prosp}). The images in the top-left panels show the HST-F160W imaging, with an overlay of the NIRSpec MSA slit orientation for one nod position.}
\figsetgrpend

\figsetgrpstart
\figsetgrpnum{4.4}
\figsetgrptitle{Spectrum of target 130040}
\figsetplot{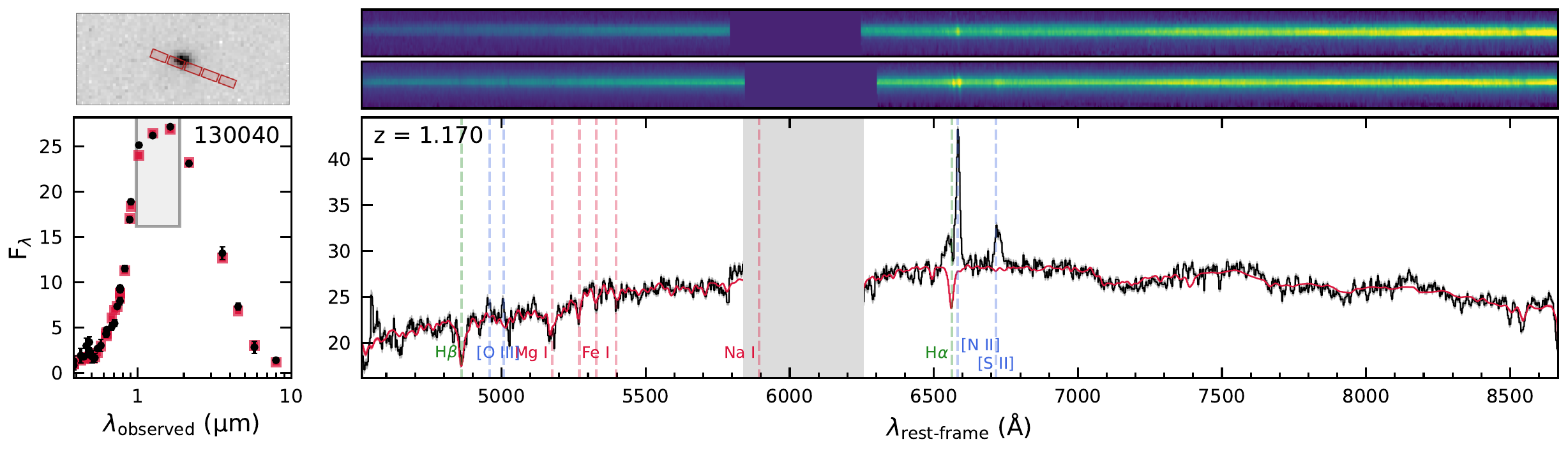}
\figsetgrpnote{Overview of UltraVISTA photometric SEDs (bottom-left), NIRSpec spectra (right), and HST image (top-left) of two example quiescent galaxies in our survey. The 2D spectra for the two observed dithers and the combined 1D spectrum are shown in the top-right and bottom-right rows, respectively. Flux densities ($F_{\lambda}$) are in $10^{-19}$\,erg\,s$^{-1}$\,cm$^{-2}$\,\AA$^{-1}$. The coverage of the NIRSpec spectra are indicated in the SED panels by the gray rectangles. The best-fit \texttt{Prospector} models to the photometry and spectra are shown in red (see Section \ref{sec:prosp}). The images in the top-left panels show the HST-F160W imaging, with an overlay of the NIRSpec MSA slit orientation for one nod position.}
\figsetgrpend

\figsetgrpstart
\figsetgrpnum{4.5}
\figsetgrptitle{Spectrum of target 128452}
\figsetplot{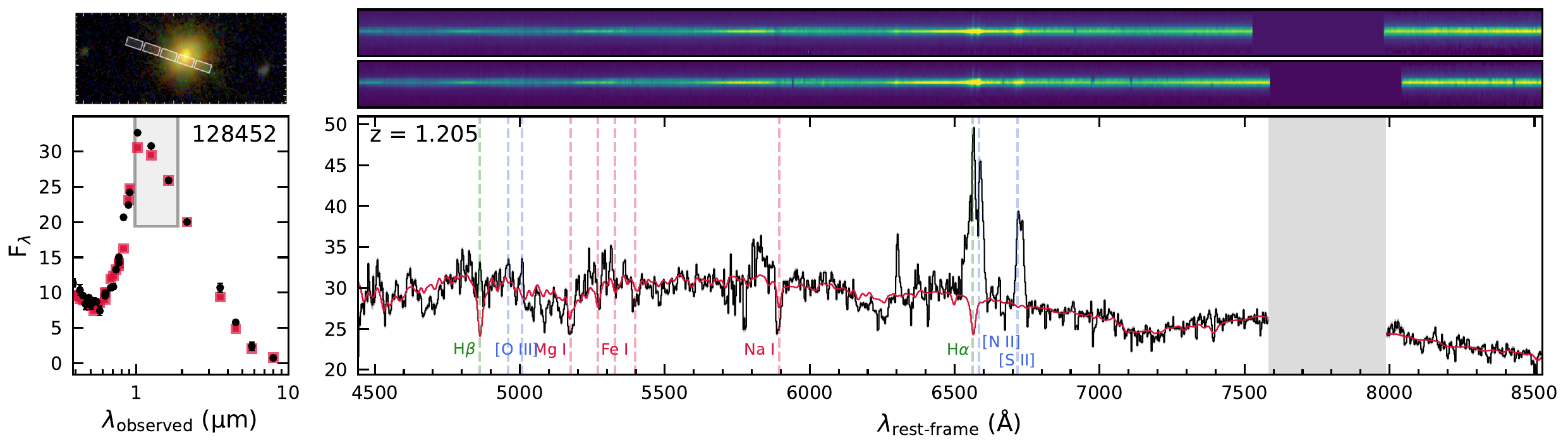}
\figsetgrpnote{Overview of UltraVISTA photometric SEDs (bottom-left), NIRSpec spectra (right), and HST image (top-left) of two example quiescent galaxies in our survey. The 2D spectra for the two observed dithers and the combined 1D spectrum are shown in the top-right and bottom-right rows, respectively. Flux densities ($F_{\lambda}$) are in $10^{-19}$\,erg\,s$^{-1}$\,cm$^{-2}$\,\AA$^{-1}$. The coverage of the NIRSpec spectra are indicated in the SED panels by the gray rectangles. The best-fit \texttt{Prospector} models to the photometry and spectra are shown in red (see Section \ref{sec:prosp}). The images in the top-left panels show the HST-F160W imaging, with an overlay of the NIRSpec MSA slit orientation for one nod position.}
\figsetgrpend

\figsetgrpstart
\figsetgrpnum{4.6}
\figsetgrptitle{Spectrum of target 127154}
\figsetplot{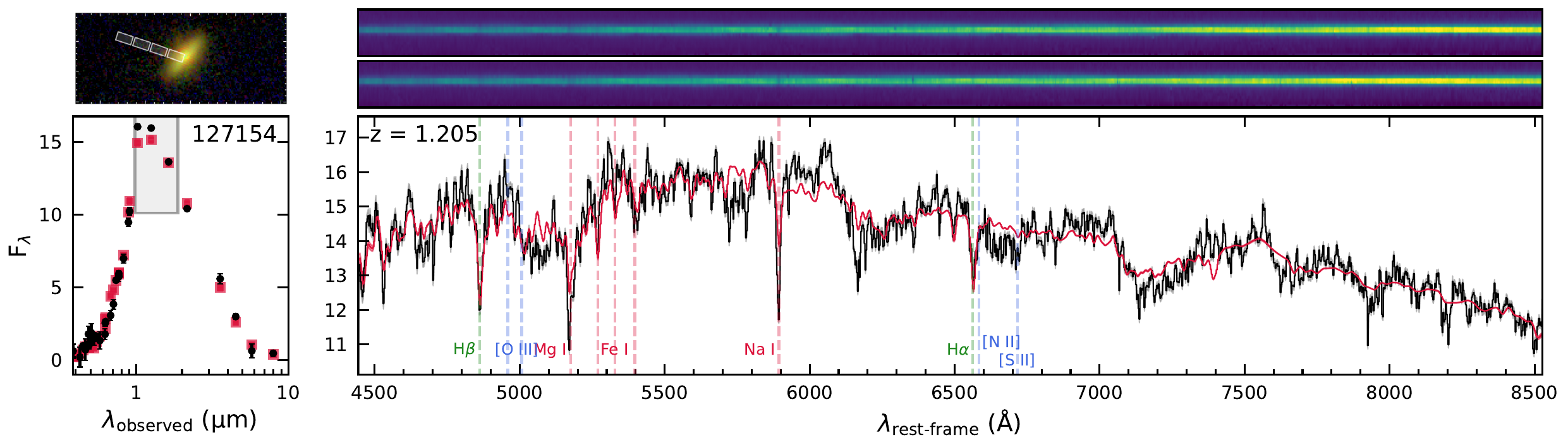}
\figsetgrpnote{Overview of UltraVISTA photometric SEDs (bottom-left), NIRSpec spectra (right), and HST image (top-left) of two example quiescent galaxies in our survey. The 2D spectra for the two observed dithers and the combined 1D spectrum are shown in the top-right and bottom-right rows, respectively. Flux densities ($F_{\lambda}$) are in $10^{-19}$\,erg\,s$^{-1}$\,cm$^{-2}$\,\AA$^{-1}$. The coverage of the NIRSpec spectra are indicated in the SED panels by the gray rectangles. The best-fit \texttt{Prospector} models to the photometry and spectra are shown in red (see Section \ref{sec:prosp}). The images in the top-left panels show the HST-F160W imaging, with an overlay of the NIRSpec MSA slit orientation for one nod position.}
\figsetgrpend

\figsetgrpstart
\figsetgrpnum{4.7}
\figsetgrptitle{Spectrum of target 130208}
\figsetplot{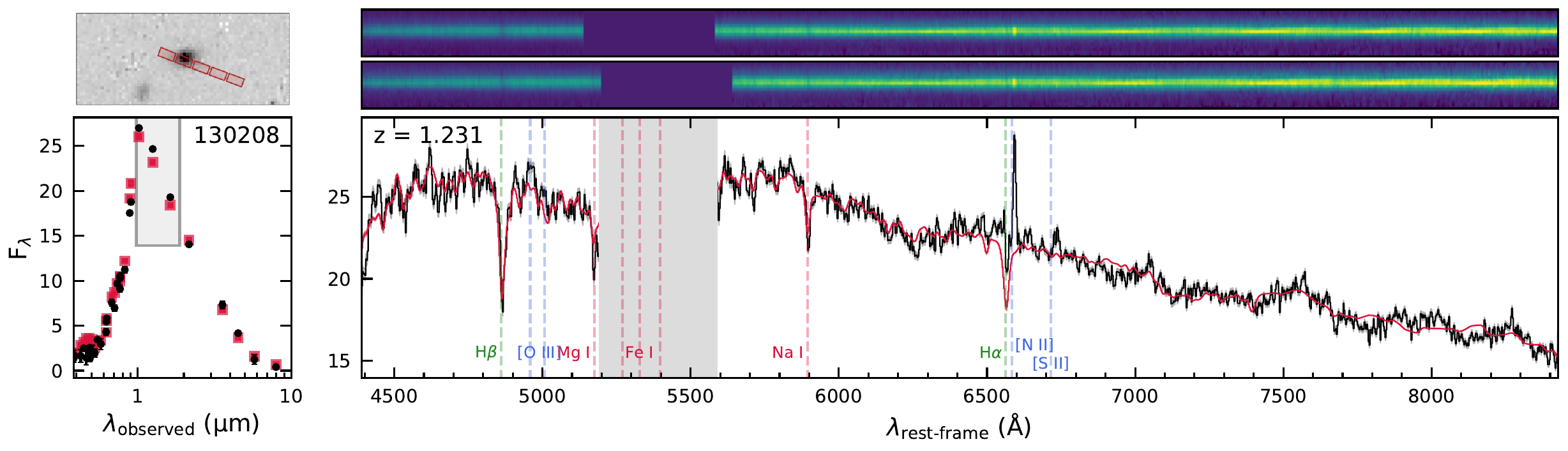}
\figsetgrpnote{Overview of UltraVISTA photometric SEDs (bottom-left), NIRSpec spectra (right), and HST image (top-left) of two example quiescent galaxies in our survey. The 2D spectra for the two observed dithers and the combined 1D spectrum are shown in the top-right and bottom-right rows, respectively. Flux densities ($F_{\lambda}$) are in $10^{-19}$\,erg\,s$^{-1}$\,cm$^{-2}$\,\AA$^{-1}$. The coverage of the NIRSpec spectra are indicated in the SED panels by the gray rectangles. The best-fit \texttt{Prospector} models to the photometry and spectra are shown in red (see Section \ref{sec:prosp}). The images in the top-left panels show the HST-F160W imaging, with an overlay of the NIRSpec MSA slit orientation for one nod position.}
\figsetgrpend

\figsetgrpstart
\figsetgrpnum{4.8}
\figsetgrptitle{Spectrum of target 129982}
\figsetplot{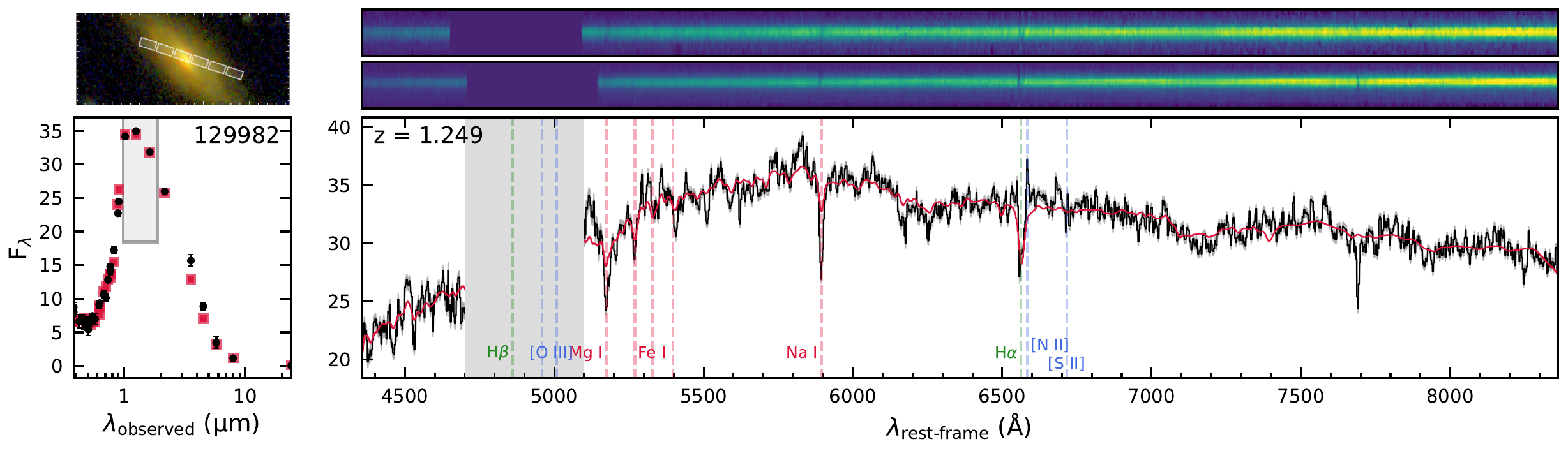}
\figsetgrpnote{Overview of UltraVISTA photometric SEDs (bottom-left), NIRSpec spectra (right), and HST image (top-left) of two example quiescent galaxies in our survey. The 2D spectra for the two observed dithers and the combined 1D spectrum are shown in the top-right and bottom-right rows, respectively. Flux densities ($F_{\lambda}$) are in $10^{-19}$\,erg\,s$^{-1}$\,cm$^{-2}$\,\AA$^{-1}$. The coverage of the NIRSpec spectra are indicated in the SED panels by the gray rectangles. The best-fit \texttt{Prospector} models to the photometry and spectra are shown in red (see Section \ref{sec:prosp}). The images in the top-left panels show the HST-F160W imaging, with an overlay of the NIRSpec MSA slit orientation for one nod position.}
\figsetgrpend

\figsetgrpstart
\figsetgrpnum{4.9}
\figsetgrptitle{Spectrum of target 127108}
\figsetplot{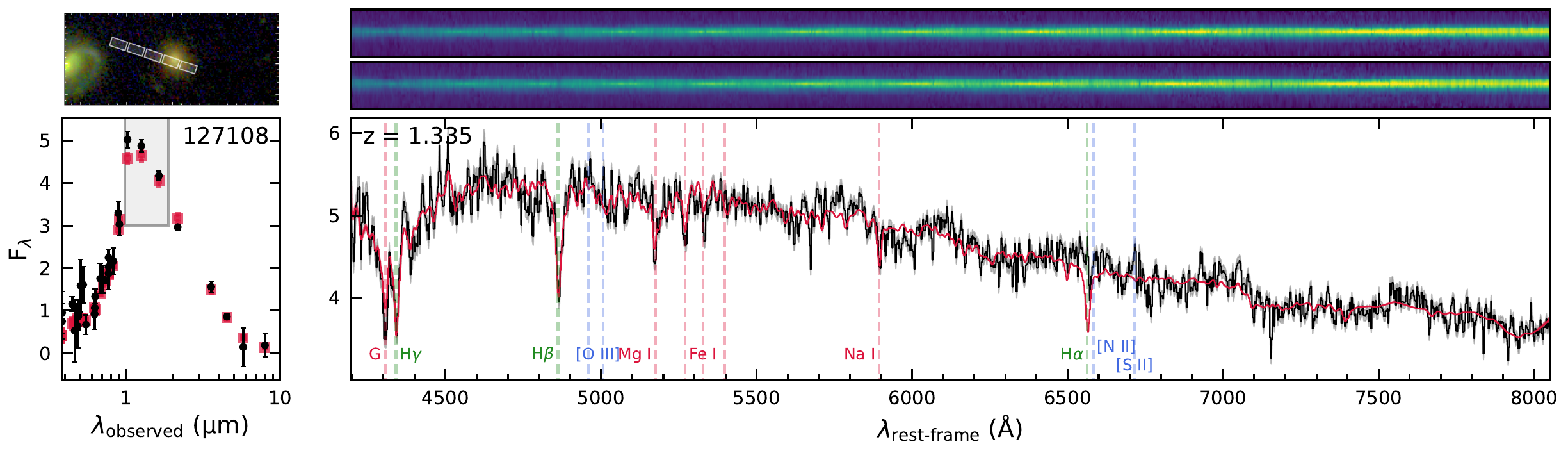}
\figsetgrpnote{Overview of UltraVISTA photometric SEDs (bottom-left), NIRSpec spectra (right), and HST image (top-left) of two example quiescent galaxies in our survey. The 2D spectra for the two observed dithers and the combined 1D spectrum are shown in the top-right and bottom-right rows, respectively. Flux densities ($F_{\lambda}$) are in $10^{-19}$\,erg\,s$^{-1}$\,cm$^{-2}$\,\AA$^{-1}$. The coverage of the NIRSpec spectra are indicated in the SED panels by the gray rectangles. The best-fit \texttt{Prospector} models to the photometry and spectra are shown in red (see Section \ref{sec:prosp}). The images in the top-left panels show the HST-F160W imaging, with an overlay of the NIRSpec MSA slit orientation for one nod position.}
\figsetgrpend

\figsetgrpstart
\figsetgrpnum{4.10}
\figsetgrptitle{Spectrum of target 129197}
\figsetplot{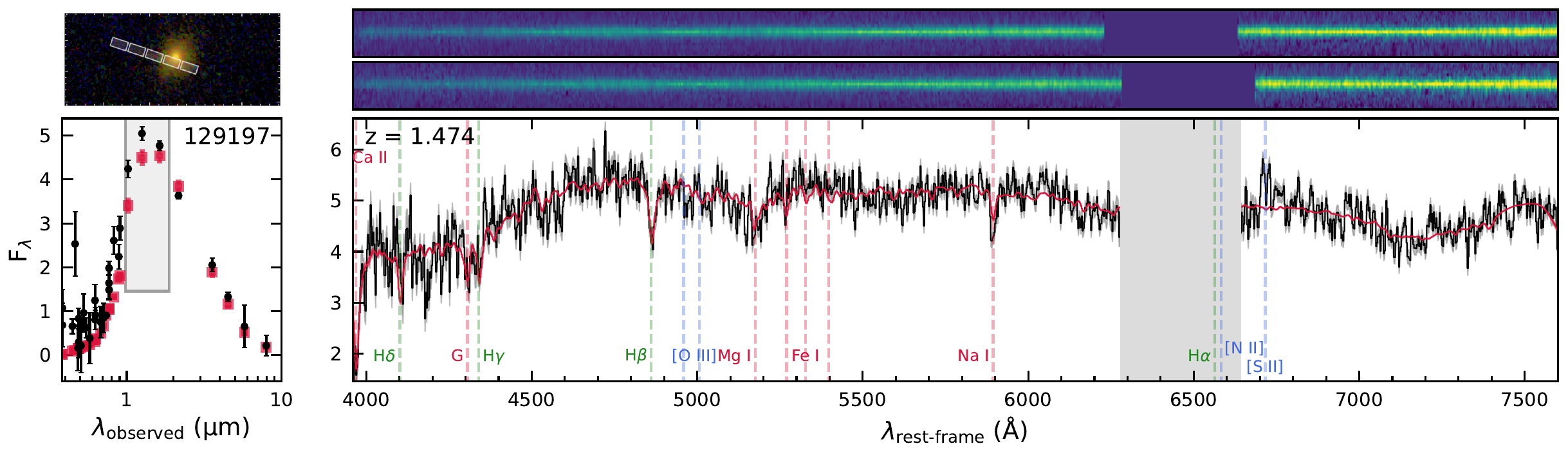}
\figsetgrpnote{Overview of UltraVISTA photometric SEDs (bottom-left), NIRSpec spectra (right), and HST image (top-left) of two example quiescent galaxies in our survey. The 2D spectra for the two observed dithers and the combined 1D spectrum are shown in the top-right and bottom-right rows, respectively. Flux densities ($F_{\lambda}$) are in $10^{-19}$\,erg\,s$^{-1}$\,cm$^{-2}$\,\AA$^{-1}$. The coverage of the NIRSpec spectra are indicated in the SED panels by the gray rectangles. The best-fit \texttt{Prospector} models to the photometry and spectra are shown in red (see Section \ref{sec:prosp}). The images in the top-left panels show the HST-F160W imaging, with an overlay of the NIRSpec MSA slit orientation for one nod position.}
\figsetgrpend

\figsetgrpstart
\figsetgrpnum{4.11}
\figsetgrptitle{Spectrum of target 130647}
\figsetplot{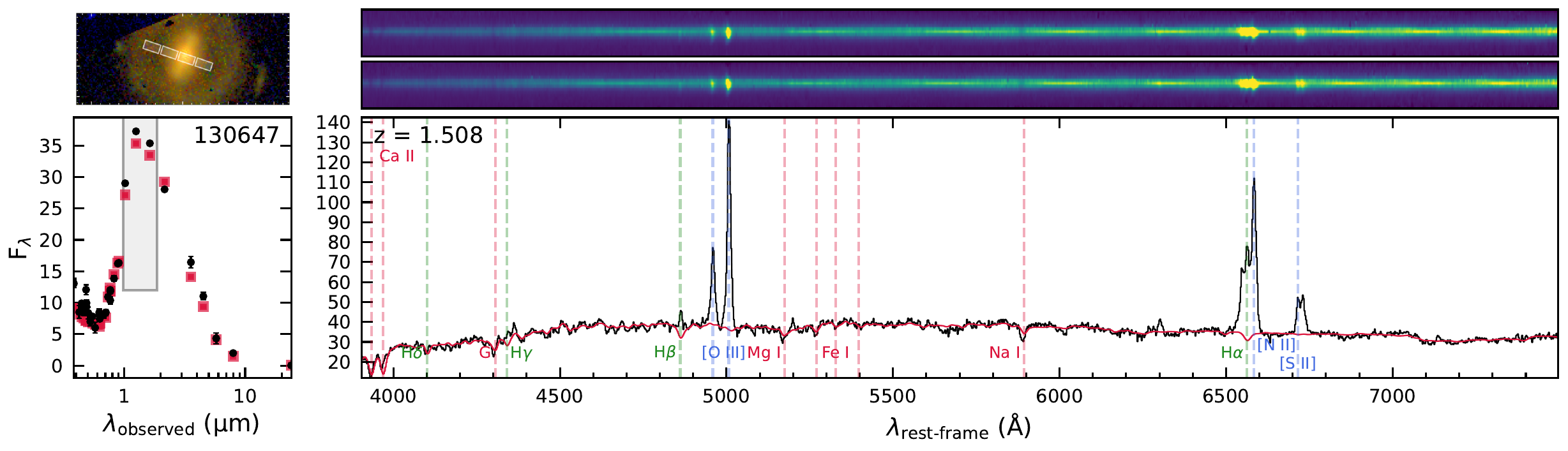}
\figsetgrpnote{Overview of UltraVISTA photometric SEDs (bottom-left), NIRSpec spectra (right), and HST image (top-left) of two example quiescent galaxies in our survey. The 2D spectra for the two observed dithers and the combined 1D spectrum are shown in the top-right and bottom-right rows, respectively. Flux densities ($F_{\lambda}$) are in $10^{-19}$\,erg\,s$^{-1}$\,cm$^{-2}$\,\AA$^{-1}$. The coverage of the NIRSpec spectra are indicated in the SED panels by the gray rectangles. The best-fit \texttt{Prospector} models to the photometry and spectra are shown in red (see Section \ref{sec:prosp}). The images in the top-left panels show the HST-F160W imaging, with an overlay of the NIRSpec MSA slit orientation for one nod position.}
\figsetgrpend

\figsetgrpstart
\figsetgrpnum{4.12}
\figsetgrptitle{Spectrum of target 130934}
\figsetplot{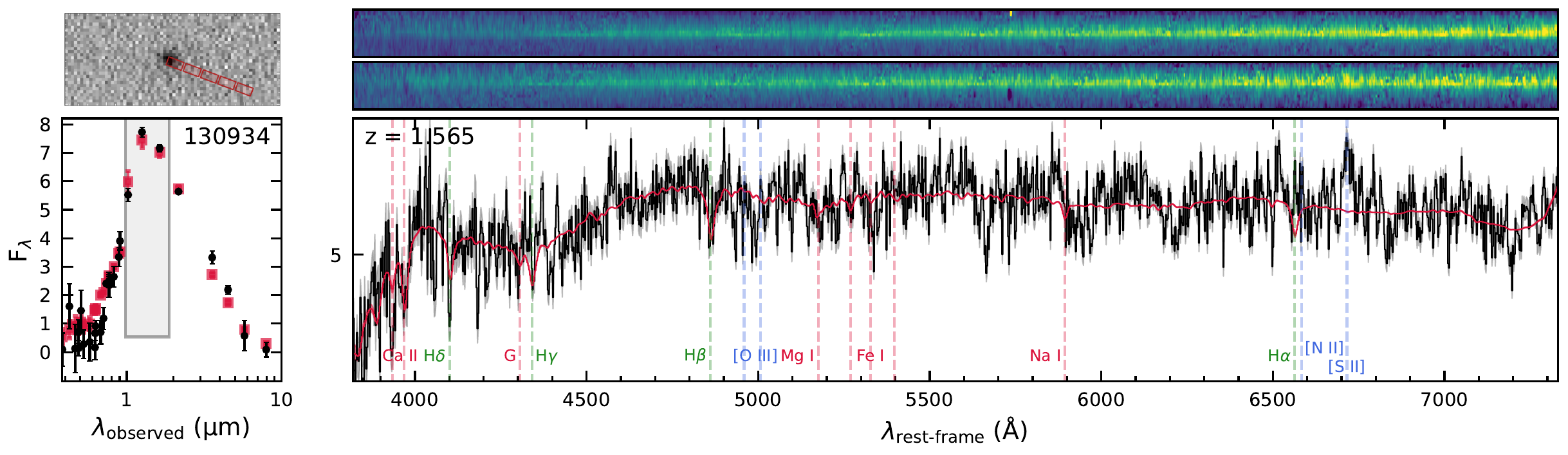}
\figsetgrpnote{Overview of UltraVISTA photometric SEDs (bottom-left), NIRSpec spectra (right), and HST image (top-left) of two example quiescent galaxies in our survey. The 2D spectra for the two observed dithers and the combined 1D spectrum are shown in the top-right and bottom-right rows, respectively. Flux densities ($F_{\lambda}$) are in $10^{-19}$\,erg\,s$^{-1}$\,cm$^{-2}$\,\AA$^{-1}$. The coverage of the NIRSpec spectra are indicated in the SED panels by the gray rectangles. The best-fit \texttt{Prospector} models to the photometry and spectra are shown in red (see Section \ref{sec:prosp}). The images in the top-left panels show the HST-F160W imaging, with an overlay of the NIRSpec MSA slit orientation for one nod position.}
\figsetgrpend

\figsetgrpstart
\figsetgrpnum{4.13}
\figsetgrptitle{Spectrum of target 130183}
\figsetplot{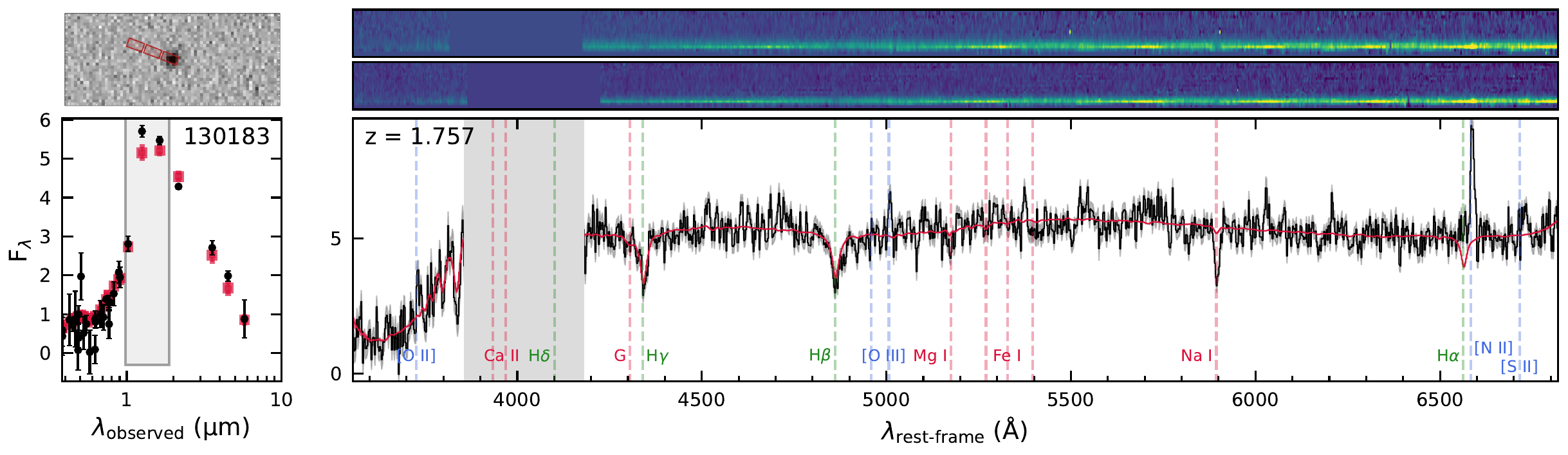}
\figsetgrpnote{Overview of UltraVISTA photometric SEDs (bottom-left), NIRSpec spectra (right), and HST image (top-left) of two example quiescent galaxies in our survey. The 2D spectra for the two observed dithers and the combined 1D spectrum are shown in the top-right and bottom-right rows, respectively. Flux densities ($F_{\lambda}$) are in $10^{-19}$\,erg\,s$^{-1}$\,cm$^{-2}$\,\AA$^{-1}$. The coverage of the NIRSpec spectra are indicated in the SED panels by the gray rectangles. The best-fit \texttt{Prospector} models to the photometry and spectra are shown in red (see Section \ref{sec:prosp}). The images in the top-left panels show the HST-F160W imaging, with an overlay of the NIRSpec MSA slit orientation for one nod position.}
\figsetgrpend

\figsetgrpstart
\figsetgrpnum{4.14}
\figsetgrptitle{Spectrum of target 128041}
\figsetplot{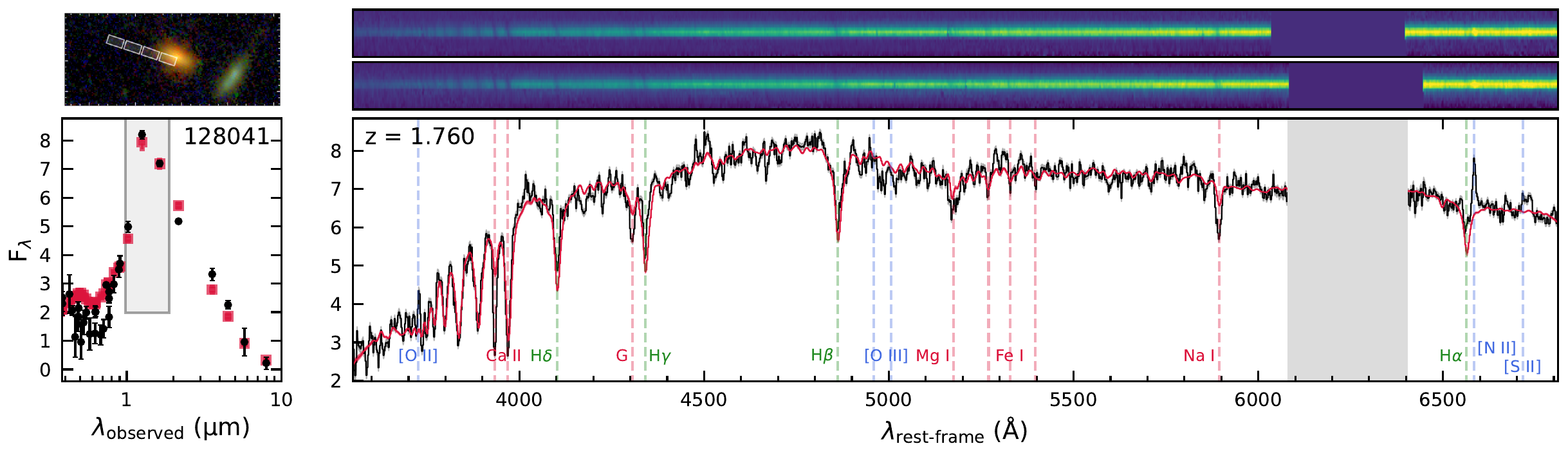}
\figsetgrpnote{Overview of UltraVISTA photometric SEDs (bottom-left), NIRSpec spectra (right), and HST image (top-left) of two example quiescent galaxies in our survey. The 2D spectra for the two observed dithers and the combined 1D spectrum are shown in the top-right and bottom-right rows, respectively. Flux densities ($F_{\lambda}$) are in $10^{-19}$\,erg\,s$^{-1}$\,cm$^{-2}$\,\AA$^{-1}$. The coverage of the NIRSpec spectra are indicated in the SED panels by the gray rectangles. The best-fit \texttt{Prospector} models to the photometry and spectra are shown in red (see Section \ref{sec:prosp}). The images in the top-left panels show the HST-F160W imaging, with an overlay of the NIRSpec MSA slit orientation for one nod position.}
\figsetgrpend

\figsetgrpstart
\figsetgrpnum{4.15}
\figsetgrptitle{Spectrum of target 127700}
\figsetplot{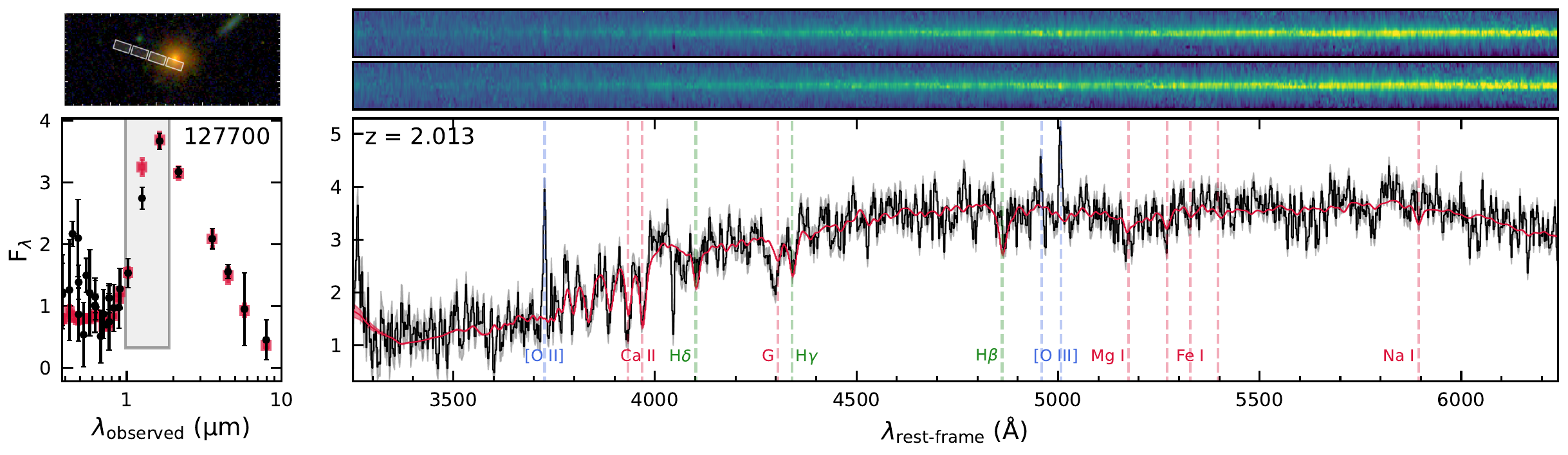}
\figsetgrpnote{Overview of UltraVISTA photometric SEDs (bottom-left), NIRSpec spectra (right), and HST image (top-left) of two example quiescent galaxies in our survey. The 2D spectra for the two observed dithers and the combined 1D spectrum are shown in the top-right and bottom-right rows, respectively. Flux densities ($F_{\lambda}$) are in $10^{-19}$\,erg\,s$^{-1}$\,cm$^{-2}$\,\AA$^{-1}$. The coverage of the NIRSpec spectra are indicated in the SED panels by the gray rectangles. The best-fit \texttt{Prospector} models to the photometry and spectra are shown in red (see Section \ref{sec:prosp}). The images in the top-left panels show the HST-F160W imaging, with an overlay of the NIRSpec MSA slit orientation for one nod position.}
\figsetgrpend

\figsetgrpstart
\figsetgrpnum{4.16}
\figsetgrptitle{Spectrum of target 129133}
\figsetplot{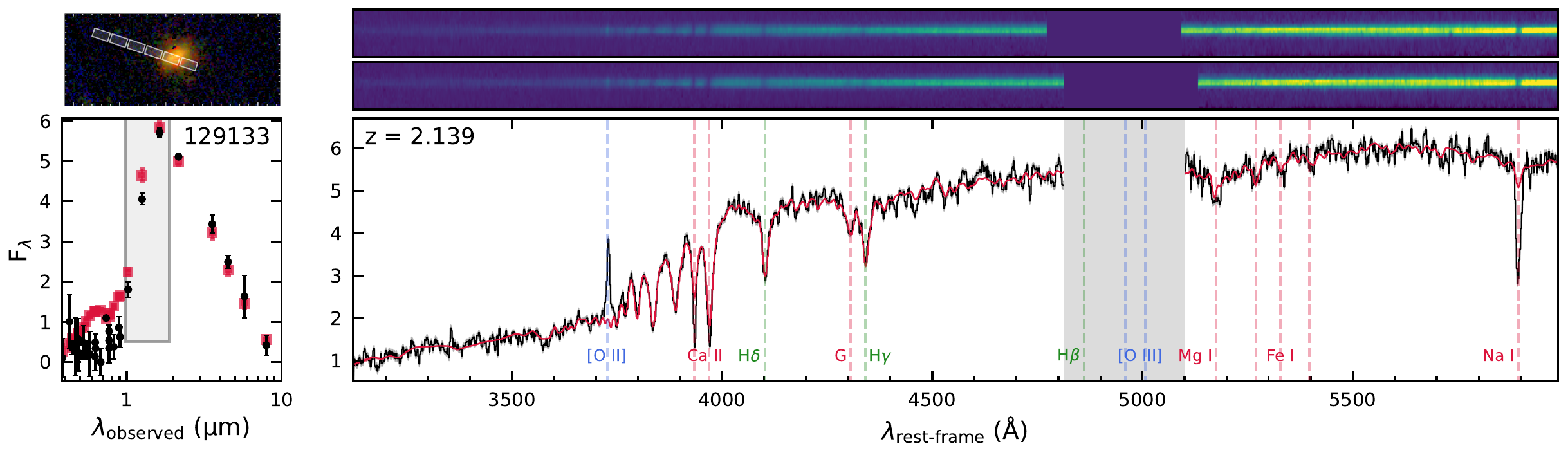}
\figsetgrpnote{Overview of UltraVISTA photometric SEDs (bottom-left), NIRSpec spectra (right), and HST image (top-left) of two example quiescent galaxies in our survey. The 2D spectra for the two observed dithers and the combined 1D spectrum are shown in the top-right and bottom-right rows, respectively. Flux densities ($F_{\lambda}$) are in $10^{-19}$\,erg\,s$^{-1}$\,cm$^{-2}$\,\AA$^{-1}$. The coverage of the NIRSpec spectra are indicated in the SED panels by the gray rectangles. The best-fit \texttt{Prospector} models to the photometry and spectra are shown in red (see Section \ref{sec:prosp}). The images in the top-left panels show the HST-F160W imaging, with an overlay of the NIRSpec MSA slit orientation for one nod position.}
\figsetgrpend

\figsetgrpstart
\figsetgrpnum{4.17}
\figsetgrptitle{Spectrum of target 127941}
\figsetplot{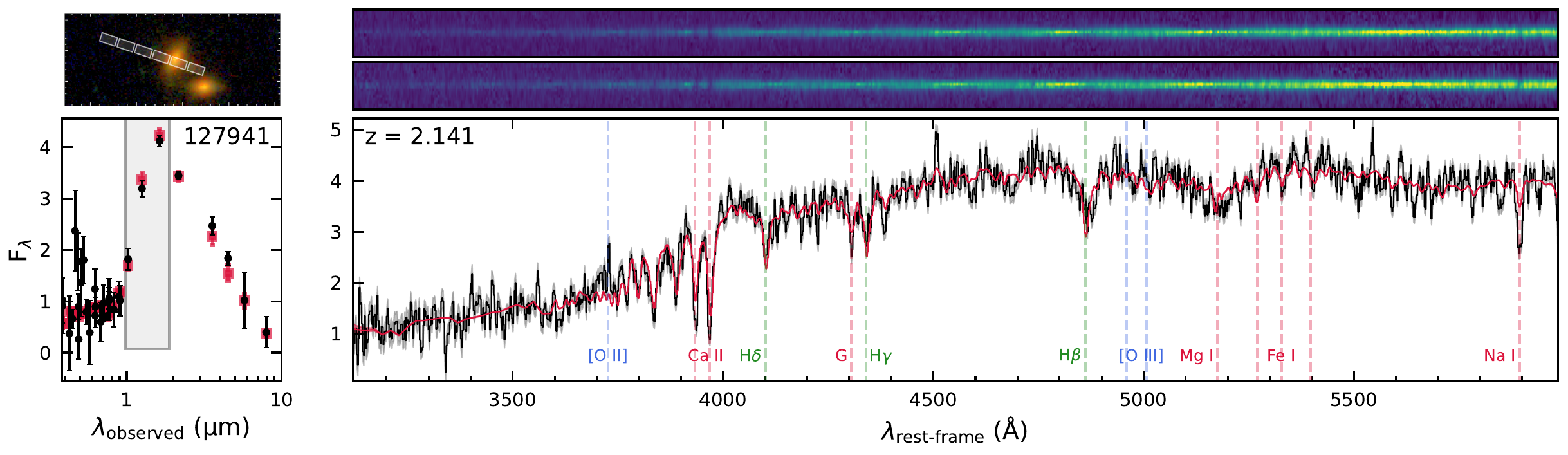}
\figsetgrpnote{Overview of UltraVISTA photometric SEDs (bottom-left), NIRSpec spectra (right), and HST image (top-left) of two example quiescent galaxies in our survey. The 2D spectra for the two observed dithers and the combined 1D spectrum are shown in the top-right and bottom-right rows, respectively. Flux densities ($F_{\lambda}$) are in $10^{-19}$\,erg\,s$^{-1}$\,cm$^{-2}$\,\AA$^{-1}$. The coverage of the NIRSpec spectra are indicated in the SED panels by the gray rectangles. The best-fit \texttt{Prospector} models to the photometry and spectra are shown in red (see Section \ref{sec:prosp}). The images in the top-left panels show the HST-F160W imaging, with an overlay of the NIRSpec MSA slit orientation for one nod position.}
\figsetgrpend

\figsetgrpstart
\figsetgrpnum{4.18}
\figsetgrptitle{Spectrum of target 128913}
\figsetplot{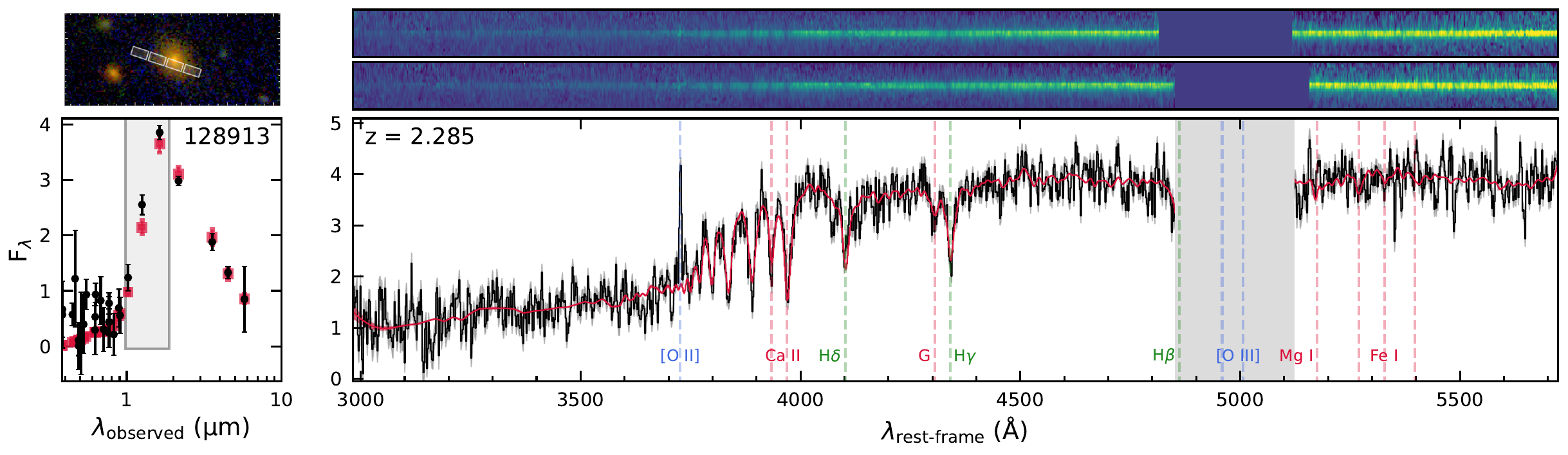}
\figsetgrpnote{Overview of UltraVISTA photometric SEDs (bottom-left), NIRSpec spectra (right), and HST image (top-left) of two example quiescent galaxies in our survey. The 2D spectra for the two observed dithers and the combined 1D spectrum are shown in the top-right and bottom-right rows, respectively. Flux densities ($F_{\lambda}$) are in $10^{-19}$\,erg\,s$^{-1}$\,cm$^{-2}$\,\AA$^{-1}$. The coverage of the NIRSpec spectra are indicated in the SED panels by the gray rectangles. The best-fit \texttt{Prospector} models to the photometry and spectra are shown in red (see Section \ref{sec:prosp}). The images in the top-left panels show the HST-F160W imaging, with an overlay of the NIRSpec MSA slit orientation for one nod position.}
\figsetgrpend

\figsetgrpstart
\figsetgrpnum{4.19}
\figsetgrptitle{Spectrum of target 130725}
\figsetplot{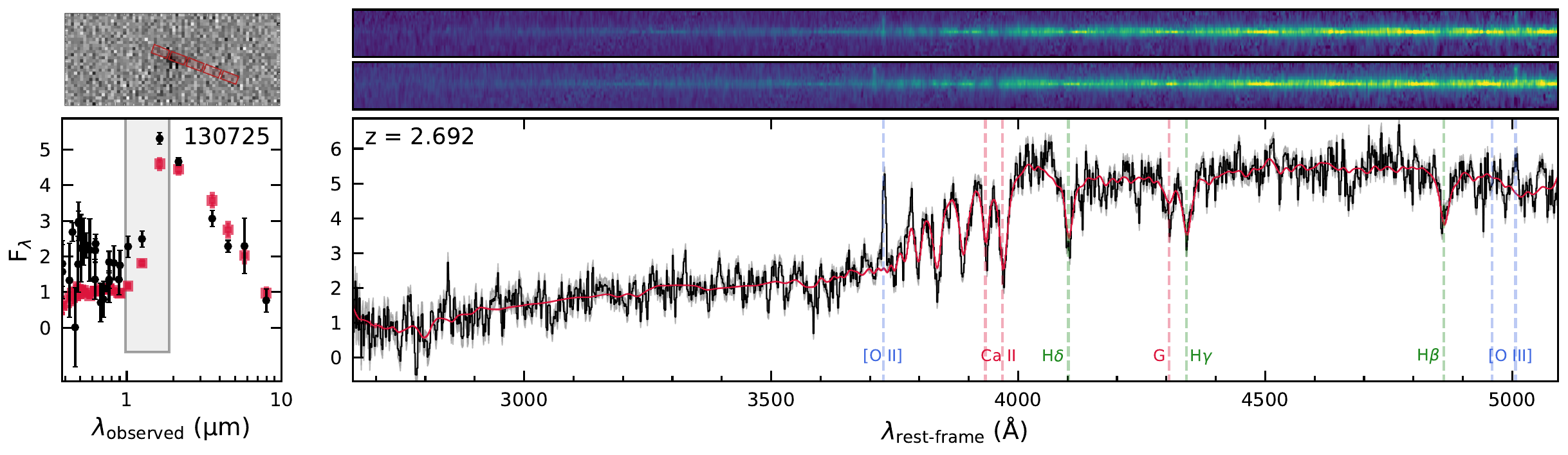}
\figsetgrpnote{Overview of UltraVISTA photometric SEDs (bottom-left), NIRSpec spectra (right), and HST image (top-left) of two example quiescent galaxies in our survey. The 2D spectra for the two observed dithers and the combined 1D spectrum are shown in the top-right and bottom-right rows, respectively. Flux densities ($F_{\lambda}$) are in $10^{-19}$\,erg\,s$^{-1}$\,cm$^{-2}$\,\AA$^{-1}$. The coverage of the NIRSpec spectra are indicated in the SED panels by the gray rectangles. The best-fit \texttt{Prospector} models to the photometry and spectra are shown in red (see Section \ref{sec:prosp}). The images in the top-left panels show the HST-F160W imaging, with an overlay of the NIRSpec MSA slit orientation for one nod position.}
\figsetgrpend

\figsetgrpstart
\figsetgrpnum{4.20}
\figsetgrptitle{Spectrum of target 129966}
\figsetplot{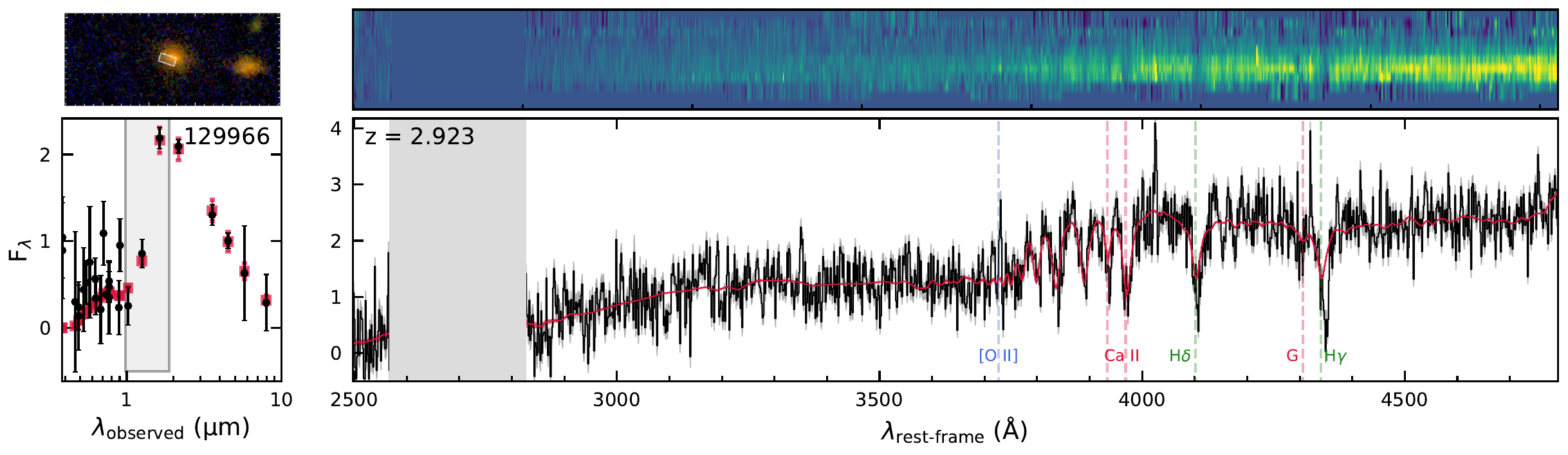}
\figsetgrpnote{Overview of UltraVISTA photometric SEDs (bottom-left), NIRSpec spectra (right), and HST image (top-left) of two example quiescent galaxies in our survey. The 2D spectra for the two observed dithers and the combined 1D spectrum are shown in the top-right and bottom-right rows, respectively. Flux densities ($F_{\lambda}$) are in $10^{-19}$\,erg\,s$^{-1}$\,cm$^{-2}$\,\AA$^{-1}$. The coverage of the NIRSpec spectra are indicated in the SED panels by the gray rectangles. The best-fit \texttt{Prospector} models to the photometry and spectra are shown in red (see Section \ref{sec:prosp}). The images in the top-left panels show the HST-F160W imaging, with an overlay of the NIRSpec MSA slit orientation for one nod position.}
\figsetgrpend
\figsetend
\begin{figure*}
\begin{center}
    \centering
    \includegraphics[width = 1.\textwidth]{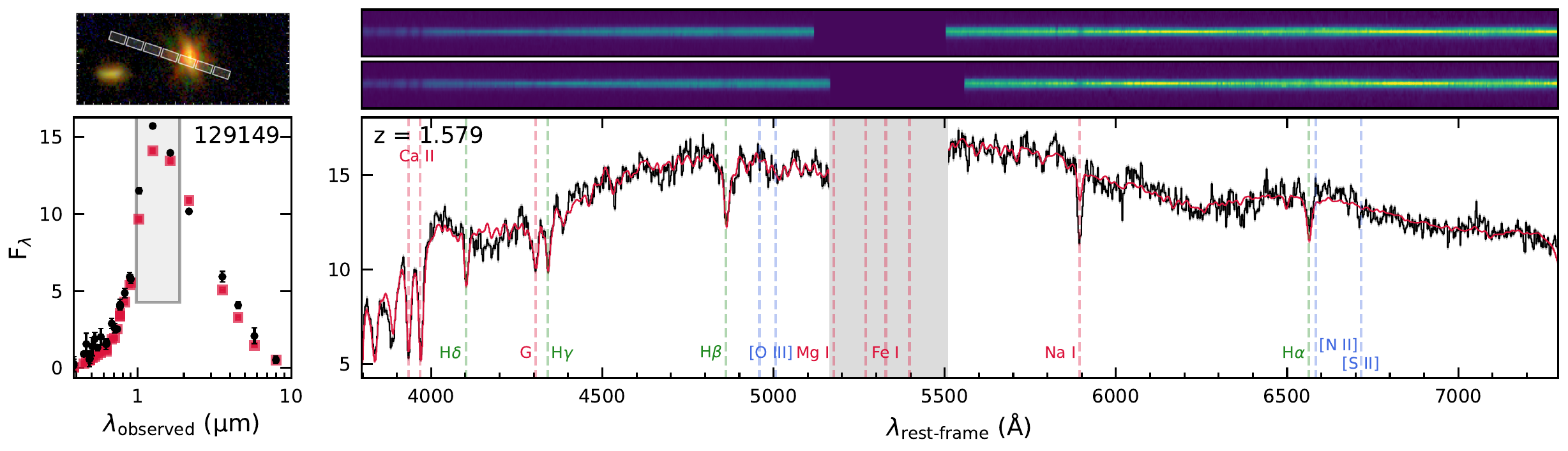}
    \includegraphics[width = 1.01\textwidth]{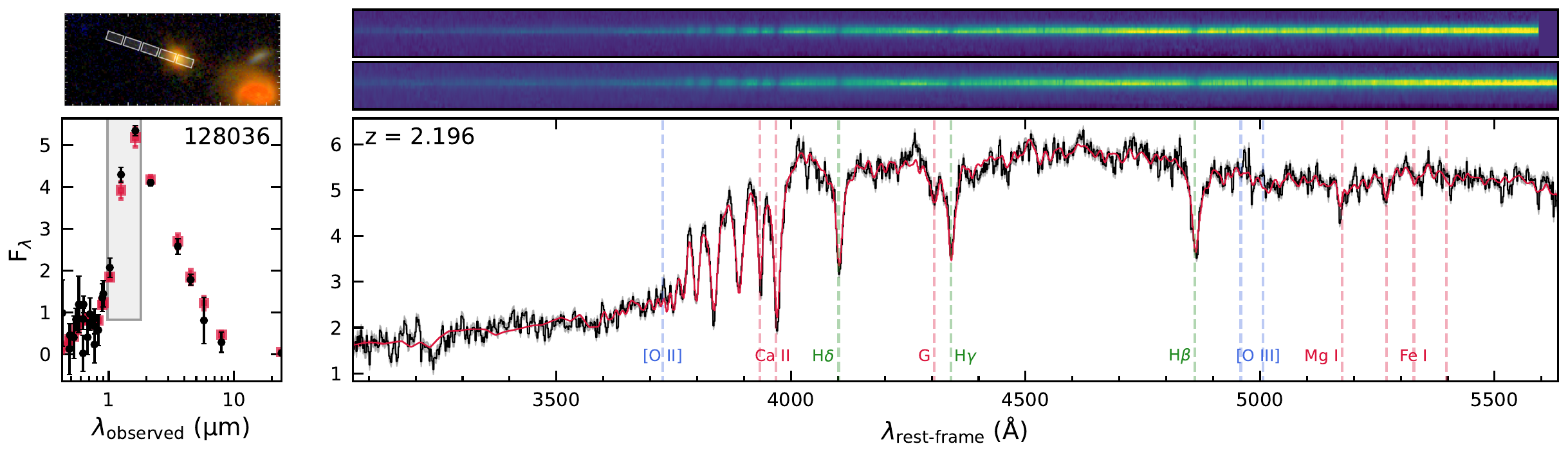}
    \caption{Overview of UltraVISTA photometric SEDs (bottom-left), NIRSpec spectra (right), and JWST COSMOS-Web RGB imaging \citep[top-left;][]{CCasey2023} of two example quiescent galaxies in our survey. The 2D spectra for the two observed dithers and the combined 1D spectrum are shown in the top-right and bottom-right rows, respectively. Flux densities ($F_{\lambda}$) are in $10^{-19}$\,erg\,s$^{-1}$\,cm$^{-2}$\,\AA$^{-1}$. The coverage of the NIRSpec spectra are indicated in the SED panels by the gray rectangles. The best-fit \texttt{Prospector} models to the photometry and spectra are shown in red (see Section \ref{sec:prosp}). The images in the top-left panels show the COSMOS-Web RGB imaging, constructed from the F115W, F277W and F444W filters \citep{CCasey2023}, with an overlay of the NIRSpec MSA slit orientation for one nod position. The spectra of all quiescent galaxies in the sample (20 images) are available as a figure set in the online journal.}
    \label{fig:example_q_spec}
\end{center}
\end{figure*}
We show an overview of the rest-frame normalized 1D spectra for all 20 quiescent targets in Figure \ref{fig:all_q_spec}, ordered by spectroscopic redshift (see Section \ref{sec:prosp}). In Figure \ref{fig:example_q_spec} we present the detailed observed-frame 1D spectra (bottom-right panels), 2D spectra for both dither positions (top-right panels), UltraVISTA photometry (bottom-left panels) and the F160W image from COSMOS-DASH \citep{LMowla2018} for two example galaxies. We also show the position of the NIRSpec MSA shutters for the first nod on the image. The spectra of all quiescent galaxies in the sample are available as a figure set in the online journal.

The rest-frame 1D spectra in Figure \ref{fig:all_q_spec} show that we observe multiple Balmer absorption lines (green dotted lines) for nearly all quiescent targets, and the Mg\,I absorption line (red dotted line) is observed for 16 of our targets. Two of our quiescent targets are at too low redshifts to capture the targeted Mg\,I line, and for two galaxies the line falls in the detector gap. For the twelve highest redshift targets we observe the 4000\,\AA\ break region, including the two Ca\,II lines (red). Additionally, we detect Na\,I, Fe\,I, and Carbon G-band absorption lines for the majority of targets. For some quiescent targets we also detect emission lines; for seven sources with rest-frame coverage below $3800$\,\AA\ we observe [O\,{\sc ii}]$\lambda\lambda$3727,3730 emission. Furthermore, for seven of the galaxies we also observe [N\,{\sc ii}]$\lambda\lambda$6550,6585, [O\,{\sc iii}]$\lambda\lambda$4960,5000, and/or Balmer emission lines. These emission lines in quiescent galaxies are thought to originate either from AGN or hot evolved stars \citep[e.g.][]{RYan2006,FBelfiore2016,MMaseda2021,SBelli2021}. Furthermore, the emission line ratios ([N\,II]/H$\alpha$, [O\,{\sc iii}]/H$\beta$) are also indicative of a non-star-forming origin. Thus, these lines likely originate from ionization by an AGN, shocks, or post-AGB stars \citep[e.g.][]{RYan2012}. A detailed investigation into the origin of these emission features will be the subject of a future study.

\begin{figure*}
    \centering
    \includegraphics[width = 0.75\textwidth]{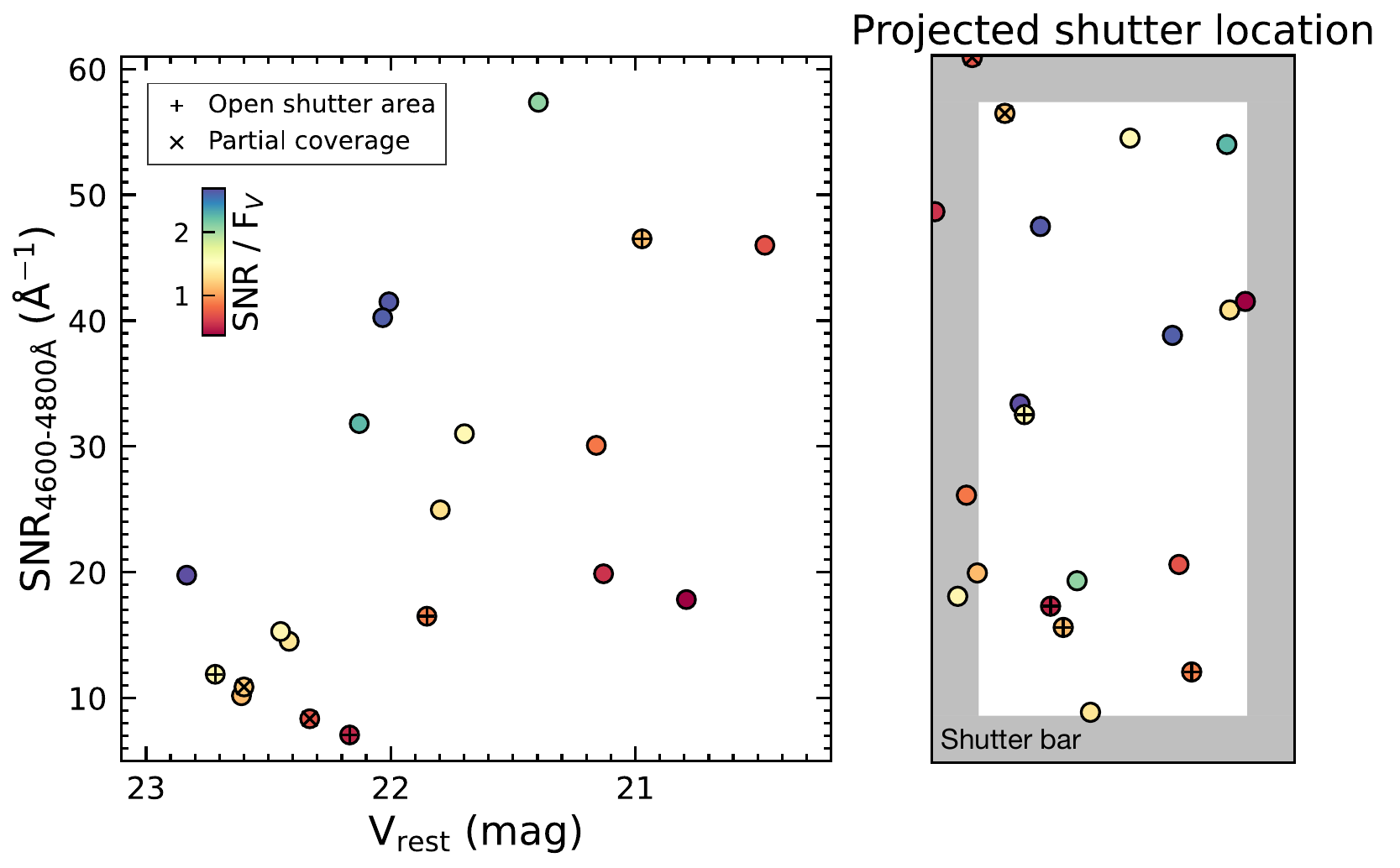}
    \caption{Median spectral signal-to-noise ratio (SNR) between rest-frame $4600-4800$\,\AA\ against rest-frame $V$-band magnitude (left panel), and projected position on the shutter (right panel) for all quiescent targets. We color the symbols in both panels by the ratio of the SNR to rest-frame $V$-band flux. Sources that were observed in an area of the MSA that is affected by failed open shutters are indicated with a plus, and crosses represent sources that were not covered by all four nod/dither configurations. The gray area in the right panel shows the area of the shutter that is covered by part of the bar separating two shutters.}
    \label{fig:SNR}
\end{figure*}
In Figure \ref{fig:SNR} we plot the median SNR per rest-frame \AA\ between $4600-4800$\,\AA\ as a function of rest-frame $V$-band magnitude. We use this spectral range to compute the SNR as all quiescent targets cover this wavelength region, and it captures no strong emission or absorption features that will bias the calculated SNR for individual galaxies. Overall, the targets that are brightest in the rest-frame $V$ band also have the highest SNRs, but for some targets the signal is lower than expected based on their magnitude. For the two sources that were not covered by all four nod/dither configurations (indicated with cross symbols), the lower SNR can be explained from their lower on-source integration time. Secondly, to include more quiescent targets in our sample we allowed objects to fall in areas of the detector that are affected by stray light from failed open shutters in the MSA. These targets are indicated with a plus symbol, and indeed have lower SNRs than sources of a comparable magnitude. 

Additionally, to further maximize our sample size we allowed sources to fall anywhere in a shutter when designing the mask, including behind the bars separating neighboring shutters. In the right panel of Figure \ref{fig:SNR} we show an MSA shutter with the projected positions of the central coordinates of our quiescent targets, plotted using the same symbols as in the left panel. 

From this panel we conclude that most of the sources with relatively low SNR lie at the edge of the shutters, or behind the shutter bars. For these less-optimally centered sources a large fraction of their light is not captured in the shutter, while the measured $V$-band flux is calculated using the galaxy's entire profile. However, since the typical size of quiescent galaxies in our redshift range at a rest-frame wavelength of $\sim$5000\,\AA\ is $\sim$2\,kpc \citep{SCutler2022}, our targets are sufficiently extended to reach SNRs $> 10$ for all sources (with full coverage and not affected by the open shutter areas) that are at the shutter edges. We also note that the difference in redshifts, sizes, Sersic indices, and alignment of our objects will increase the scatter in our SNRs, since the luminosity and observed fraction of each galaxy will differ. 

In summary, the observed SNRs show that our choices for designing the MSA mask affected the obtained flux levels of our spectra. Nonetheless, the final SNRs ($\gtrsim 10$) we obtain are sufficient for the goals of this survey. If we had been more conservative in the mask design by not allowing sources to fall behind the shutter bars and not using shutters affected by failed open shutters, our sample would have decreased from 20 to eleven quiescent targets. Overall, this shows that by allowing more flexibility in the mask-design we have greatly increased the scientific potential of the survey, while still achieving sufficiently high SNRs. However, we also note that for surveys that have less extended sources or require very high SNRs, it is necessary to be more conservative in the MSA design.

\section{Analysis}\label{sec:analysis}
\subsection{Redshifts and Stellar Population Modeling}\label{sec:prosp}
We derive redshifts and stellar population properties for all quiescent targets by fitting the 1D spectra for the quiescent targets in addition to their UltraVISTA DR3 photometry \citep{AMuzzin2013a} using the \texttt{Prospector} code \citep{JLeja2019, BJohnson2021}. \texttt{Prospector} uses the Flexible Stellar Population Synthesis \citep[FSPS;][]{CConroy2009, CConroy2010} library, the MILES spectral library \citep{JFalcon2011}, and the MIST isochrones \citep{JChoi2016, ADotter2016} to construct stellar population models. We used the \texttt{dynesty} dynamic nested sampling package \citep{JSpeagle2020} to sample the posterior distributions. We do not include the option in \texttt{Prospector} that fits for nebular emission lines in the spectrum, as the models used to fit for these emission lines are often not flexible enough to capture the observed line ratios due to the influence of, e.g., AGN, abundance ratios, or radiative transfer through dust. Instead, we masked out the spectrum within 50\,\AA\ of the [O\,{\sc ii}]$\lambda\lambda$3727,3730 doublet, and within 100\,\AA\ of the [O\,{\sc iii}]$\lambda\lambda$4960,5000, [N\,{\sc ii}]$\lambda\lambda$6550,6585, and [S\,{\sc ii}]$\lambda\lambda$6718, 6733 doublets. For galaxies with strong emission lines (127700, 128451, 130040, 130647) we also correct the photometry for the contribution of emission lines before fitting. We also found that the deep Na D absorption feature biased the best-fit metallicity in some galaxies in our sample. We thus also mask out the spectrum within 50\,\AA\ of this feature. We describe the free parameters and priors that we used in our modeling below, and summarize them in Table \ref{tab:prosp_priors}.

\begin{table*}[t]
\centering
\caption{Parameters and priors used in the \texttt{Prospector} fits\label{tab:prosp_priors}}
\begin{tabular}{p{0.12\textwidth} p{0.14\textwidth} p{0.3\textwidth} p{0.35\textwidth}}
\hline \hline
& Parameter & Description & Prior \\
\hline
Global    & log($M_{*}/M_{\odot}$) & Total stellar mass formed & Uniform: min = 9.5, max = 12.5\\[2pt]
\white{.} & log($Z_{*}/Z_{\odot}$) & Stellar metallicity & Uniform: min = -1.4, max = 0.19\\[2pt]
\white{.} & $\sigma$ & Stellar velocity dispersion & Uniform: min = 5, max = 600\,km\,s$^{-1}$\\[2pt]
\white{.} & $z$ & Redshift & Gaussian: $\mu$ = estimated $z_{\text{spec}}$, $\sigma$ = 0.05 \\[2pt]
Dust      & $\hat{\tau}_{\lambda,2}$ & Diffuse dust optical depth& Uniform: min = 0.0, max = 2.5 mag \\[2pt]
\white{.} & $\hat{\tau}_{\lambda,1}$ & Birth-cloud dust optical depth& Fixed to $\hat{\tau}_{\lambda,2}$ \\[2pt]
\white{.} & $n$ &Slope of Kriek \& Conroy dust law&Uniform: min = $-1.0$, max = 0.4\\[2pt]
SFH       & log\,SFR$_{\text{ratio}}$ & Ratio of SFR in adjacent bins & Student $t$ (8-vector): $\tau$ = 0.3, $\nu$ = 1\\[2pt]
SFH: PSB  & log\,SFR$_{\text{ratio,young}}$ & Ratio of SFR in youngest bin to the last flex bin & Student $t$: $\tau$ = 0.3, $\nu$ = 1\\[2pt]
\white{.} & log\,SFR$_{\text{ratio,old}}$ & Ratio of SFR in old bins to the first flex bin & Student $t$ (4-vector): $\tau$ = 0.3, $\nu$ = 1\\[2pt]
\white{.} & log\,SFR$_{\text{ratio}}$ & Ratio of SFR in adjacent flex bins &  Student $t$ (4-vector): $\tau$ = 0.3, $\nu$ = 1\\[2pt]
\white{.} & $t_{\text{last}}$ & Width of youngest bin & Uniform: min = 0.01\,Gyr, max = 0.3 $t_{\text{univ}}$\\[2pt]
Noise     & $j_{\text{spec}}$ & Spectroscopic jitter term & Uniform: min = 1, max = 5\\[2pt]
\white{.} & $f_{\text{out}}$ & Fraction of pixels in spectrum that are outliers & Uniform: min = 0.0, max = 0.5\\[2pt]
\white{.} & $s_{\text{out}}$ & Inflation of the nominal noise for outlier pixels & Fixed to 50\\[2pt]
\hline 
\multicolumn{4}{p{0.95\textwidth}}{\textbf{Note.} The Student-$t$ priors for the SFHs are centered at the UniverseMachine SFH predictions for a quiescent galaxy of similar mass and redshift, as described in the text.}
\end{tabular}
\end{table*}

\subsubsection{Star-Formation History Models}
We fit two models with different non-parametric SFHs to our data. The first SFH is the fixed-bin model described by \citet{JLeja2019_b}. In this SFH there are $N$ user-defined fixed time-bins, all of which have a constant SFR that is varied in the fit. In our model we used 9 time-bins, where the first two bins go from a lookback time of 0 -- 30\,Myr and 30 -- 100\,Myr, and the last bin covers 0.85\,$t_{\text{univ}}$ -- $t_{\text{univ}}$ at the redshift of the galaxy. The remaining 6 bins are spaced evenly in log lookback time. The second SFH we fit is the post-starburst (PSB) model described by \citet{KSuess2022a}. This SFH model allows for more flexibility in the SFH, enabling it to capture short, recent bursts of star formation and rapid quenching. This is achieved by fitting a SFH with 9 time-bins, where the 4 oldest ($t > 0.6\,t_{\text{univ}}$) bins have fixed bin-edges, while the bin-edges of the 5 youngest bins are varied in the fit. The 4 fixed bins have variable SFRs, whereas the 5 bins with flexible edges all form an equal amount of stellar mass. To capture low levels of star formation after the burst, the last bin has both a variable SFR and a variable bin edge.

For both SFH models the SFR in each bin is determined by fitting the log of the ratio of the SFR in adjacent bins (log\,SFR$_{\text{ratio}}$) using a Student-$t$ prior. Following \citet{KSuess2022b}, we used predicted SFHs for quiescent galaxies from the UniverseMachine public data \citep{PBehroozi2019} to set physically motivated priors. To this end, we first found the UniverseMachine predicted SFH for quiescent galaxies at the redshift of the galaxy to be fit, for a stellar mass of log\,$M_{*}/M_{\odot} = 10.8$, corresponding to the UniverseMachine mass-bin closest to the estimated stellar masses of galaxies in our sample. We then calculated the log\,SFR$_{\text{ratio}}$ that would produce this SFH, and used these values to set a Student-$t$ prior with a width of 0.3 dex and a degree of freedom of one.

For the fixed-bin SFH model we fit the spectroscopic redshift of the galaxy using a Gaussian prior. We estimated the redshift of each galaxy by hand first, and set this as the mean of the prior, and set the width to 0.05. The best-fit spectroscopic redshift of the fixed-bin model is then used as a fixed parameter in the PSB model. All other parameters in the fit are set up identically for the two models.

After fitting both SFH models we determined for each galaxy which model best fits the data based on the lowest reduced $\chi^2$ statistic. The difference in the $\chi^2$ value between the two best-fit models ranges between $\sim1-15\%$ for the galaxies in our sample. We note that for galaxies with a very small difference between the two residuals, the best-fit SFHs and stellar population parameters of the two models are nearly identical.

\subsubsection{General Model Set-Up}
We assumed the \citet{GChabrier2003} initial mass function, and added nebular continuum and dust emission using the standard \texttt{Prospector} parameters. In addition to the SFH shape, we also fit for the total stellar mass formed, metallicity, velocity dispersion, and dust attenuation. We set a flat prior on log\,$M_{*}/M_{\odot}$ between 9.5 and 12.5, and let log\,$Z / Z_{\odot}$ vary between $-1.4$ and $0.19$, corresponding to the metallicity limits of the MILES spectral library \citep{JFalcon2011}. The velocity dispersion was varied between 5 and 600\,km\,s$^{-1}$. The prior range for the velocity dispersion was chosen to be this extended in order to correct for uncertainties in the effective spectral resolution of NIRSpec/MSA, as we describe in more detail below. 

We use the \citet{MKriek2013} dust law, in which the UV dust bump and dust slope are correlated. For the dust model we set the optical depth and dust index as free parameters. We set a uniform prior on the dust index between $-1$ and 0.4 \citep{SNoll2009}. For the diffuse dust optical depth we used a flat prior from 0.0 to 2.5\,mag, and fixed the birth-cloud dust optical depth such that young stars are attenuated twice as much as old stars \citep{VWild2020}. 

\subsubsection{Noise Modeling and Calibration}
To ensure that bad pixels and calibration issues did not heavily affect our fits, we fit for noise and calibration parameters. We included the \texttt{Prospector} pixel outlier model that inflates the uncertainties of a fraction $f_{\text{outlier}}$ of pixels by a factor $s_{\text{outlier}}$. We set a uniform prior on $f_{\text{outlier}}$ between 0.0 and 0.5, and fix $s_{\text{outlier}}$ to 50. Secondly, we fit for an increase of the noise of the spectrum by including the noise model described in \citet{BJohnson2021}, with a free spectroscopic jitter term that we varied between 1 and 5. Lastly, to account for poor calibration of the spectra we include \texttt{Prospector}'s polynomial SED model, which optimizes out a polynomial in the model spectrum. We use a 12th order polynomial, and find that the average scale of the polynomial correction is $\sim5-10\%$.

\subsubsection{Line-Spread Function}
To include the wavelength-dependent spectral resolution of NIRSpec in the fit, we broadened the model spectra by the line-spread function (LSF) calculated from the instrumental resolution curve provided in JWST User Documentation (JDox)\footnote{\url{https://jwst-docs.stsci.edu/jwst-near-infrared-spectrograph/nirspecinstrumentation/nirspec-dispersers-and-filters}}. However, these resolution curves are pre-launch estimates for a uniformly illuminated slit, and recent results from e.g. \citet{AdeGraaff2023}, \citet{DNidever2024}, and \citet{TNanayakkara2024} indicate that the true spectral resolution is significantly higher. 

We model the true spectral resolution in the final rectified 2D frames for our sources based on their morphology and slit placement, and find that the true resolution is a factor $\sim1.22-1.38$ higher than the reported resolution (see Appendix \ref{ap:LSF}). This factor is independent of wavelength. In our modeling we thus multiply the JDox resolution curve by the scaling factor we find for each galaxy. For galaxies for which no morphological measurements are available (6 out of 20 quiescent targets), we cannot model the exact LSF and instead assume a conservative factor of 1.2, corresponding to our modeled LSF of a uniformly illuminated slit.\\

We show the resulting best-fit model to the spectra and photometry for two representative quiescent galaxies in Figure \ref{fig:example_q_spec}. The spectra of all quiescent galaxies in the sample (20 images) are available as a figure set in the online journal, and we report the best-fit stellar population parameters in Table \ref{tab:sample}. The stellar masses we report are the surviving stellar masses, converted from the total stellar mass formed using the surviving mass fraction obtained in the \texttt{Prospector} fit.\\

For the star-forming galaxy targets we derive spectroscopic redshifts by fitting Gaussians to the observed emission lines (see Appendix \ref{ap:SF_gals}). Multiple emission lines were detected for most filler galaxies, yielding robust spectroscopic redshifts for 46 out of 53 star-forming filler galaxies. For a few of the higher-redshift galaxies, only one emission feature was observed ([O\,{\sc ii}]). Nonetheless, due to the consistency between the spectroscopic and photometric redshift, we were confident about the line identification. For seven star-forming filler galaxies no spectroscopic features were detected; these galaxies were not included in Appendix \ref{ap:SF_gals} and omitted from the rest of this study. Though the redshifts for all star-forming galaxies have been derived from emission lines, we note that we detect continuum emission and absorption lines for many star-forming galaxies as well.

\subsection{Evaluation of Best-Fit SPS Models}
For all galaxies the fit to the spectrum and photometry looks reasonable (see Figure set \ref{fig:example_q_spec}). However, for some spectral regions the models do not properly capture the spectrum due to the fact that the resolution of the MILES stellar library varies with wavelength; the resolution is high between $\sim$3750--7200\,\AA, but outside this range the resolution is lower than the spectral resolution of our data. For targets with $z \gtrsim 2$ this affects the fits to the spectra at the blue end, where the models follow the overall shape of the observed spectra, but cannot capture details. At the upper wavelength limit the low-resolution models affect galaxies in our sample with redshifts below $z \sim 1.5$. The spectra of these sources cover the rest-frame spectral region of $\lambda \gtrsim 6200$\AA\ where TiO and CN absorption bands are present. These absorption bands are characteristic of evolved, carbon and oxygen-rich stars \citep[e.g.][]{SLu2024}, and are generally only visible for evolved stellar populations where young, blue stars do not dominate over this population \citep[e.g.][]{FAllard2000, JSánchez2012}. Because of the low spectral resolution of the models, all galaxies in our sample with TiO absorption bands in the spectra have best-fit models that fit this spectral region poorly, both in overall spectral shape and in specific absorption features.

Constraining spectral models of distant galaxies outside the $3750-7200$\,\AA\ wavelength range was very challenging in the pre-JWST era, since high-resolution ground-based observations have strong telluric absorption bands at these wavelengths. The spectra we present in this work show that JWST observations will provide the medium-resolution spectra needed to calibrate models blue and red-ward of the current spectral coverage. Furthermore, these spectra can also be used to improve models for the influence of evolved stars in stellar populations with ages around $\sim1$\,Gyr \citep[e.g.,][]{CMaraston2005, MKriek2010, SZibetti2013, SLu2024}.

We also note that for some galaxies strong absorption lines (e.g. Mg\,I and Na\,I in galaxy 129149) are not captured by the best-fit model. One possible explanation for this discrepancy between the observed and model spectra is the fact that we fit the spectrum and photometry simultaneously. Currently, photometric data needs to be included in the modeling in order to fit for inaccurate calibration of the spectra. However, by including photometry there is less flexibility for the model to accurately fit the observed spectrum. A second explanation for the inaccurate fit of specific absorption lines like Mg\,I and Na\,I is the fact that \texttt{Prospector} only uses the Solar chemical abundance pattern in its models. However, observations have shown that quiescent galaxies at high redshifts have non-Solar abundances \citep[e.g.,][]{JChoi2014, RMaiolino2019, ABeverage2024}, leading to different relative depths of absorption lines in a spectrum. To properly constrain all absorption lines in the spectra, \citet{ABeverage2024b} has fit the abundance of individual elements in the galaxies. We refer to this paper for more details on the differences between the \texttt{Prospector} fits presented in this work and the results from chemical abundance modeling.

Another explanation for why \texttt{Prospector} fails to reproduce the very deep Na\,{\sc id}$\lambda\lambda$5891,5897\AA\ absorption feature in some galaxies (e.g. 129133), is the fact that this line is also a tracer of neutral gas in the interstellar medium (ISM) as well as AGN outflows.
Although a detailed analysis of the contributions of the ISM and outflows is beyond the scope of this research, we note that the contribution of neutral gas could be used to explain the excess Na\,{\sc id} absorption in distant quiescent galaxies \citep[e.g.,][]{MJafariyazani2020, SBelli2023, RDavies2024}.

Lastly, it is possible that for some galaxies none of the stellar spectral templates used by \texttt{Prospector} describe the observed spectrum well, for example due to poor modeling of the influence of evolved stars, chemical abundances, or AGN. Indeed, for galaxies for which we find a high ($>3$) best-fit spectroscopic jitter term (targets 128452 and 127154) it is likely that the templates do not match the observed data, resulting in a best-fit spectrum that does not fit the data well.

\subsection{Sample Characteristics}\label{sec:samp_chars}
Our quiescent galaxy candidate targets were selected to be at $z>1.1$, are relatively bright, and fall in the quiescent box of the $UVJ$ diagram. However, our selection criteria were only based on photometric information, leading to uncertainties in the redshifts and resulting rest-frame $U-V$ and $V-J$ colors. In this section we present the sample characteristics we find when we include the spectroscopic information.

\begin{figure}
\begin{center}
    \centering
    \includegraphics[width = .47\textwidth]{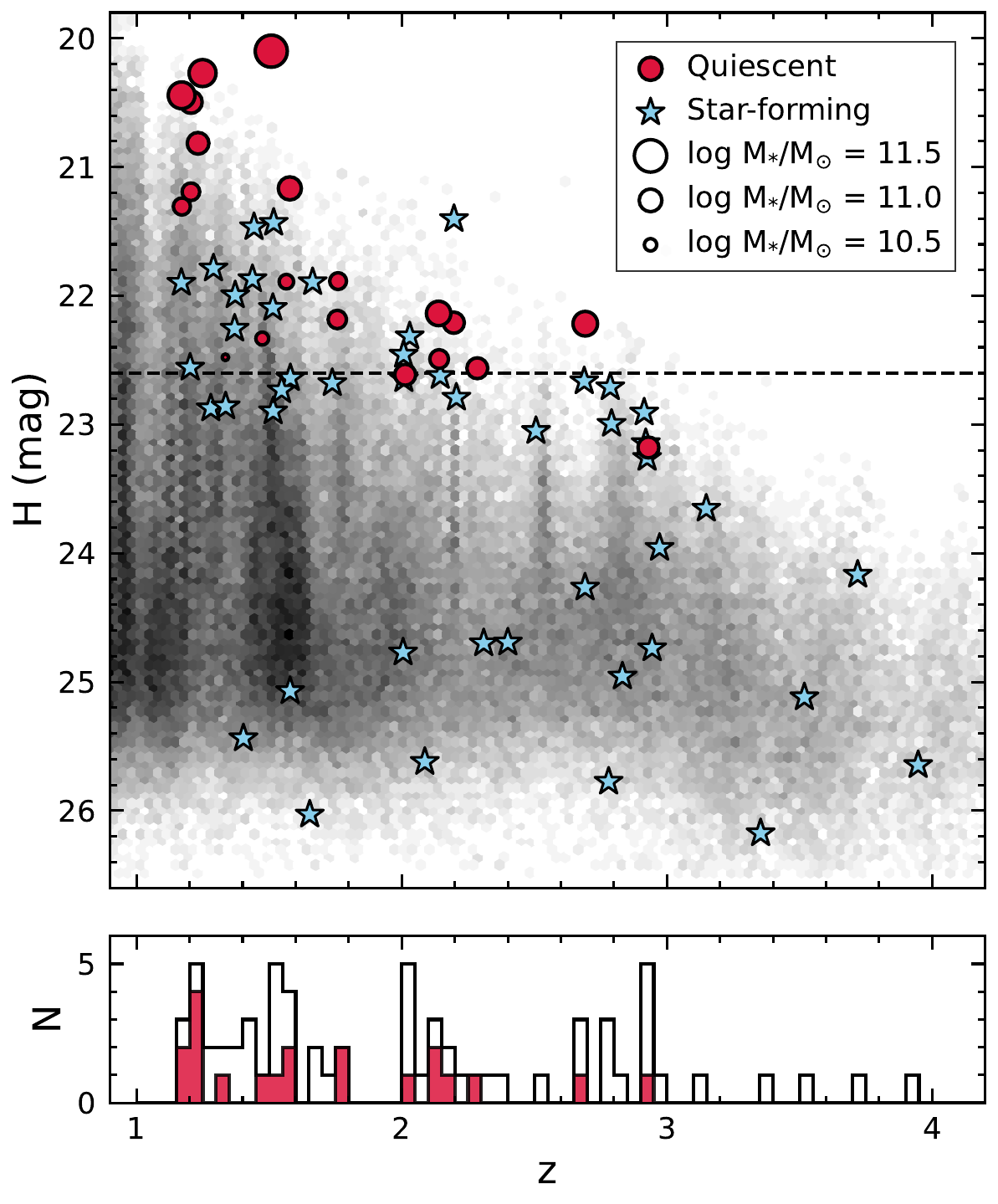}
    \caption{Top panel: $H$-band magnitude against redshift for all quiescent targets (red circles) and the filler targets for which we measured a spectroscopic redshift (blue stars). The symbol sizes of the quiescent targets are scaled by their \texttt{Prospector} best-fit stellar mass. The dashed line at $H = 22.6$ shows the magnitude limit we initially used to select our quiescent galaxy targets. One of our quiescent targets is fainter than this limit, and was not prioritized when designing the MSA configuration. In the background we show the parent UltraVISTA DR3 catalog from which our sample was drawn. Bottom panel: Redshift distribution of all targets for which a spectroscopic redshift was measured (white shading). The red shaded histogram shows the redshift distribution of only the quiescent targets.}
    \label{fig:z_Hmag}
\end{center}
\end{figure}

\begin{figure*}[!t]
\begin{center}
    \centering
    \includegraphics[width = 1.\textwidth]{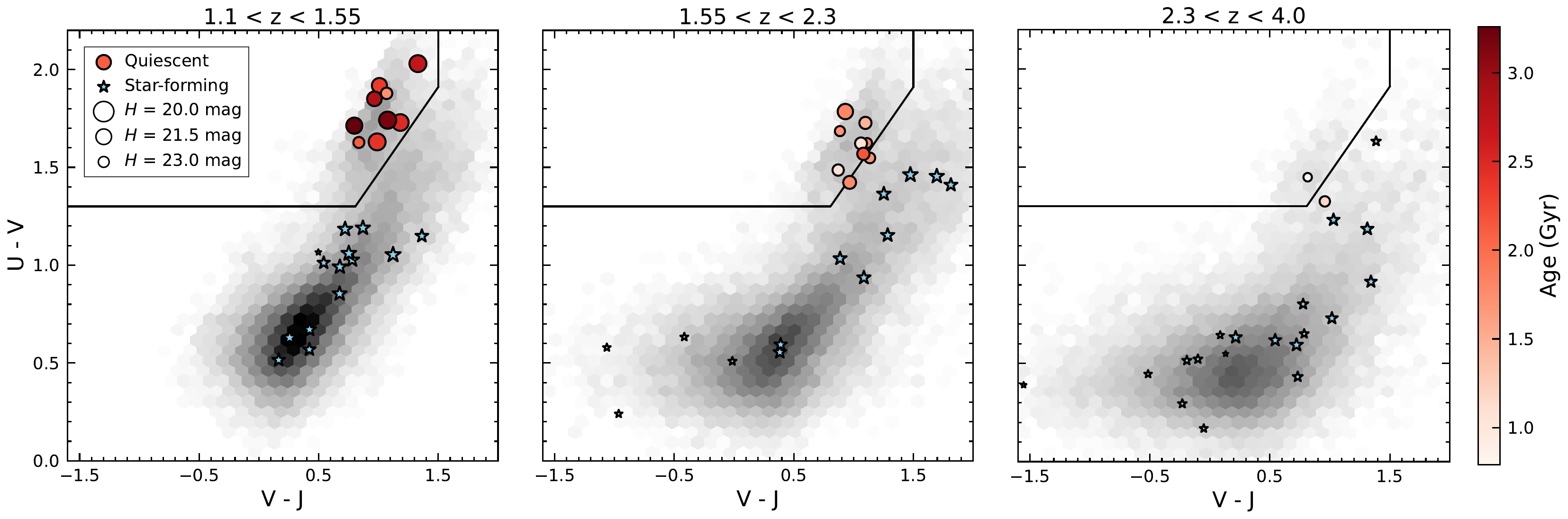}
    \caption{Rest-frame $U - V$ against $V - J$ colors for all galaxies observed in SUSPENSE, divided in three redshift intervals. Prior to the observations, candidate quiescent galaxies (red circles) were selected based on their rest-frame colors (obtained from photometric redshifts) using the indicated selection box \citep{AMuzzin2013a}. However, when using spectroscopic redshifts several candidate quiescent targets scattered just outside of this selection box. Nonetheless, all of these galaxies have confirmed quiescent stellar populations (see Section \ref{sec:SFHs}). We color the quiescent targets by their \texttt{Prospector} mass-weighted stellar age. The star-forming filler galaxies for which a spectroscopic redshift was measured are indicated by the blue stars. The size of each data-point is scaled to the $H$-band magnitudes of the corresponding galaxy. In the background, we show the colors of galaxies in the parent UltraVISTA catalog for each redshift interval.}
    \label{fig:UVJ}
\end{center}
\end{figure*}

In Figure \ref{fig:z_Hmag} we show the quiescent and star-forming targets in magnitude-redshift space relative to the parent distribution from the full UltraVISTA catalog. We only show the 46 out of 53 filler galaxies for which we could measure a spectroscopic redshift. This figure shows that nearly all quiescent targets are brighter than $H<22.6$; our criteria to ensure that our spectra are sufficiently deep to achieve our science goals. Only one quiescent galaxy in our sample is fainter; this galaxy was initially added as a filler target when creating the MSA mask, but since the observed spectrum has a high enough SNR we now include it in our quiescent galaxy sample (see Section \ref{sec:survey}). Since we prioritized bright star-forming galaxies when selecting filler targets, the majority of the star-forming targets are at the bright end of the parent distribution, and only $\sim 1/5$ of the star-forming galaxies are relatively faint compared to the complete population. We note that the magnitude-redshift diagram for the parent catalog is based on photometric redshifts only, which introduces (systematic) uncertainties on the background distribution.

The histogram in Figure \ref{fig:z_Hmag} shows that the redshifts of the quiescent and star-forming targets in our sample are not significantly clustered, except for a slight peak at $z \sim 1.2$. This peak may suggest that there could be a potential overdensity at this redshift, but, at face value, most of the galaxies in our sample do not appear to be part of an overdensity. This observation may be somewhat surprising, as we selected our pointing for the extraordinarily large amount of quiescent galaxies we could observe within one MSA configuration. A detailed investigation into potential overdensities in our sample is beyond the scope of this paper.

In Figure \ref{fig:UVJ} we show the rest-frame $U-V$ vs $V-J$ colors of all quiescent and star-forming targets, separated into three redshift bins. The $UVJ$ colors were calculated from the photometry using \texttt{eazy} \citep{GBrammer2008}, with the redshift set to the spectroscopic redshift. The top-left boxes in each panel indicate the region we used to select the quiescent targets \citep[red symbols;][]{AMuzzin2013a}, while star-forming targets lie outside of this box. For the initial selection of our quiescent targets we used $UVJ$ colors based on the photometric redshift. Three of our quiescent targets (IDs 127941, 128041, and 130725) scattered out of the box when using the spectroscopic redshift. However, since these galaxies lie very close to the selection box they may still be recently quenched or post-starburst galaxies with quiescent stellar populations \citep[e.g.,][]{SBelli2019, KSuess2021, MPark2023}. We will investigate the star-formation properties of these galaxies in more detail in the next section.

The quiescent galaxy sample spans across the full quiescent parent distribution in $UVJ$ space, indicating that our sample is representative of the full distant quiescent galaxy population. Specifically, in the low redshift bin our sample contains old and red quiescent galaxies, while at higher redshifts the quiescent galaxy sample shifts towards bluer colors. This is consistent with the younger ages we find for the bluer galaxies, as indicated by the colors of the symbols. This shift to younger ages and bluer galaxies is expected, as the universe is much younger at this redshift.

\subsection{Star-Formation Properties}\label{sec:SFHs}
We have selected our quiescent galaxy sample based on their location in $UVJ$ space. However, this selection was only based on photometric colors and redshifts, and galaxies with incorrect photometric redshifts may have erroneously scattered into the quiescent region, or the other way around. In the previous section we indeed showed that after including spectroscopic redshifts, the $UVJ$ colors of some galaxies shifted outside of the quiescent galaxy selection box. In order to assess whether our galaxies are indeed quiescent, we derived the star-formation properties of our quiescent sample, using the detailed SFHs obtained from the \texttt{Prospector} fits described in Section \ref{sec:prosp}.

\begin{figure}
\begin{center}
    \centering
    \includegraphics[width = .49\textwidth]{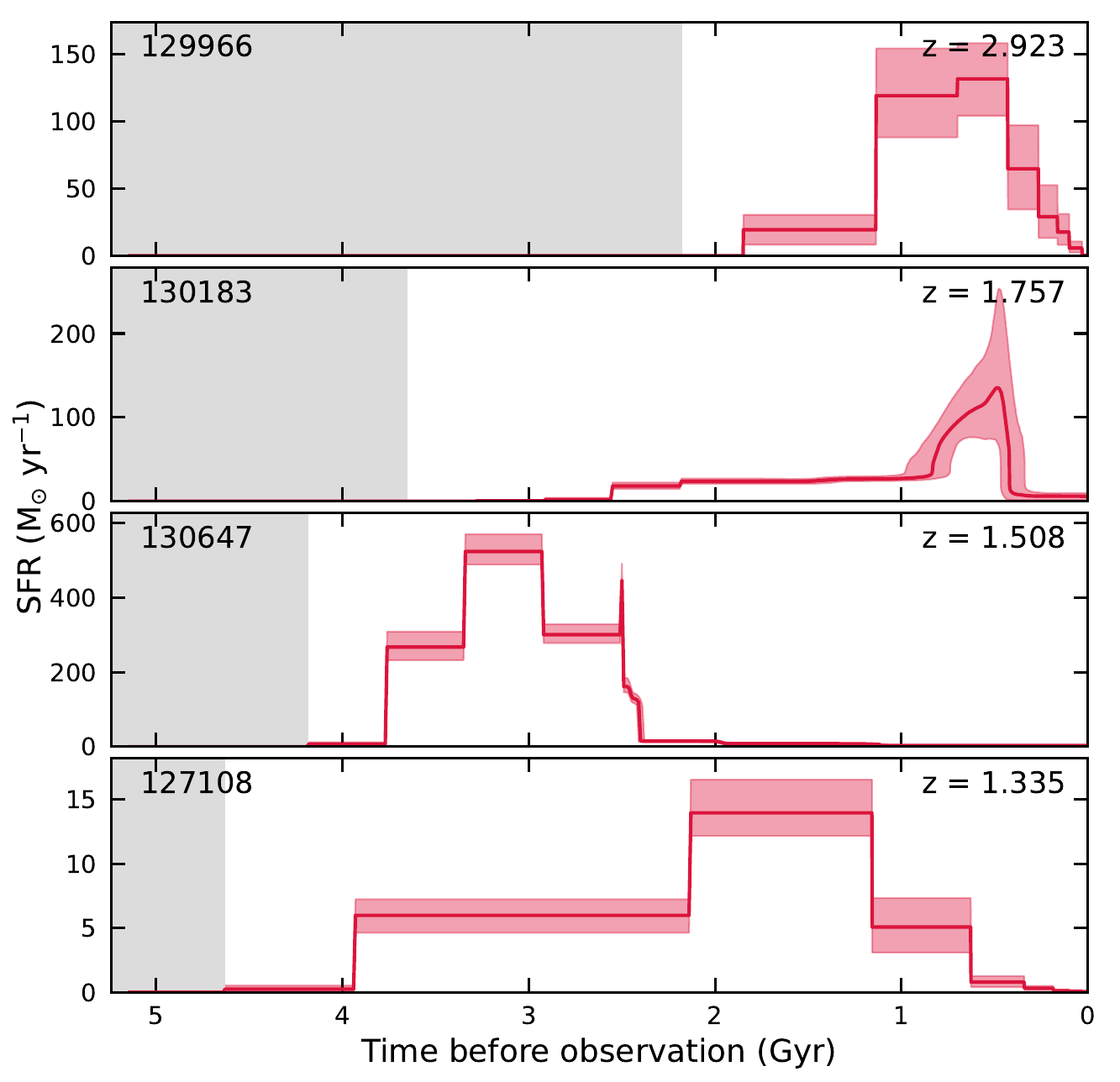}
    \caption{Four example best-fit SFHs for galaxies in the quiescent galaxy sample. The SFHs are ordered by decreasing redshift, as indicated in the top right of each panel. The red shaded areas represent the 16th-84th percentile confidence intervals of the SFHs. Targets 130183 and 130647 are best fit by a PSB-SFH, while 129966 and 127108 have fixed-bin SFHs. The gray shaded area in each panel corresponds to ages that are older than the age of the universe for the observed redshift.}
    \label{fig:SFHs_example}
\end{center}
\end{figure}

In Figure \ref{fig:SFHs_example} we show the best-fit SFHs for four example galaxies in the quiescent sample. These four SFHs illustrate the diversity of SFHs in our sample. Galaxies 129966 and 127108 are best fit by the fixed-bin SFH model, and have SFHs with long star-formation timescales and gradual quenching. In contrast, 130183 and 130647 both have SFHs that are better fit by the PSB model, with a gradually increasing amount of star-formation ending in a short burst, after which most of the star formation is suppressed. In total, four out of twenty primary targets (indicated in Table \ref{tab:sample}) have a SFH best described by the bursty PSB model, while sixteen galaxies have more gradual fixed-bin SFHs. From this variety of SFHs we find that the star-formation timescales of galaxies in our sample range between $\sim1-4$\,Gyr. We show the SFHs for all quiescent targets in Appendix \ref{ap:all_SFHs}. 

\begin{figure}[!t]
\begin{center}
    \centering
    \includegraphics[width = .47\textwidth]{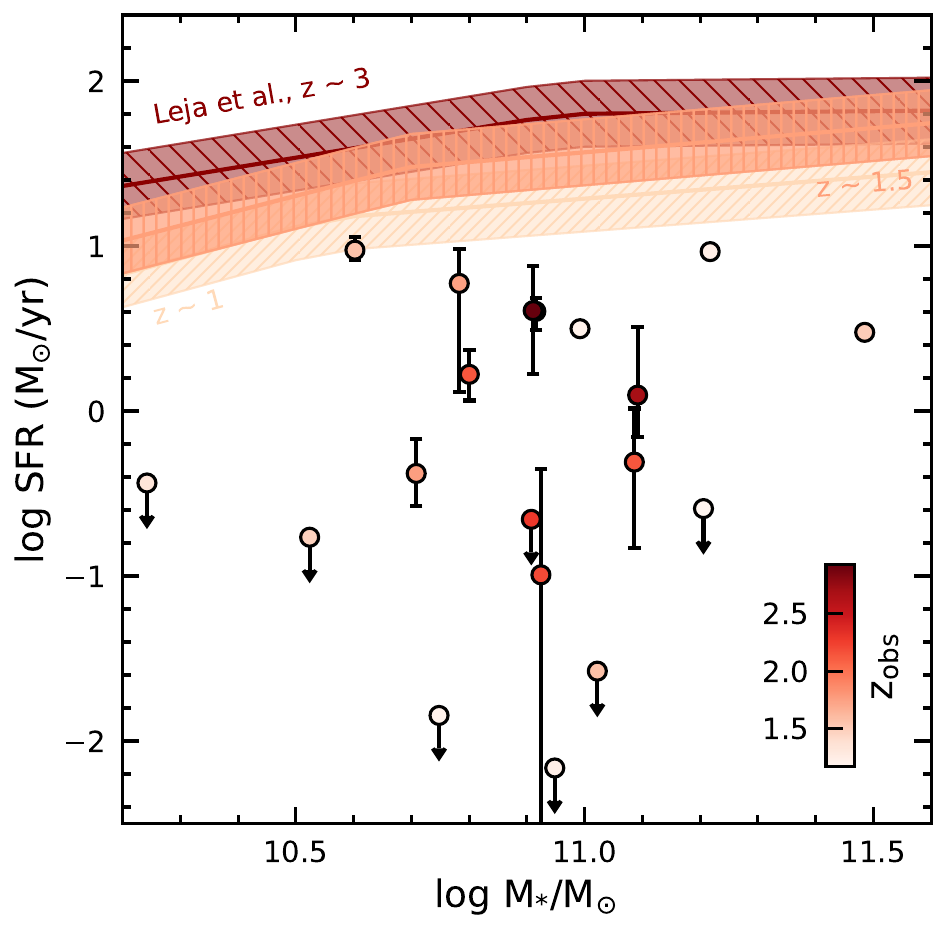}
    \caption{Stellar mass against SFR of the primary targets in our sample. For galaxies with log(SFR/$\text{M}_{\odot}~\text{yr}^{-1}$) $<-1$ we show a 3$\sigma$ upper limit to decrease the dynamic range of the figure. The symbols are colored by the redshift at observation, and we show the \citet{JLeja2022} star-forming main sequence for $z\sim1$ (yellow shaded area),  $z\sim1.5$ (orange shaded area), and $z\sim3$ (red shaded area). All candidate quiescent galaxies in our sample lie at least $2\sigma$ below the star-forming main sequence at their redshift.}
    \label{fig:mass_SFR}
\end{center}
\end{figure}

To assess whether our quiescent targets truly have quiescent stellar populations, we obtain their current SFRs from the \texttt{Prospector} best-fit SFHs (see Table \ref{tab:sample}). In Figure \ref{fig:mass_SFR} we show the SFRs against stellar mass of our quiescent sample. We also plot the \citet{JLeja2022} star-forming main sequence (SFMS) ridge model at $z\sim1$, $z\sim1.5$ and $z\sim3$. To decrease the dynamic range of this figure we show the $3\sigma$ upper limit for sources with log\,SFR $<-1$. All data-points lie $>2\sigma_{\text{SFMS}}$ below the SFMS at their respective redshifts, which shows that all primary targets indeed have quenched star formation. 

In the previous section we found that three primary targets (IDs 127941, 128041, and 130725) had shifted outside the quiescent selection box of the $UVJ$ diagram after including spectroscopic information. However, this selection box was empirically selected \citep{AMuzzin2013a}, and will not be a hard limit for quiescence. Indeed, all of the galaxies outside the selection box lie $> 7\sigma_{\text{SFMS}}$ below the star-forming sequence, meaning their SFRs are strongly suppressed, and we include these galaxies in our quiescent sample.\\

\begin{figure*}
\begin{center}
    \centering
    \includegraphics[width = \textwidth]{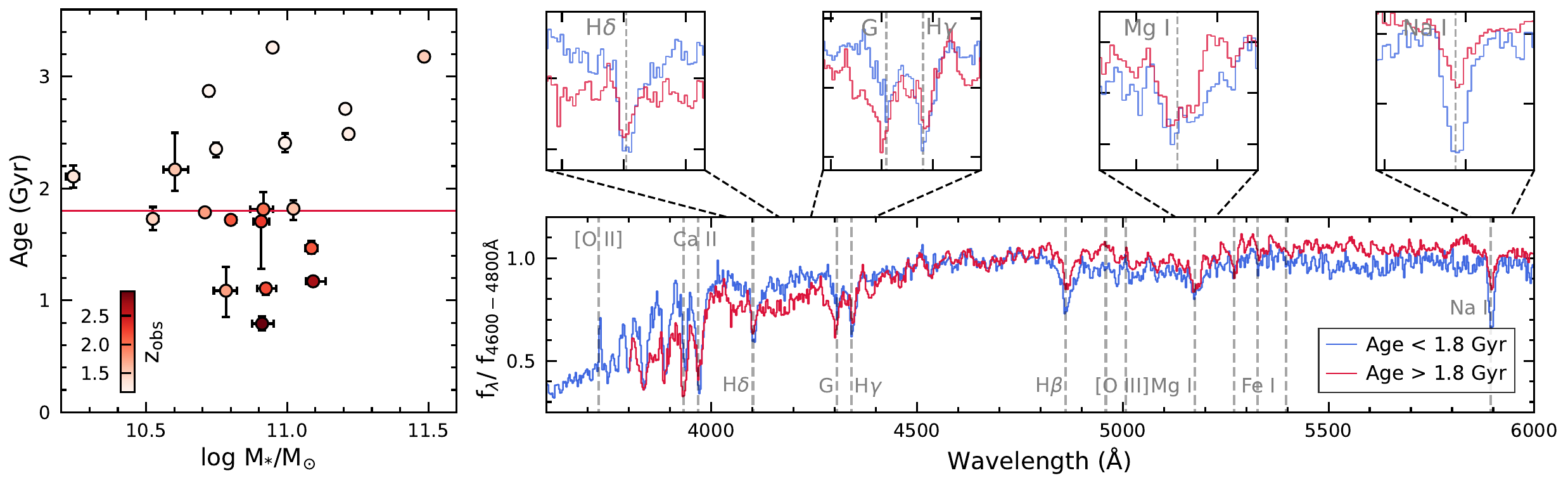}
    \caption{Left panel: Mass-weighted age against stellar mass of all quiescent galaxies in our sample. The symbols are colored by the redshift of the target. The red line at 1.8\,Gyr indicates the age we use to separate the sample in two age bins. Right panel: median-stacked spectrum of all galaxies in each age bin. The dashed vertical lines indicate Balmer, absorption and emission features. The top-right panels zoom in on the spectral regions around H$\delta$, H$\gamma$, Mg\,{\sc i}, and Na\,{\sc i}, to illustrate how these spectral features differ between the two age bins.}
    \label{fig:mass_age}
\end{center}
\end{figure*}

\subsection{Ages and Formation Times}\label{sec:ages_formation_times}
We calculate the mass-weighted age of all primary galaxies from their SFHs (see Table \ref{tab:sample}). We show these mean stellar ages as a function of stellar mass in the left panel of Figure \ref{fig:mass_age}. The quiescent galaxies in our sample have ages of $0.8-3.3$\,Gyr, and show a trend with observed redshift; the youngest galaxies in our sample were observed at the highest redshift. To illustrate how different spectral features are affected by the stellar age of a galaxy, we divide the sample into two age bins separated at a mass-weighted age of 1.8\,Gyr. We create a median stack of the spectra of the galaxies in each age bin at a rest-frame wavelength range of $3600-6000$\,\AA\ (right panel in Figure \ref{fig:mass_age}). Before stacking we first normalized the spectra of each galaxy by its mean flux at $4600-4800$\,\AA, and resampled the spectra to a common wavelength grid.

From the stacked spectra we can see clear differences in spectral features between the two age bins. For example, the 4000\,\AA\ break in the $>1.75$\,Gyr stacked spectrum is more pronounced compared to the young age bin. Furthermore, the Balmer absorption lines of the young age-bin are significantly deeper compared to continuum levels. This is most clear for the H$\delta$ and H$\gamma$ lines, which are sensitive to recently quenched star formation, while the difference in the depth of the H$\beta$ absorption line is less distinct between the two bins. This is likely because the youngest, most recently quenched galaxies still have some H$\beta$ emission, leading to a less deep absorption feature. The Mg\,{\sc i} and Fe\,{\sc i} metal absorption features, on the other hand, are significantly deeper for the old age-bin due to the fact that lower-mass, cooler stars with more prominent metal absorption features start to dominate the spectrum at older ages. The Na\,{\sc i} absorption line is instead deeper for the younger galaxies, likely due to the fact that this feature is also sensitive to neutral gas in the ISM, as well as AGN outflows. The distinct differences in the stacked spectra of the two age bins illustrate that a several individual spectral features are strongly affected by the age of a galaxy, and can be used to constrain detailed SFHs.

\begin{figure}
\begin{center}
    \centering
    \includegraphics[width = .49\textwidth]{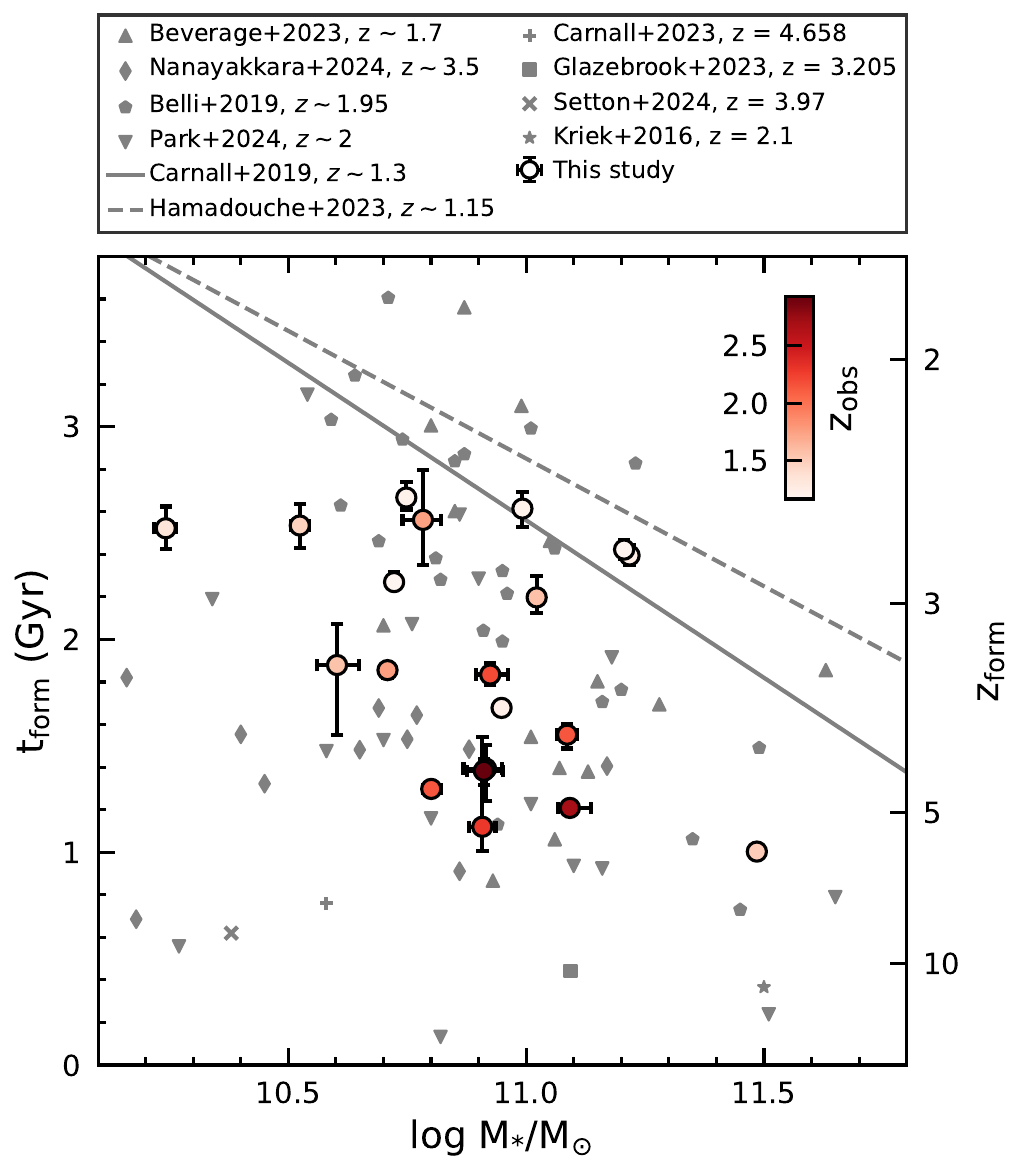}
    \caption{Age of the universe where 50\% of the stellar mass in the galaxy has formed ($t_{\text{form}}$) as a function of stellar mass for all quiescent galaxies in our sample. The symbols are colored by the redshift at observation. We show the \citet{TNanayakkara2024} $z\sim3-4$ (gray diamonds),  \citet{ABeverage2024} $z\sim1.7$ (gray triangles), \citet{SBelli2019} $z\sim1.9$ (gray pentagons), and \citet{MPark2024} $z\sim2$ (gray downward pointing triangles) samples. Furthermore, we plot the mean relations from \citet{ACarnall2019} (gray line) and \citet{MHamadouche2023} (dashed gray line) at $z\sim1.3$ and $z\sim1.15$, respectively. We also indicate four quiescent galaxies with early-onset star formation identified by \citet{ACarnall2023a} (gray plus), \citet{KGlazebrook2023} (gray square), \citet{DSetton2024} (gray cross), and \citet{MKriek2016} (gray star).}
    \label{fig:mass_tform}
\end{center}
\end{figure}

From the mass-weighted ages we also calculate the formation times ($t_{\text{form}}$) of the quiescent galaxies in our sample. Here $t_{\text{form}}$ is defined as the age of the universe at which 50\% of the stellar mass in the galaxies has formed. In Figure \ref{fig:mass_tform} we show $t_{\text{form}}$ as a function of stellar mass for our sample, with the symbols colored by the observed redshift. The galaxies in our sample have an average $t_{\text{form}}$ of $\sim1.0-2.7$\,Gyr, equivalent to a formation redshift between $z_{\text{form}} \sim 5.5-2.5$. We see no strong trend between average stellar formation time and stellar mass for our sample. For our massive galaxies, this is consistent with the findings of \citet{ACarnall2019} and \citet{AGallazzi2014} at $z\sim1.1$ and $z\sim0.7$, respectively, who found that while there is a trend between mass and formation time at lower masses, this trend flattens out at log\,$M_*/M_{\odot} \gtrsim 11$. 

In Figure \ref{fig:mass_tform}, we also show the spectroscopic samples by \citet{TNanayakkara2024} at $z_{\text{obs}} \sim 3-4$, \citet{SBelli2019} at $z_{\text{obs}} \sim 1.95$, \citet{MPark2024} at $z\sim2$, and \citet{ABeverage2024} at $z_{\text{obs}}\sim1.7$. Furthermore, we show the mean relations from \citet{ACarnall2019} ($z_{\text{obs}}\sim1.3$) and \citet{MHamadouche2023} ($z_{\text{obs}}\sim1.15$). The formation redshifts for most galaxies in these samples are consistent with those found for our sample at comparable redshifts.

Additionally, we show four high-redshift quiescent galaxies for which very early-onset star formation was identified from their spectra \citep{MKriek2016,ACarnall2023a, KGlazebrook2023, DSetton2024}, with inferred formation redshifts of $z \sim 8$. The \citet{TNanayakkara2024}, \citet{MPark2024}, and \citet{ABeverage2024} samples also contain galaxies with formation redshifts $\gtrsim 6$. These recent observations of early star formation in high redshift quiescent targets suggest that the build up of massive galaxies in the early universe was much more rapid compared to the local universe. Interestingly, while four galaxies in our sample were already quiescent by $z=3$, none of the quiescent galaxies in our sample have SFHs that are consistent with very early formation times.

\begin{figure}[t]
    \centering
    \includegraphics[width = 0.5\textwidth]{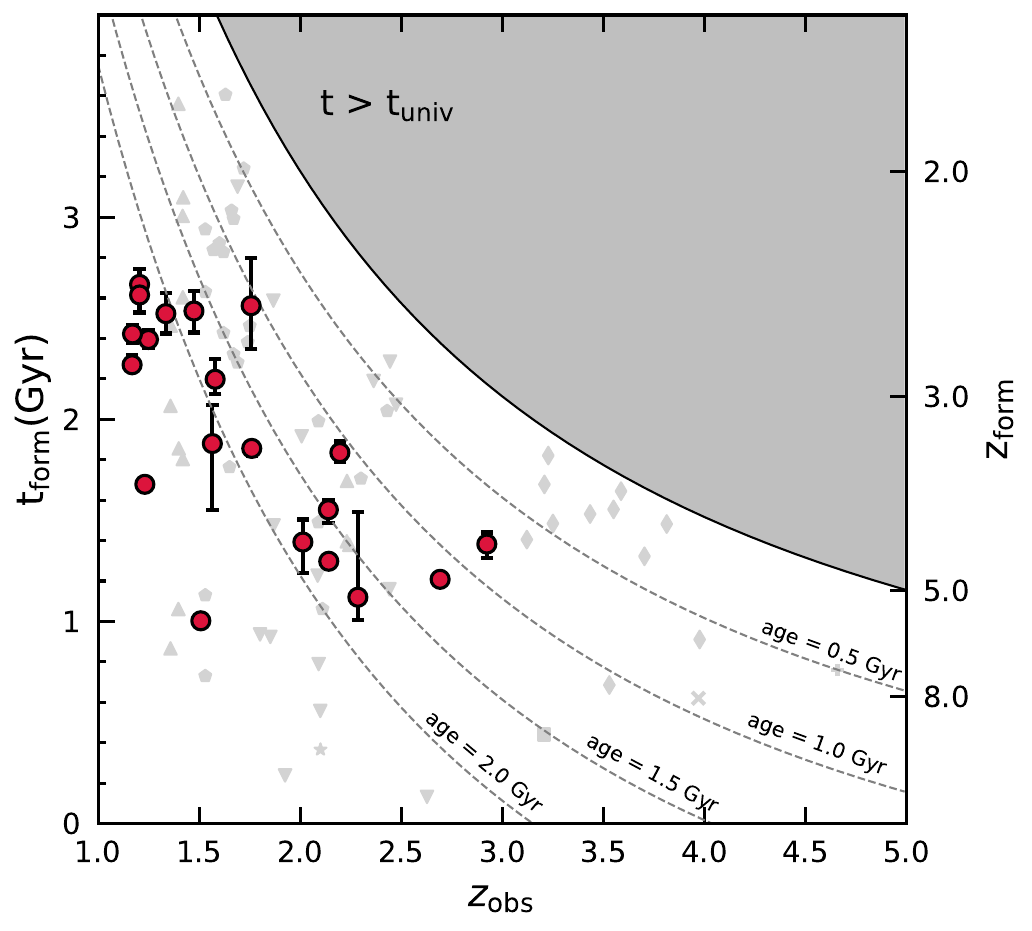}
    \caption{Age of the universe where 50\% of the stellar mass in the galaxy has formed ($t_{\text{form}}$) as a function of the observed redshift ($z_{\text{obs}}$) for all quiescent galaxies in our sample. The gray shaded area indicates the age of the universe as a function of redshift, and the dashed gray lines indicate ages of 0.5, to 2.0 Gyrs. We show different samples from literature using the same symbols as in Figure \ref{fig:mass_tform}.}
    \label{fig:z_t_form}
\end{figure}

Figure \ref{fig:mass_tform} shows a clear trend with redshift and mean formation time for our sample, with galaxies at lower redshift generally forming later than the higher redshift galaxies. In Figure \ref{fig:z_t_form} we show the formation time of our sample as a function of redshift, and find that on average, the galaxies at $z_{\text{obs}}>1.75$ formed 50\% of their stellar mass by $z \sim 4$, while for $z_{\text{obs}} < 1.75$ galaxies the mean formation redshift is $z\sim3$. Furthermore, when we compare our sample with other samples at similar and higher redshifts, we see that our sample covers representative ages of the population across our redshift range, although we sample relatively more old galaxies compared to other studies.

\subsection{Average SFHs}
\begin{figure}[t!]
\begin{center}
    \centering
    \includegraphics[width = .47\textwidth]{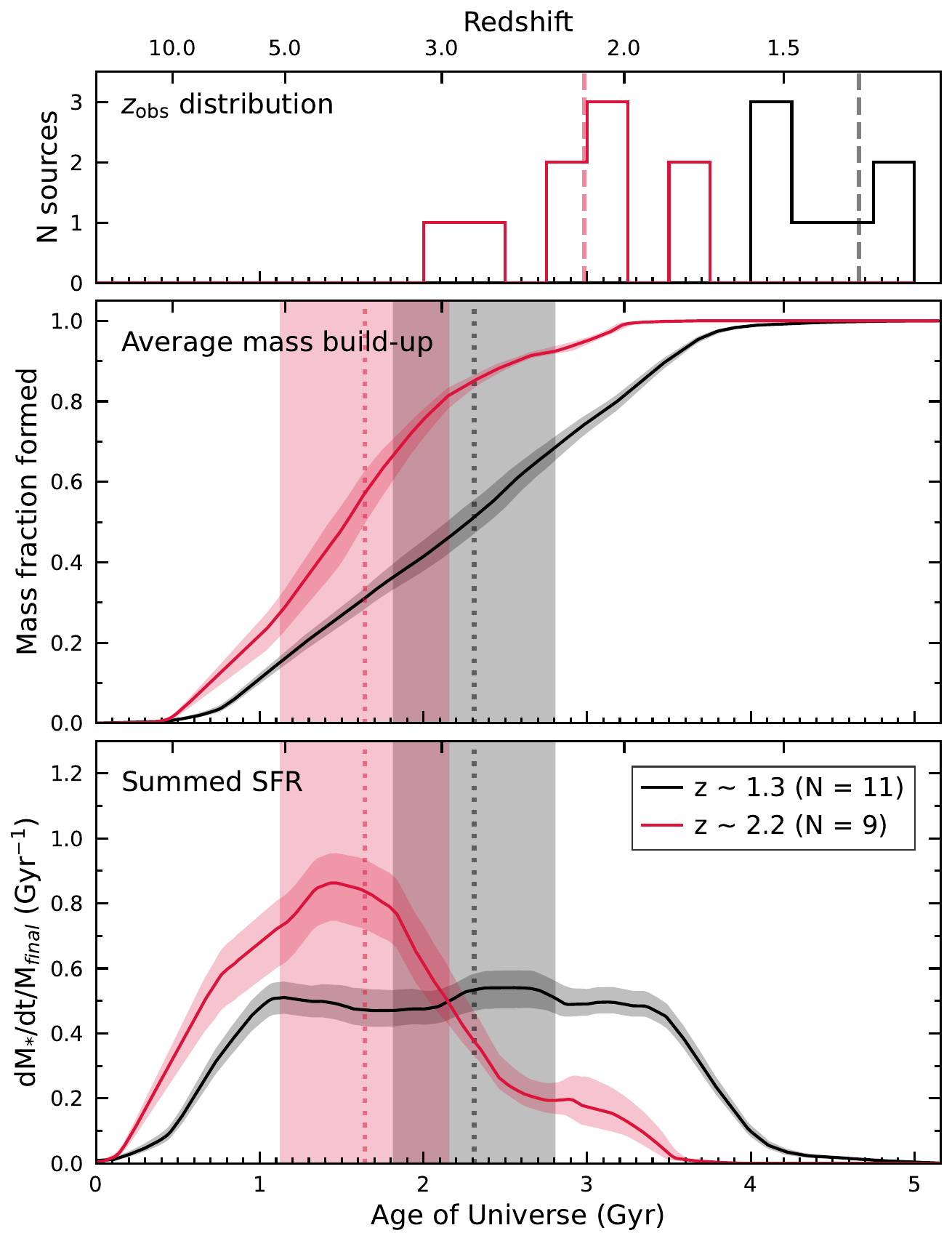}
    \caption{Top panel: Distribution of the observed redshifts, separated in two redshift bins. The dashed lines indicate the mean redshift of each bin. Middle panel: the average SFH of all quiescent targets, separated into two redshift bins. Here the SFH corresponds to the total mass fraction formed as a function of age of the universe. Bottom panel: the summed SFR of all quiescent galaxies as a function of age of the universe, separated into two redshift bins. Before summing the SFRs of the galaxies we first normalize them by their total stellar mass. The shaded regions around the mean SFHs indicate the error on the mean. In the middle and bottom panels, the dotted vertical lines indicate the average $t_{\text{form}}$ of each redshift bin, with the vertical bands corresponding to the standard deviation of $t_{\text{form}}$.}
    \label{fig:SFH_binned}
\end{center}
\end{figure}
In Figure \ref{fig:SFH_binned} we show the average SFHs of our galaxy sample, split into two redshift bins ($z_{\text{obs, lim}} = 1.75$) with mean redshifts of $z_{\text{obs}} = 1.3$ and $z_{\text{obs}} = 2.2$.
The middle panel of this figure shows the average mass build up over time, calculated by taking the mean cumulative mass fraction as a function of time for all galaxies in the bin. The bottom panel of Figure \ref{fig:SFH_binned} shows the summed SFRs as a function of age of the universe. Before summing the SFRs, all galaxies were normalized by their total stellar mass. We then smooth the summed SFRs of the redshift bins with a box kernel to reduce the noise from scatter in the individual galaxies' SFHs. Figure \ref{fig:SFH_binned} shows that the summed star-formation timescale is much shorter for galaxies at $z_{\text{obs}} \sim 2.2$ compared to the $z_{\text{obs}} \sim 1.3$ sample, consistent with the higher mean formation redshift we find for the galaxies at higher redshift. Furthermore, we also see that the onset of star-formation is slightly later for the low redshift bin. 

Only one galaxy in the low-$z$ sample (130647) already had a significantly suppressed SFR at the redshifts at which the high-$z$ sample was observed (see Figure \ref{fig:all_sfhs}); all the other galaxies quench at later times. This galaxy forms its stars on a timescale that is comparable with galaxies in the high-$z$ bin, while all other galaxies in the low-$z$ bin form their stars over a longer period.

\begin{figure}[t!]
\begin{center}
    \centering
    \includegraphics[width = .47\textwidth]{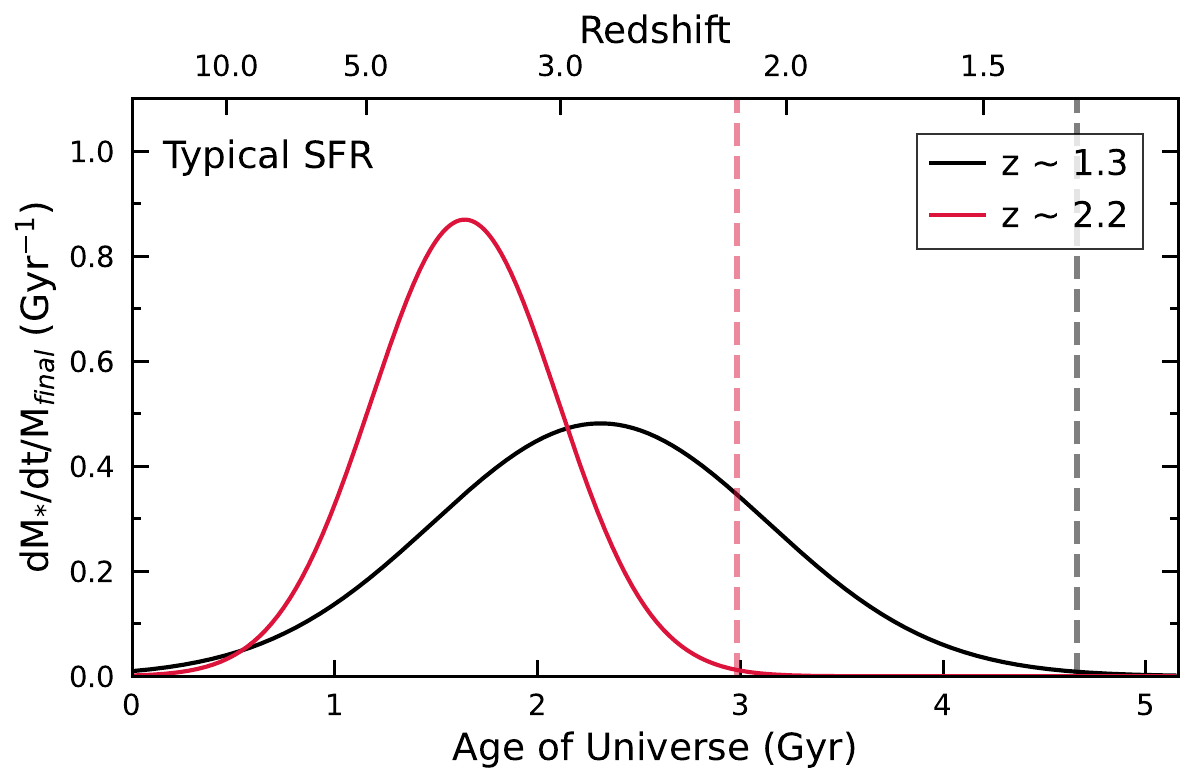}
    \caption{Illustrative ``typical'' SFH of quiescent galaxies at $z_{\text{obs}} \sim 1.3$ and $z_{\text{obs}} \sim 2.2$, assuming a simplistic Gaussian model. We take the mean of the Gaussian as the mean $t_{\text{form}}$ of each sample, and the FWHM of the Gaussian correspond to the median FWHM of the SFHs of the individual galaxies in each redshift bin. The median age of the universe for the two redshift bins are indicated by the dashed line. On average, quiescent galaxies at high redshift formed earlier and more rapidly than at low redshifts.}
    \label{fig:typical_SFH}
\end{center}
\end{figure}

The galaxies within the two bins in Figure \ref{fig:SFH_binned} have been observed at different redshifts and have different formation times, which will broaden the average star formation timescale of each bin. To account for this effect and compare the average intrinsic star formation timescale, we show an illustrative model for the typical SFHs for the two redshift bins, parameterized as a Gaussian in Figure \ref{fig:typical_SFH}. The FWHM of the Gaussian corresponds to the star-formation timescale, and is estimated from the median FWHM of the star-formation timescales of the individual galaxies in the redshift bin. The mean of the typical SFH Gaussian is set to the median $t_{\text{form}}$ of the redshift bin. It is important to note that this parameterization does not reflect the true shape of the SFHs, but was chosen for its simplicity, and should be taken as an illustration only. Nonetheless, there is a clear difference between the typical SFHs at $z_{\text{obs}}\sim1.3$ and $z_{\text{obs}}\sim2.2$, with higher-redshift quiescent galaxies forming more rapidly (FWHM = 1.0\,Gyr) than their low-redshift counterparts (FWHM = 1.9\,Gyr). 

Interestingly, when we compare the inferred typical SFHs for the SUSPENSE sample with a sample of quiescent galaxies at similar redshifts ($z_{\text{obs}} \sim 1.4$ and $z_{\text{obs}}\sim2.1$) from \citet{ABeverage2024}, we see that our star-formation timescales are a factor $\sim$2 and $\sim$6 longer, for the low and high-redshift bins respectively. Similarly, \citet{ABeverage2024b} measures detailed chemical abundances of the galaxies in our sample and finds that while the mean stellar ages from this chemical abundance fitting are in good agreement with our findings, their star-formation timescales are a factor $\sim$6 shorter. This difference likely reflects the different methods to derive the star-formation timescales; in \citet{ABeverage2024} and \citet{ABeverage2024b} these are inferred from chemical abundances. These chemical abundance studies show that distant quiescent galaxies are carbon and iron deficient, implying they quenched before significant enrichment by AGB stars and Type-Ia supernovae, consistent with star formation scales of $\lesssim200$\,Myr.
In this study, on the other hand, we model the full SFH using a non-parametric model, which are likely is biased to older, more extended SFHs by design \citep{JLeja2019,JLeja2019_b,ACarnall2024}. We thus note that our timescales may be biased to be longer, though the trend with redshift that we find appears to be real. For a more detailed discussion on the comparison between the star-formation timescales we refer to \citet{ABeverage2024b}.

\section{Discussion}\label{sec:implications}
\subsection{Implications for the evolution of the massive quiescent galaxy population}
In the previous section, we showed that the star-formation timescales of massive quiescent galaxies become significantly shorter from $z_{\text{obs}} \sim 1.3$ to $z_{\text{obs}} \sim 2.2$. This indicates that the mass build-up of distant quiescent galaxies is much more rapid compared to later times. This result is consistent with the findings from \citet{ABeverage2024}, who show that this trend extends from $z \sim 2.1$ up to $z \sim 0$. Furthermore, observations of quiescent galaxies out to $z \sim 5$ \citep[e.g.,][]{BForrest2020b,KGlazebrook2023,ACarnall2023b, DSetton2024} suggest that the evolution of star-formation timescale extends to even earlier times. In this section we discuss the implications for the evolution of the quiescent galaxy population from these observations.

One explanation for the observed evolution in the star-formation timescale across redshift is progenitor bias \citep{PvanDokkum2001a}; between $z_{\text{obs}} \sim 1.3$ and $z_{\text{obs}} \sim 2.2$ the quiescent galaxy population grows in number as more star-forming galaxies are quenched. Thus in our low redshift bin there are more galaxies that formed more gradually (i.e. longer star formation timescales) and entered the quiescent population at a later time.

A second explanation for the evolution of the SFHs are mergers, which change the properties of individual galaxies. In this scenario, major and minor mergers add new stellar populations to the galaxy, with SFHs that are different to those from the stars that are formed in situ. If the SFH of the accreted galaxy is more extended, or if the stars in that galaxy are younger, we would find that the SFH is broadened. The importance of minor mergers in the evolution of the quiescent galaxies has been supported by observations of the evolution in size and color gradients \citep[e.g.,][]{RBezanson2009, TNaab2009, PvanDokkum2010, KSuess2020, KSuess2021, TMiller2023}. Furthermore, observations have shown that distant quiescent galaxies are surrounded by numerous small companions, which may merge into the central galaxy at later times \citep{ANewman2012,KSuess2023}.

If progenitor bias, rather than mergers and late-time star formation, is the main driver behind the observed evolution of the star-formation timescale, this would imply that quiescent galaxies that formed rapidly in the early universe should still exist as rare relics in the population at lower redshifts. Directly calculating the expected number of these remnant galaxies in our low-redshift sample is challenging due to the fact that we have observed only one pointing, selected for its exceptionally high amount of distant quiescent galaxies. However, since all targets that we initially selected to be quiescent were indeed confirmed to be quiescent, and had broadly consistent spectroscopic and photometric redshifts, we can infer that the success rate of photometrically identifying distant quiescent galaxies at $1<z<3$ is extremely high (100\% for our survey).

Thus, we can compare the number densities of quiescent galaxies with $10.50 < $ log$(M_*/M_{\odot}) < 11.25$ at $z = 2.5$ and $z = 1.5$ from \citet{DMcLeod2021}, and find that the number densities increase by a factor of $\sim$3.5. This implies that, if we assume no evolution of individual galaxies, 1 out of 3.5 quiescent galaxies at $z\sim1.5$ was already quiescent at $z\sim2.5$. Naively, we would thus expect to find that $\sim$3 out of 11 sources in our low-redshift bin are remnant galaxies. However, in our sample only one galaxy (130647) in the $z_{\text{obs}} \sim 1.3$ bin was already quiescent at $z\sim2.5$. The reason that we find fewer remnant galaxies than expected could thus be easily explained by the late-time evolution of galaxies that quenched by $z\sim2.5$, through late-time star-formation or mergers. On the other hand, as our sample is small and the selection is based on magnitude we should be cautious when interpreting these results, since it is possible that we have missed fainter, older remnant galaxies in our sample. 

Thus, our observations suggest that population growth alone likely cannot explain the observed evolution of the star-formation timescale at $z>1$, and instead mergers and late-time star formation should also contribute to this evolution. On the other hand, the short star formation timescales (200\,Myr) and similar formation times that \citet{ABeverage2024b} find for the galaxies in the $z_{\text{obs}} \sim 1.3$  bin imply that all galaxies in our sample would have been quiescent by $z\sim2.5$. This result is difficult to reconcile with the evolving number densities of quiescent galaxies between the low and high redshift bins. However, since the reported star formation timescales of \citet{ABeverage2024b} are based on sample averages, caution is needed to interpret these results. Thus, larger samples for which (resolved) chemical abundance, and kinematic measurements and detailed SFHs are constrained are needed to solve this tension, and will allow us to investigate the influence of minor and major mergers on the observed evolution in star-formation timescales in more detail.

\subsection{Implications for ``maximally old'' distant quiescent galaxies}
Several spectroscopic studies of distant quiescent galaxies have shown indications of massive galaxies that have already formed nearly all their stars by $z = 5$ \citep[e.g.][]{MKriek2016, BForrest2020b, KGlazebrook2023, ACarnall2023b, DSetton2024, ABeverage2024}. In our sample, the most extreme galaxy has an average formation redshift of $z\sim5.5$ and formed nearly all its stars by $z\sim3$. Thus, our galaxies are not as extreme as the ones found by the above studies. 

The lack of maximally old galaxies in our sample could in part be explained by the fact that the observation redshift of our sample is lower compared to most of the studies mentioned above. If the evolution of the quiescent galaxy population is driven by progenitor bias, there would be relatively fewer extremely old galaxies at lower redshifts as newly quenched galaxies are added to the population. On the other hand, if merging is the main driver of the evolution of the quiescent population, galaxies that quenched early in the universe would likely merge with lower-mass galaxies after they become quiescent. These mergers would then shift the SFH of this initial early quencher to a later formation time and more extended SFH. Thus, if mergers are dominant, it is unlikely that we detect galaxies with a formation redshift of $z>10$ in our sample at $z\sim3$.

Irrespective of the evolutionary scenario we will thus, by construction, have fewer maximally old galaxies in our sample, as the low-redshift population, on average, quenched (and thus formed) at a later time compared to samples at higher redshifts. Furthermore, while previously detected maximally old galaxies are often extremely massive and/or bright, our sample is more representative of the complete quiescent galaxy population (see Figure \ref{fig:UVJ}), without a strong bias for extremely bright or massive galaxies. Compared to previous studies of maximally old galaxies, our galaxy sample thus presents a more complete overview of the quiescent galaxy population at its redshift, making the detection of maximally old galaxies less likely.

Furthermore, the low number densities of these ``maximally old'' sources make it unlikely that our sample includes a maximally old galaxy. \citet{KGlazebrook2023} calculate a number density of $6\times10^{-7}$~Mpc$^{-3}$ for this type of galaxy, which corresponds to $\sim0.05$ maximally old galaxies per MSA pointing. With our single pointing it is thus not surprising that our sample does not include such objects.

We also note that the methods for obtaining the stellar age differ between our study and other studies presenting maximally old galaxies, leading to uncertainties in comparing the mean formation times of these samples. Firstly, the ages of the galaxies presented by \citet{BForrest2020b, KGlazebrook2023, TNanayakkara2024} and \citet{DSetton2024} were inferred from low-resolution spectra for which individual metal absorption features cannot be resolved. Without measurements of individual metal absorption features it is difficult to break the age-metallicity degeneracy, leading to large uncertainties in the reported stellar ages for these galaxies. Furthermore, while the \citet{ABeverage2024} and \citet{MKriek2016} samples modeled the individual metal absorption features, the final age was modeled as a simple single or two component age, rather than a full SFH. This simplification of the SFHs of these samples compared to the non-parametric SFHs used in this study makes comparison between the our study and these studies more challenging. 

We note that our current study also suffers from modeling uncertainties, particularly as we model the photometric and spectroscopic data simultaneously. Interestingly, \citet{ABeverage2024b} fit our spectra with \texttt{alf} \citep{CConroy2012d, CConroy2018} to constrain individual abundances and ages simultaneously, while not taking photometry into account, and find ages that are consistent with the ages inferred from \texttt{Prospector}, suggesting that different chemical abundance patterns and including photometry do not significantly bias the best-fit ages. However, we do note that our inferred SFHs could be biased to late-time star formation, as younger stellar populations will have more weight than the oldest stars in the galaxy, leading to more extended and younger SFHs.

\section{Summary}\label{sec:summary}
In this paper we present an overview and first results of the Cycle 1 JWST-SUSPENSE program. This ultradeep spectroscopic survey targeted 20 distant quiescent galaxy candidates at $z = 1-3$, as well as 53 star-forming galaxies at similar and higher redshifts. The observations were executed with the NIRSpec MSA, using the G140M-F100LP dispersion-filter combination resulting in a rest-frame wavelength coverage of $\sim3000-7000$\,\AA\ at the targeted redshift.
We use $\sim$5 shutter slitlets and a two-shutter nod, to observe the full extent of our primary targets and avoid self-subtraction.
With 16 hours of on-source integration time, the SUSPENSE program has obtained the largest sample of quiescent galaxies with ultra-deep (18 galaxies with continuum SNR$>10$\,\AA$^{-1}$) medium-resolution spectra at $z > 1$ to date.

The spectra of all quiescent targets show clear Balmer and/or metal absorption features. Our quiescent galaxy sample is representative of the full quiescent galaxy population in our redshift range in $UVJ$ space. We fit spectroscopic redshifts, stellar population parameters, and detailed SFHs for all quiescent targets in our sample using spectrophotometric fitting with \texttt{Prospector}. All galaxies in our sample are massive, with stellar masses ranging from $10^{10.24}-10^{11.49}$\,M$_{\odot}$. We used the best-fit SFHs to determine the SFRs and find that all 20 of our primary targets lie $>2\sigma$ below the star-forming main sequence at their redshift, indicating that they are indeed quiescent galaxies. Eleven of our targets show emission lines in their spectra (primarily [O\,{\sc ii}], and [N\,{\sc ii}] emission), likely originating from ionization due to an AGN or hot evolved stars. These emission lines will be the subject of a future study.

The mass-weighted ages of the galaxies in our sample range from $0.8-3.3$\,Gyr, with star-formation timescales between $\sim0.5-4$\,Gyr. Four galaxies in our sample were already quiescent by $z = 3$.
On average, the high redshift galaxies formed earlier than their low-redshift counterparts; 50\% of the stellar mass of galaxies in the $z\sim2.2$ bin formed at $z_{\text{form}} \sim 4$, while for $z\sim1.3$ bin the average formation redshift is $z_{\text{form}} \sim 3$. This difference in formation redshift is consistent with the typically shorter star formation timescales we find for galaxies at $z\sim2.2$ compared to $z\sim1.3$.

The observed evolution in star-formation histories can be explained by the growth of the quiescent galaxy population as the universe ages; at lower redshifts the addition of new galaxies with more extended SFHs alters the \textit{average} SFH properties of the quiescent population. However, number density calculations show that mergers and/or late-time star formation likely also contribute to the observed evolution. 

In future studies we will extend upon this analysis by studying the (resolved) stellar, chemical, and kinematic properties of our quiescent targets. Thus, SUSPENSE presents a unique sample that will help us further unravel how these massive quiescent galaxies formed, when and why their star formation quenched, and how they evolved into the local quiescent population. Moreover, the spectra from the SUSPENSE program illustrate the power of JWST to obtain the deep, high-resolution spectroscopy needed to study distant, massive quiescent galaxies in great detail. Thus, JWST has marked a new era, in which ultra-deep data will revolutionize our understanding of the formation and evolution of quiescent galaxies over cosmic time. \\

This work is based on observations made with the NASA/ESA/CSA James Webb Space Telescope. The data were obtained from the Mikulski Archive for Space Telescopes at the Space Telescope Science Institute, which is operated by the Association of Universities for Research in Astronomy, Inc., under NASA contract NAS 5-03127 for JWST. These observations are associated with program JWST-GO-2110. Support for program JWST-GO-2110 was provided by NASA through a grant from the Space Telescope Science Institute, which is operated by the Association of Universities for Research in Astronomy, Inc., under NASA contract NAS 5-03127. MK acknowledges funding from the Dutch Research Council (NWO) through the award of the Vici grant VI.C.222.047 (project 2010007169). PEMP acknowledges the support from the Dutch Research Council (NWO) through the Veni grant VI.Veni.222.364.

\software{\texttt{Prospector} \citep{JLeja2019_b, BJohnson2021}, \texttt{MSAEXP} (\url{https://github.com/gbrammer/msaexp}), \texttt{EAZY} \citep{GBrammer2008}, \texttt{Source Extractor} \citep{EBertin1996}, \texttt{grizli} \citep{grizli}, \texttt{MSAFIT} \citep{AdeGraaff2023}}

\facilities{JWST (NIRSpec)}

The JWST data presented in this article were obtained from the Mikulski Archive for Space Telescopes (MAST) at the Space Telescope Science Institute. The specific observations analyzed can be accessed via \dataset[doi: 10.17909/6wjp-qb35]{https://doi.org/10.17909/6wjp-qb35}.

\bibliography{mybib}{}

\begin{thebibliography}{}
\expandafter\ifx\csname natexlab\endcsname\relax\def\natexlab#1{#1}\fi
\providecommand{\url}[1]{\href{#1}{#1}}
\providecommand{\dodoi}[1]{doi:~\href{http://doi.org/#1}{\nolinkurl{#1}}}
\providecommand{\doeprint}[1]{\href{http://ascl.net/#1}{\nolinkurl{http://ascl.net/#1}}}
\providecommand{\doarXiv}[1]{\href{https://arxiv.org/abs/#1}{\nolinkurl{https://arxiv.org/abs/#1}}}

\bibitem[{{Allard} {et~al.}(2000){Allard}, {Hauschildt}, \& {Schwenke}}]{FAllard2000}
{Allard}, F., {Hauschildt}, P.~H., \& {Schwenke}, D. 2000, \apj, 540, 1005, \dodoi{10.1086/309366}

\bibitem[{{Almaini} {et~al.}(2017){Almaini}, {Wild}, {Maltby}, {Hartley}, {Simpson}, {Hatch}, {McLure}, {Dunlop}, \& {Rowlands}}]{OAlmaini2017}
{Almaini}, O., {Wild}, V., {Maltby}, D.~T., {et~al.} 2017, \mnras, 472, 1401, \dodoi{10.1093/mnras/stx1957}

\bibitem[{Almeida {et~al.}(2012)Almeida, Terlevich, Terlevich, Fernandes, \& Morales-Luis}]{JSánchez2012}
Almeida, J.~S., Terlevich, R., Terlevich, E., Fernandes, R.~C., \& Morales-Luis, A.~B. 2012, The Astrophysical Journal, 756, 163, \dodoi{10.1088/0004-637X/756/2/163}

\bibitem[{{Antwi-Danso} {et~al.}(2023){Antwi-Danso}, {Papovich}, {Esdaile}, {Nanayakkara}, {Glazebrook}, {Hutchison}, {Whitaker}, {Marsan}, {Diaz}, {Marchesini}, {Muzzin}, {Tran}, {Setton}, {Kaushal}, {Speagle}, \& {Cole}}]{JAntwi-Danso2023}
{Antwi-Danso}, J., {Papovich}, C., {Esdaile}, J., {et~al.} 2023, arXiv e-prints, arXiv:2307.09590, \dodoi{10.48550/arXiv.2307.09590}

\bibitem[{{Asplund} {et~al.}(2009){Asplund}, {Grevesse}, {Sauval}, \& {Scott}}]{MAsplund2009}
{Asplund}, M., {Grevesse}, N., {Sauval}, A.~J., \& {Scott}, P. 2009, \araa, 47, 481, \dodoi{10.1146/annurev.astro.46.060407.145222}

\bibitem[{{Behroozi} {et~al.}(2019){Behroozi}, {Wechsler}, {Hearin}, \& {Conroy}}]{PBehroozi2019}
{Behroozi}, P., {Wechsler}, R.~H., {Hearin}, A.~P., \& {Conroy}, C. 2019, \mnras, 488, 3143, \dodoi{10.1093/mnras/stz1182}

\bibitem[{{Belfiore} {et~al.}(2016){Belfiore}, {Maiolino}, {Maraston}, {Emsellem}, {Bershady}, {Masters}, {Yan}, {Bizyaev}, {Boquien}, {Brownstein}, {Bundy}, {Drory}, {Heckman}, {Law}, {Roman-Lopes}, {Pan}, {Stanghellini}, {Thomas}, {Weijmans}, \& {Westfall}}]{FBelfiore2016}
{Belfiore}, F., {Maiolino}, R., {Maraston}, C., {et~al.} 2016, \mnras, 461, 3111, \dodoi{10.1093/mnras/stw1234}

\bibitem[{{Belli} {et~al.}(2017){Belli}, {Newman}, \& {Ellis}}]{SBelli2017}
{Belli}, S., {Newman}, A.~B., \& {Ellis}, R.~S. 2017, \apj, 834, 18, \dodoi{10.3847/1538-4357/834/1/18}

\bibitem[{{Belli} {et~al.}(2019){Belli}, {Newman}, \& {Ellis}}]{SBelli2019}
---. 2019, \apj, 874, 17, \dodoi{10.3847/1538-4357/ab07af}

\bibitem[{{Belli} {et~al.}(2021){Belli}, {Contursi}, {Genzel}, {Tacconi}, {F{\"o}rster-Schreiber}, {Lutz}, {Combes}, {Neri}, {Garc{\'\i}a-Burillo}, {Schuster}, {Herrera-Camus}, {Tadaki}, {Davies}, {Davies}, {Johnson}, {Lee}, {Leja}, {Nelson}, {Price}, {Shangguan}, {Shimizu}, {Tacchella}, \& {{\"U}bler}}]{SBelli2021}
{Belli}, S., {Contursi}, A., {Genzel}, R., {et~al.} 2021, \apjl, 909, L11, \dodoi{10.3847/2041-8213/abe6a6}

\bibitem[{{Belli} {et~al.}(2023){Belli}, {Park}, {Davies}, {Mendel}, {Johnson}, {Conroy}, {Benton}, {Bugiani}, {Emami}, {Leja}, {Li}, {Maheson}, {Mathews}, {Naidu}, {Nelson}, {Tacchella}, {Terrazas}, \& {Weinberger}}]{SBelli2023}
{Belli}, S., {Park}, M., {Davies}, R.~L., {et~al.} 2023, arXiv e-prints, arXiv:2308.05795, \dodoi{10.48550/arXiv.2308.05795}

\bibitem[{{Bertin} \& {Arnouts}(1996)}]{EBertin1996}
{Bertin}, E., \& {Arnouts}, S. 1996, \aaps, 117, 393

\bibitem[{{Beverage} {et~al.}(2023){Beverage}, {Kriek}, {Suess}, {Conroy}, {Price}, {Barro}, {Bezanson}, {Franx}, {Lorenz}, {Ma}, {Mowla}, {Pasha}, {van Dokkum}, \& {Weisz}}]{ABeverage2024}
{Beverage}, A.~G., {Kriek}, M., {Suess}, K.~A., {et~al.} 2023, arXiv e-prints, arXiv:2312.05307, \dodoi{10.48550/arXiv.2312.05307}

\bibitem[{{Beverage} {et~al.}(2024){Beverage}, {Slob}, {Kriek}, {Conroy}, {Barro}, {Bezanson}, {Brammer}, {Cheng}, {de Graaff}, {F{\"o}rster Schreiber}, {Franx}, {Lorenz}, {Mancera Pi{\~n}a}, {Marchesini}, {Muzzin}, {Newman}, {Price}, {Shapley}, {Stefanon}, {Suess}, {van Dokkum}, {Weinberg}, \& {Weisz}}]{ABeverage2024b}
{Beverage}, A.~G., {Slob}, M., {Kriek}, M., {et~al.} 2024, arXiv e-prints, arXiv:2407.02556, \dodoi{10.48550/arXiv.2407.02556}

\bibitem[{{Bezanson} {et~al.}(2009){Bezanson}, {van Dokkum}, {Tal}, {Marchesini}, {Kriek}, {Franx}, \& {Coppi}}]{RBezanson2009}
{Bezanson}, R., {van Dokkum}, P.~G., {Tal}, T., {et~al.} 2009, \apj, 697, 1290, \dodoi{10.1088/0004-637X/697/2/1290}

\bibitem[{{Bournaud} {et~al.}(2007){Bournaud}, {Jog}, \& {Combes}}]{FBournaud2007}
{Bournaud}, F., {Jog}, C.~J., \& {Combes}, F. 2007, \aap, 476, 1179, \dodoi{10.1051/0004-6361:20078010}

\bibitem[{Brammer(2023)}]{grizli}
Brammer, G. 2023, grizli, 1.8.3,  Zenodo, \dodoi{10.5281/zenodo.7767790}

\bibitem[{{Brammer} {et~al.}(2008){Brammer}, {van Dokkum}, \& {Coppi}}]{GBrammer2008}
{Brammer}, G.~B., {van Dokkum}, P.~G., \& {Coppi}, P. 2008, \apj, 686, 1503, \dodoi{10.1086/591786}

\bibitem[{Bushouse {et~al.}(2023)Bushouse, Eisenhamer, Dencheva, Davies, Greenfield, Morrison, Hodge, Simon, Grumm, Droettboom, Slavich, Sosey, Pauly, Miller, Jedrzejewski, Hack, Davis, Crawford, Law, Gordon, Regan, Cara, MacDonald, Bradley, Shanahan, Jamieson, Teodoro, Williams, \& Pena-Guerrero}]{bushouse_2023_10022973}
Bushouse, H., Eisenhamer, J., Dencheva, N., {et~al.} 2023, JWST Calibration Pipeline, 1.12.5,  Zenodo, \dodoi{10.5281/zenodo.10022973}

\bibitem[{{Carnall} {et~al.}(2019){Carnall}, {McLure}, {Dunlop}, {Cullen}, {McLeod}, {Wild}, {Johnson}, {Appleby}, {Dav{\'e}}, {Amorin}, {Bolzonella}, {Castellano}, {Cimatti}, {Cucciati}, {Gargiulo}, {Garilli}, {Marchi}, {Pentericci}, {Pozzetti}, {Schreiber}, {Talia}, \& {Zamorani}}]{ACarnall2019}
{Carnall}, A.~C., {McLure}, R.~J., {Dunlop}, J.~S., {et~al.} 2019, \mnras, 490, 417, \dodoi{10.1093/mnras/stz2544}

\bibitem[{{Carnall} {et~al.}(2022){Carnall}, {McLure}, {Dunlop}, {Hamadouche}, {Cullen}, {McLeod}, {Begley}, {Amorin}, {Bolzonella}, {Castellano}, {Cimatti}, {Fontanot}, {Gargiulo}, {Garilli}, {Mannucci}, {Pentericci}, {Talia}, {Zamorani}, {Calabro}, {Cresci}, \& {Hathi}}]{ACarnall2022}
---. 2022, \apj, 929, 131, \dodoi{10.3847/1538-4357/ac5b62}

\bibitem[{{Carnall} {et~al.}(2023{\natexlab{a}}){Carnall}, {McLure}, {Dunlop}, {McLeod}, {Wild}, {Cullen}, {Magee}, {Begley}, {Cimatti}, {Donnan}, {Hamadouche}, {Jewell}, \& {Walker}}]{ACarnall2023a}
---. 2023{\natexlab{a}}, \nat, 619, 716, \dodoi{10.1038/s41586-023-06158-6}

\bibitem[{{Carnall} {et~al.}(2023{\natexlab{b}}){Carnall}, {McLeod}, {McLure}, {Dunlop}, {Begley}, {Cullen}, {Donnan}, {Hamadouche}, {Jewell}, {Jones}, {Pollock}, \& {Wild}}]{ACarnall2023b}
{Carnall}, A.~C., {McLeod}, D.~J., {McLure}, R.~J., {et~al.} 2023{\natexlab{b}}, \mnras, 520, 3974, \dodoi{10.1093/mnras/stad369}

\bibitem[{{Carnall} {et~al.}(2024){Carnall}, {Cullen}, {McLure}, {McLeod}, {Begley}, {Donnan}, {Dunlop}, {Shapley}, {Rowlands}, {Almaini}, {Arellano-C{\'o}rdova}, {Barrufet}, {Cimatti}, {Ellis}, {Grogin}, {Hamadouche}, {Illingworth}, {Koekemoer}, {Leung}, {Lovell}, {P{\'e}rez-Gonz{\'a}lez}, {Santini}, {Stanton}, \& {Wild}}]{ACarnall2024}
{Carnall}, A.~C., {Cullen}, F., {McLure}, R.~J., {et~al.} 2024, arXiv e-prints, arXiv:2405.02242, \dodoi{10.48550/arXiv.2405.02242}

\bibitem[{{Carollo} {et~al.}(2013){Carollo}, {Bschorr}, {Renzini}, {Lilly}, {Capak}, {Cibinel}, {Ilbert}, {Onodera}, {Scoville}, {Cameron}, {Mobasher}, {Sanders}, \& {Taniguchi}}]{MCarollo2013}
{Carollo}, C.~M., {Bschorr}, T.~J., {Renzini}, A., {et~al.} 2013, \apj, 773, 112, \dodoi{10.1088/0004-637X/773/2/112}

\bibitem[{{Casey} {et~al.}(2023){Casey}, {Kartaltepe}, {Drakos}, {Franco}, {Harish}, {Paquereau}, {Ilbert}, {Rose}, {Cox}, {Nightingale}, {Robertson}, {Silverman}, {Koekemoer}, {Massey}, {McCracken}, {Rhodes}, {Akins}, {Allen}, {Amvrosiadis}, {Arango-Toro}, {Bagley}, {Bongiorno}, {Capak}, {Champagne}, {Chartab}, {Ch{\'a}vez Ortiz}, {Chworowsky}, {Cooke}, {Cooper}, {Darvish}, {Ding}, {Faisst}, {Finkelstein}, {Fujimoto}, {Gentile}, {Gillman}, {Gould}, {Gozaliasl}, {Hayward}, {He}, {Hemmati}, {Hirschmann}, {Jahnke}, {Jin}, {Khostovan}, {Kokorev}, {Lambrides}, {Laigle}, {Larson}, {Leung}, {Liu}, {Liaudat}, {Long}, {Magdis}, {Mahler}, {Mainieri}, {Manning}, {Maraston}, {Martin}, {McCleary}, {McKinney}, {McPartland}, {Mobasher}, {Pattnaik}, {Renzini}, {Rich}, {Sanders}, {Sattari}, {Scognamiglio}, {Scoville}, {Sheth}, {Shuntov}, {Sparre}, {Suzuki}, {Talia}, {Toft}, {Trakhtenbrot}, {Urry}, {Valentino}, {Vanderhoof}, {Vardoulaki}, {Weaver}, {Whitaker}, {Wilkins}, {Yang}, \& {Zavala}}]{CCasey2023}
{Casey}, C.~M., {Kartaltepe}, J.~S., {Drakos}, N.~E., {et~al.} 2023, \apj, 954, 31, \dodoi{10.3847/1538-4357/acc2bc}

\bibitem[{{Cecchi} {et~al.}(2019){Cecchi}, {Bolzonella}, {Cimatti}, \& {Girelli}}]{RCecchi2019}
{Cecchi}, R., {Bolzonella}, M., {Cimatti}, A., \& {Girelli}, G. 2019, \apjl, 880, L14, \dodoi{10.3847/2041-8213/ab2c80}

\bibitem[{{Chabrier}(2003)}]{GChabrier2003}
{Chabrier}, G. 2003, \pasp, 115, 763, \dodoi{10.1086/376392}

\bibitem[{{Chang} {et~al.}(2013){Chang}, {van der Wel}, {Rix}, {Holden}, {Bell}, {McGrath}, {Wuyts}, {H{\"a}ussler}, {Barden}, {Faber}, {Mozena}, {Ferguson}, {Guo}, {Galametz}, {Grogin}, {Kocevski}, {Koekemoer}, {Dekel}, {Huang}, {Hathi}, \& {Donley}}]{YChang2013}
{Chang}, Y.-Y., {van der Wel}, A., {Rix}, H.-W., {et~al.} 2013, \apj, 773, 149, \dodoi{10.1088/0004-637X/773/2/149}

\bibitem[{{Choi} {et~al.}(2014){Choi}, {Conroy}, {Moustakas}, {Graves}, {Holden}, {Brodwin}, {Brown}, \& {van Dokkum}}]{JChoi2014}
{Choi}, J., {Conroy}, C., {Moustakas}, J., {et~al.} 2014, \apj, 792, 95, \dodoi{10.1088/0004-637X/792/2/95}

\bibitem[{{Choi} {et~al.}(2016){Choi}, {Dotter}, {Conroy}, {Cantiello}, {Paxton}, \& {Johnson}}]{JChoi2016}
{Choi}, J., {Dotter}, A., {Conroy}, C., {et~al.} 2016, \apj, 823, 102, \dodoi{10.3847/0004-637X/823/2/102}

\bibitem[{{Cimatti} {et~al.}(2004){Cimatti}, {Daddi}, {Renzini}, {Cassata}, {Vanzella}, {Pozzetti}, {Cristiani}, {Fontana}, {Rodighiero}, {Mignoli}, \& {Zamorani}}]{ACimatti2004}
{Cimatti}, A., {Daddi}, E., {Renzini}, A., {et~al.} 2004, \nat, 430, 184, \dodoi{10.1038/nature02668}

\bibitem[{{Conroy} \& {Gunn}(2010)}]{CConroy2010}
{Conroy}, C., \& {Gunn}, J.~E. 2010, \apj, 712, 833, \dodoi{10.1088/0004-637X/712/2/833}

\bibitem[{{Conroy} {et~al.}(2009){Conroy}, {Gunn}, \& {White}}]{CConroy2009}
{Conroy}, C., {Gunn}, J.~E., \& {White}, M. 2009, \apj, 699, 486, \dodoi{10.1088/0004-637X/699/1/486}

\bibitem[{{Conroy} \& {van Dokkum}(2012)}]{CConroy2012d}
{Conroy}, C., \& {van Dokkum}, P.~G. 2012, \apj, 760, 71, \dodoi{10.1088/0004-637X/760/1/71}

\bibitem[{{Conroy} {et~al.}(2018){Conroy}, {Villaume}, {van Dokkum}, \& {Lind}}]{CConroy2018}
{Conroy}, C., {Villaume}, A., {van Dokkum}, P.~G., \& {Lind}, K. 2018, \apj, 854, 139, \dodoi{10.3847/1538-4357/aaab49}

\bibitem[{{Cutler} {et~al.}(2022){Cutler}, {Whitaker}, {Mowla}, {Brammer}, {van der Wel}, {Marchesini}, {van Dokkum}, {Momcheva}, {Song}, {Akhshik}, {Nelson}, {Bezanson}, {Franx}, {Kriek}, {Lange-Vagle}, {Leja}, {MacKenty}, {Muzzin}, \& {Shipley}}]{SCutler2022}
{Cutler}, S.~E., {Whitaker}, K.~E., {Mowla}, L.~A., {et~al.} 2022, \apj, 925, 34, \dodoi{10.3847/1538-4357/ac341c}

\bibitem[{{Daddi} {et~al.}(2005){Daddi}, {Renzini}, {Pirzkal}, {Cimatti}, {Malhotra}, {Stiavelli}, {Xu}, {Pasquali}, {Rhoads}, {Brusa}, {di Serego Alighieri}, {Ferguson}, {Koekemoer}, {Moustakas}, {Panagia}, \& {Windhorst}}]{EDaddi2005}
{Daddi}, E., {Renzini}, A., {Pirzkal}, N., {et~al.} 2005, \apj, 626, 680, \dodoi{10.1086/430104}

\bibitem[{{Davies} {et~al.}(2024){Davies}, {Belli}, {Park}, {Mendel}, {Johnson}, {Conroy}, {Benton}, {Bugiani}, {Emami}, {Leja}, {Li}, {Maheson}, {Mathews}, {Naidu}, {Nelson}, {Tacchella}, {Terrazas}, \& {Weinberger}}]{RDavies2024}
{Davies}, R.~L., {Belli}, S., {Park}, M., {et~al.} 2024, \mnras, 528, 4976, \dodoi{10.1093/mnras/stae327}

\bibitem[{{de Graaff} {et~al.}(2024{\natexlab{a}}){de Graaff}, {Setton}, {Brammer}, {Cutler}, {Suess}, {Labbe}, {Leja}, {Weibel}, {Maseda}, {Whitaker}, {Bezanson}, {Boogaard}, {Cleri}, {De Lucia}, {Franx}, {Greene}, {Hirschmann}, {Matthee}, {McConachie}, {Naidu}, {Oesch}, {Price}, {Rix}, {Valentino}, {Wang}, \& {Williams}}]{AdeGraaff2024}
{de Graaff}, A., {Setton}, D.~J., {Brammer}, G., {et~al.} 2024{\natexlab{a}}, arXiv e-prints, arXiv:2404.05683, \dodoi{10.48550/arXiv.2404.05683}

\bibitem[{{de Graaff} {et~al.}(2024{\natexlab{b}}){de Graaff}, {Rix}, {Carniani}, {Suess}, {Charlot}, {Curtis-Lake}, {Arribas}, {Baker}, {Boyett}, {Bunker}, {Cameron}, {Chevallard}, {Curti}, {Eisenstein}, {Franx}, {Hainline}, {Hausen}, {Ji}, {Johnson}, {Jones}, {Maiolino}, {Maseda}, {Nelson}, {Parlanti}, {Rawle}, {Robertson}, {Tacchella}, {{\"U}bler}, {Williams}, {Willmer}, \& {Willott}}]{AdeGraaff2023}
{de Graaff}, A., {Rix}, H.-W., {Carniani}, S., {et~al.} 2024{\natexlab{b}}, \aap, 684, A87, \dodoi{10.1051/0004-6361/202347755}

\bibitem[{{D'Eugenio} {et~al.}(2023){D'Eugenio}, {Perez-Gonzalez}, {Maiolino}, {Scholtz}, {Perna}, {Circosta}, {Uebler}, {Arribas}, {Boeker}, {Bunker}, {Carniani}, {Charlot}, {Chevallard}, {Cresci}, {Curtis-Lake}, {Jones}, {Kumari}, {Lamperti}, {Looser}, {Parlanti}, {Rix}, {Robertson}, {Rodriguez Del Pino}, {Tacchella}, {Venturi}, \& {Willott}}]{FDEugenio2023}
{D'Eugenio}, F., {Perez-Gonzalez}, P., {Maiolino}, R., {et~al.} 2023, arXiv e-prints, arXiv:2308.06317, \dodoi{10.48550/arXiv.2308.06317}

\bibitem[{{Dotter}(2016)}]{ADotter2016}
{Dotter}, A. 2016, \apjs, 222, 8, \dodoi{10.3847/0067-0049/222/1/8}

\bibitem[{{Esdaile} {et~al.}(2021){Esdaile}, {Glazebrook}, {Labb{\'e}}, {Taylor}, {Schreiber}, {Nanayakkara}, {Kacprzak}, {Oesch}, {Tran}, {Papovich}, {Spitler}, \& {Straatman}}]{JEsdaile2021}
{Esdaile}, J., {Glazebrook}, K., {Labb{\'e}}, I., {et~al.} 2021, \apjl, 908, L35, \dodoi{10.3847/2041-8213/abe11e}

\bibitem[{{Falc{\'o}n-Barroso} {et~al.}(2011){Falc{\'o}n-Barroso}, {S{\'a}nchez-Bl{\'a}zquez}, {Vazdekis}, {Ricciardelli}, {Cardiel}, {Cenarro}, {Gorgas}, \& {Peletier}}]{JFalcon2011}
{Falc{\'o}n-Barroso}, J., {S{\'a}nchez-Bl{\'a}zquez}, P., {Vazdekis}, A., {et~al.} 2011, \aap, 532, A95, \dodoi{10.1051/0004-6361/201116842}

\bibitem[{{Ferruit} {et~al.}(2022){Ferruit}, {Jakobsen}, {Giardino}, {Rawle}, {Alves de Oliveira}, {Arribas}, {Beck}, {Birkmann}, {B{\"o}ker}, {Bunker}, {Charlot}, {de Marchi}, {Franx}, {Henry}, {Karakla}, {Kassin}, {Kumari}, {L{\'o}pez-Caniego}, {L{\"u}tzgendorf}, {Maiolino}, {Manjavacas}, {Marston}, {Moseley}, {Muzerolle}, {Pirzkal}, {Rauscher}, {Rix}, {Sabbi}, {Sirianni}, {te Plate}, {Valenti}, {Willott}, \& {Zeidler}}]{PFerruit2022}
{Ferruit}, P., {Jakobsen}, P., {Giardino}, G., {et~al.} 2022, \aap, 661, A81, \dodoi{10.1051/0004-6361/202142673}

\bibitem[{{Forrest} {et~al.}(2020{\natexlab{a}}){Forrest}, {Annunziatella}, {Wilson}, {Marchesini}, {Muzzin}, {Cooper}, {Marsan}, {McConachie}, {Chan}, {Gomez}, {Kado-Fong}, {L Barbera}, {Labb{\'e}}, {Lange-Vagle}, {Nantais}, {Nonino}, {Pe{\~n}a}, {Saracco}, {Stefanon}, \& {van der Burg}}]{BForrest2020a}
{Forrest}, B., {Annunziatella}, M., {Wilson}, G., {et~al.} 2020{\natexlab{a}}, \apjl, 890, L1, \dodoi{10.3847/2041-8213/ab5b9f}

\bibitem[{{Forrest} {et~al.}(2020{\natexlab{b}}){Forrest}, {Marsan}, {Annunziatella}, {Wilson}, {Muzzin}, {Marchesini}, {Cooper}, {Chan}, {McConachie}, {Gomez}, {Kado-Fong}, {La Barbera}, {Lange-Vagle}, {Nantais}, {Nonino}, {Saracco}, {Stefanon}, \& {van der Burg}}]{BForrest2020b}
{Forrest}, B., {Marsan}, Z.~C., {Annunziatella}, M., {et~al.} 2020{\natexlab{b}}, \apj, 903, 47, \dodoi{10.3847/1538-4357/abb819}

\bibitem[{{Forrest} {et~al.}(2022){Forrest}, {Wilson}, {Muzzin}, {Marchesini}, {Cooper}, {Marsan}, {Annunziatella}, {McConachie}, {Zaidi}, {Gomez}, {Urbano Stawinski}, {Chang}, {de Lucia}, {La Barbera}, {Lubin}, {Nantais}, {Pe{\~n}a}, {Saracco}, {Surace}, \& {Stefanon}}]{BForrest2022}
{Forrest}, B., {Wilson}, G., {Muzzin}, A., {et~al.} 2022, \apj, 938, 109, \dodoi{10.3847/1538-4357/ac8747}

\bibitem[{{Franx} {et~al.}(2003){Franx}, {Labb{\'e}}, {Rudnick}, {van Dokkum}, {Daddi}, {F{\"o}rster Schreiber}, {Moorwood}, {Rix}, {R{\"o}ttgering}, {van der Wel}, {van der Werf}, \& {van Starkenburg}}]{MFranx2003}
{Franx}, M., {Labb{\'e}}, I., {Rudnick}, G., {et~al.} 2003, \apjl, 587, L79, \dodoi{10.1086/375155}

\bibitem[{{Gallazzi} {et~al.}(2014){Gallazzi}, {Bell}, {Zibetti}, {Brinchmann}, \& {Kelson}}]{AGallazzi2014}
{Gallazzi}, A., {Bell}, E.~F., {Zibetti}, S., {Brinchmann}, J., \& {Kelson}, D.~D. 2014, \apj, 788, 72, \dodoi{10.1088/0004-637X/788/1/72}

\bibitem[{{Gargiulo} {et~al.}(2017){Gargiulo}, {Bolzonella}, {Scodeggio}, {Krywult}, {De Lucia}, {Guzzo}, {Garilli}, {Granett}, {de la Torre}, {Abbas}, {Adami}, {Arnouts}, {Bottini}, {Cappi}, {Cucciati}, {Davidzon}, {Franzetti}, {Fritz}, {Haines}, {Hawken}, {Iovino}, {Le Brun}, {Le F{\`e}vre}, {Maccagni}, {Ma{\l}ek}, {Marulli}, {Moutard}, {Polletta}, {Pollo}, {Tasca}, {Tojeiro}, {Vergani}, {Zanichelli}, {Zamorani}, {Bel}, {Branchini}, {Coupon}, {Ilbert}, {Moscardini}, \& {Peacock}}]{AGargiulo2017}
{Gargiulo}, A., {Bolzonella}, M., {Scodeggio}, M., {et~al.} 2017, \aap, 606, A113, \dodoi{10.1051/0004-6361/201630112}

\bibitem[{{Glazebrook} {et~al.}(2017){Glazebrook}, {Schreiber}, {Labb{\'e}}, {Nanayakkara}, {Kacprzak}, {Oesch}, {Papovich}, {Spitler}, {Straatman}, {Tran}, \& {Yuan}}]{KGlazebrook2017}
{Glazebrook}, K., {Schreiber}, C., {Labb{\'e}}, I., {et~al.} 2017, \nat, 544, 71, \dodoi{10.1038/nature21680}

\bibitem[{{Glazebrook} {et~al.}(2023){Glazebrook}, {Nanayakkara}, {Schreiber}, {Lagos}, {Kawinwanichakij}, {Jacobs}, {Chittenden}, {Brammer}, {Kacprzak}, {Labbe}, {Marchesini}, {Marsan}, {Oesch}, {Papovich}, {Remus}, {Tran}, {Esdaile}, \& {Chandro Gomez}}]{KGlazebrook2023}
{Glazebrook}, K., {Nanayakkara}, T., {Schreiber}, C., {et~al.} 2023, arXiv e-prints, arXiv:2308.05606, \dodoi{10.48550/arXiv.2308.05606}

\bibitem[{{Gould} {et~al.}(2023){Gould}, {Brammer}, {Valentino}, {Whitaker}, {Weaver}, {Lagos}, {Rizzo}, {Franco}, {Hsieh}, {Ilbert}, {Jin}, {Magdis}, {McCracken}, {Mobasher}, {Shuntov}, {Steinhardt}, {Strait}, \& {Toft}}]{KGould2023}
{Gould}, K. M.~L., {Brammer}, G., {Valentino}, F., {et~al.} 2023, \aj, 165, 248, \dodoi{10.3847/1538-3881/accadc}

\bibitem[{{Grogin} {et~al.}(2011){Grogin}, {Kocevski}, {Faber}, {Ferguson}, {Koekemoer}, {Riess}, {Acquaviva}, {Alexander}, {Almaini}, {Ashby}, {Barden}, {Bell}, {Bournaud}, {Brown}, {Caputi}, {Casertano}, {Cassata}, {Castellano}, {Challis}, {Chary}, {Cheung}, {Cirasuolo}, {Conselice}, {Roshan Cooray}, {Croton}, {Daddi}, {Dahlen}, {Dav{\'e}}, {de Mello}, {Dekel}, {Dickinson}, {Dolch}, {Donley}, {Dunlop}, {Dutton}, {Elbaz}, {Fazio}, {Filippenko}, {Finkelstein}, {Fontana}, {Gardner}, {Garnavich}, {Gawiser}, {Giavalisco}, {Grazian}, {Guo}, {Hathi}, {H{\"a}ussler}, {Hopkins}, {Huang}, {Huang}, {Jha}, {Kartaltepe}, {Kirshner}, {Koo}, {Lai}, {Lee}, {Li}, {Lotz}, {Lucas}, {Madau}, {McCarthy}, {McGrath}, {McIntosh}, {McLure}, {Mobasher}, {Moustakas}, {Mozena}, {Nandra}, {Newman}, {Niemi}, {Noeske}, {Papovich}, {Pentericci}, {Pope}, {Primack}, {Rajan}, {Ravindranath}, {Reddy}, {Renzini}, {Rix}, {Robaina}, {Rodney}, {Rosario}, {Rosati}, {Salimbeni}, {Scarlata}, {Siana}, {Simard}, {Smidt}, {Somerville}, {Spinrad},
  {Straughn}, {Strolger}, {Telford}, {Teplitz}, {Trump}, {van der Wel}, {Villforth}, {Wechsler}, {Weiner}, {Wiklind}, {Wild}, {Wilson}, {Wuyts}, {Yan}, \& {Yun}}]{NGrogin2011}
{Grogin}, N.~A., {Kocevski}, D.~D., {Faber}, S.~M., {et~al.} 2011, \apjs, 197, 35, \dodoi{10.1088/0067-0049/197/2/35}

\bibitem[{{Gu} {et~al.}(2022){Gu}, {Greene}, {Newman}, {Kreisch}, {Quenneville}, {Ma}, \& {Blakeslee}}]{MGu2022}
{Gu}, M., {Greene}, J.~E., {Newman}, A.~B., {et~al.} 2022, \apj, 932, 103, \dodoi{10.3847/1538-4357/ac69ea}

\bibitem[{{Hamadouche} {et~al.}(2022){Hamadouche}, {Carnall}, {McLure}, {Dunlop}, {McLeod}, {Cullen}, {Begley}, {Bolzonella}, {Buitrago}, {Castellano}, {Cucciati}, {Fontana}, {Gargiulo}, {Moresco}, {Pozzetti}, \& {Zamorani}}]{MHamadouche2022}
{Hamadouche}, M.~L., {Carnall}, A.~C., {McLure}, R.~J., {et~al.} 2022, \mnras, 512, 1262, \dodoi{10.1093/mnras/stac535}

\bibitem[{{Hamadouche} {et~al.}(2023){Hamadouche}, {Carnall}, {McLure}, {Dunlop}, {Begley}, {Cullen}, {McLeod}, {Donnan}, \& {Stanton}}]{MHamadouche2023}
---. 2023, \mnras, 521, 5400, \dodoi{10.1093/mnras/stad773}

\bibitem[{{Hartley} {et~al.}(2023){Hartley}, {Nelson}, {Suess}, {Garcia}, {Park}, {Hernquist}, {Bezanson}, {Nevin}, {Pillepich}, {Schechter}, {Terrazas}, {Torrey}, {Wellons}, {Whitaker}, \& {Williams}}]{AHartley2023}
{Hartley}, A.~I., {Nelson}, E.~J., {Suess}, K.~A., {et~al.} 2023, \mnras, 522, 3138, \dodoi{10.1093/mnras/stad1162}

\bibitem[{{Horne}(1986)}]{KHorne1986}
{Horne}, K. 1986, \pasp, 98, 609, \dodoi{10.1086/131801}

\bibitem[{{Jafariyazani} {et~al.}(2020){Jafariyazani}, {Newman}, {Mobasher}, {Belli}, {Ellis}, \& {Patel}}]{MJafariyazani2020}
{Jafariyazani}, M., {Newman}, A.~B., {Mobasher}, B., {et~al.} 2020, \apjl, 897, L42, \dodoi{10.3847/2041-8213/aba11c}

\bibitem[{{Ji} \& {Giavalisco}(2022)}]{ZJi2022}
{Ji}, Z., \& {Giavalisco}, M. 2022, \apj, 935, 120, \dodoi{10.3847/1538-4357/ac7f43}

\bibitem[{{Johnson} {et~al.}(2021){Johnson}, {Leja}, {Conroy}, \& {Speagle}}]{BJohnson2021}
{Johnson}, B.~D., {Leja}, J., {Conroy}, C., \& {Speagle}, J.~S. 2021, \apjs, 254, 22, \dodoi{10.3847/1538-4365/abef67}

\bibitem[{{Khochfar} \& {Silk}(2006)}]{SKhochfar2006}
{Khochfar}, S., \& {Silk}, J. 2006, \apjl, 648, L21, \dodoi{10.1086/507768}

\bibitem[{{Koekemoer} {et~al.}(2011){Koekemoer}, {Faber}, {Ferguson}, {Grogin}, {Kocevski}, {Koo}, {Lai}, {Lotz}, {Lucas}, {McGrath}, {Ogaz}, {Rajan}, {Riess}, {Rodney}, {Strolger}, {Casertano}, {Castellano}, {Dahlen}, {Dickinson}, {Dolch}, {Fontana}, {Giavalisco}, {Grazian}, {Guo}, {Hathi}, {Huang}, {van der Wel}, {Yan}, {Acquaviva}, {Alexander}, {Almaini}, {Ashby}, {Barden}, {Bell}, {Bournaud}, {Brown}, {Caputi}, {Cassata}, {Challis}, {Chary}, {Cheung}, {Cirasuolo}, {Conselice}, {Roshan Cooray}, {Croton}, {Daddi}, {Dav{\'e}}, {de Mello}, {de Ravel}, {Dekel}, {Donley}, {Dunlop}, {Dutton}, {Elbaz}, {Fazio}, {Filippenko}, {Finkelstein}, {Frazer}, {Gardner}, {Garnavich}, {Gawiser}, {Gruetzbauch}, {Hartley}, {H{\"a}ussler}, {Herrington}, {Hopkins}, {Huang}, {Jha}, {Johnson}, {Kartaltepe}, {Khostovan}, {Kirshner}, {Lani}, {Lee}, {Li}, {Madau}, {McCarthy}, {McIntosh}, {McLure}, {McPartland}, {Mobasher}, {Moreira}, {Mortlock}, {Moustakas}, {Mozena}, {Nandra}, {Newman}, {Nielsen}, {Niemi}, {Noeske}, {Papovich},
  {Pentericci}, {Pope}, {Primack}, {Ravindranath}, {Reddy}, {Renzini}, {Rix}, {Robaina}, {Rosario}, {Rosati}, {Salimbeni}, {Scarlata}, {Siana}, {Simard}, {Smidt}, {Snyder}, {Somerville}, {Spinrad}, {Straughn}, {Telford}, {Teplitz}, {Trump}, {Vargas}, {Villforth}, {Wagner}, {Wandro}, {Wechsler}, {Weiner}, {Wiklind}, {Wild}, {Wilson}, {Wuyts}, \& {Yun}}]{AKoekemoer2011}
{Koekemoer}, A.~M., {Faber}, S.~M., {Ferguson}, H.~C., {et~al.} 2011, \apjs, 197, 36, \dodoi{10.1088/0067-0049/197/2/36}

\bibitem[{{Kriek} \& {Conroy}(2013)}]{MKriek2013}
{Kriek}, M., \& {Conroy}, C. 2013, \apjl, 775, L16, \dodoi{10.1088/2041-8205/775/1/L16}

\bibitem[{{Kriek} {et~al.}(2006){Kriek}, {van Dokkum}, {Franx}, {Quadri}, {Gawiser}, {Herrera}, {Illingworth}, {Labb{\'e}}, {Lira}, {Marchesini}, {Rix}, {Rudnick}, {Taylor}, {Toft}, {Urry}, \& {Wuyts}}]{MKriek2006}
{Kriek}, M., {van Dokkum}, P.~G., {Franx}, M., {et~al.} 2006, \apjl, 649, L71, \dodoi{10.1086/508371}

\bibitem[{{Kriek} {et~al.}(2010){Kriek}, {Labb{\'e}}, {Conroy}, {Whitaker}, {van Dokkum}, {Brammer}, {Franx}, {Illingworth}, {Marchesini}, {Muzzin}, {Quadri}, \& {Rudnick}}]{MKriek2010}
{Kriek}, M., {Labb{\'e}}, I., {Conroy}, C., {et~al.} 2010, \apjl, 722, L64, \dodoi{10.1088/2041-8205/722/1/L64}

\bibitem[{{Kriek} {et~al.}(2016){Kriek}, {Conroy}, {van Dokkum}, {Shapley}, {Choi}, {Reddy}, {Siana}, {van de Voort}, {Coil}, \& {Mobasher}}]{MKriek2016}
{Kriek}, M., {Conroy}, C., {van Dokkum}, P.~G., {et~al.} 2016, \nat, 540, 248, \dodoi{10.1038/nature20570}

\bibitem[{{Kriek} {et~al.}(2019){Kriek}, {Price}, {Conroy}, {Suess}, {Mowla}, {Pasha}, {Bezanson}, {van Dokkum}, \& {Barro}}]{MKriek2019}
{Kriek}, M., {Price}, S.~H., {Conroy}, C., {et~al.} 2019, \apjl, 880, L31, \dodoi{10.3847/2041-8213/ab2e75}

\bibitem[{{Kriek} {et~al.}(2023){Kriek}, {Beverage}, {Price}, {Suess}, {Barro}, {Bezanson}, {Conroy}, {Cutler}, {Franx}, {Lin}, {Lorenz}, {Ma}, {Momcheva}, {Mowla}, {Pasha}, {van Dokkum}, \& {Whitaker}}]{MKriek2023}
{Kriek}, M., {Beverage}, A.~G., {Price}, S.~H., {et~al.} 2023, arXiv e-prints, arXiv:2311.16232, \dodoi{10.48550/arXiv.2311.16232}

\bibitem[{{Kroupa}(2001)}]{PKroupa2001}
{Kroupa}, P. 2001, \mnras, 322, 231, \dodoi{10.1046/j.1365-8711.2001.04022.x}

\bibitem[{Lagos {et~al.}(2018)Lagos, Stevens, Bower, Davis, Contreras, Padilla, Obreschkow, Croton, Trayford, Welker, \& Theuns}]{CLagos2018}
Lagos, C. d.~P., Stevens, A. R.~H., Bower, R.~G., {et~al.} 2018, Monthly Notices of the Royal Astronomical Society, 473, 4956–4974, \dodoi{10.1093/mnras/stx2667}

\bibitem[{{Leja} {et~al.}(2019{\natexlab{a}}){Leja}, {Carnall}, {Johnson}, {Conroy}, \& {Speagle}}]{JLeja2019_b}
{Leja}, J., {Carnall}, A.~C., {Johnson}, B.~D., {Conroy}, C., \& {Speagle}, J.~S. 2019{\natexlab{a}}, \apj, 876, 3, \dodoi{10.3847/1538-4357/ab133c}

\bibitem[{{Leja} {et~al.}(2019{\natexlab{b}}){Leja}, {Johnson}, {Conroy}, {van Dokkum}, {Speagle}, {Brammer}, {Momcheva}, {Skelton}, {Whitaker}, {Franx}, \& {Nelson}}]{JLeja2019}
{Leja}, J., {Johnson}, B.~D., {Conroy}, C., {et~al.} 2019{\natexlab{b}}, \apj, 877, 140, \dodoi{10.3847/1538-4357/ab1d5a}

\bibitem[{{Leja} {et~al.}(2022){Leja}, {Speagle}, {Ting}, {Johnson}, {Conroy}, {Whitaker}, {Nelson}, {van Dokkum}, \& {Franx}}]{JLeja2022}
{Leja}, J., {Speagle}, J.~S., {Ting}, Y.-S., {et~al.} 2022, \apj, 936, 165, \dodoi{10.3847/1538-4357/ac887d}

\bibitem[{{Lu} {et~al.}(2024){Lu}, {Daddi}, {Maraston}, {Dickinson}, {Arrabal Haro}, {Gobat}, {Renzini}, {Giavalisco}, {Bagley}, {Calabr{\`o}}, {Cheng}, {de la Vega}, {D'Eugenio}, {Elbaz}, {Finkelstein}, {G{\'o}mez-Guijarro}, {Gu}, {Hathi}, {Huertas-Company}, {Kartaltepe}, {Koekemoer}, {Le Bail}, {Lyu}, {Magnelli}, {Mobasher}, {Papovich}, {Pirzkal}, {Rich}, {Tacchella}, \& {Yung}}]{SLu2024}
{Lu}, S., {Daddi}, E., {Maraston}, C., {et~al.} 2024, arXiv e-prints, arXiv:2403.07414.
\newblock \doarXiv{2403.07414}

\bibitem[{{Maiolino} \& {Mannucci}(2019)}]{RMaiolino2019}
{Maiolino}, R., \& {Mannucci}, F. 2019, \aapr, 27, 3, \dodoi{10.1007/s00159-018-0112-2}

\bibitem[{{Maraston}(2005)}]{CMaraston2005}
{Maraston}, C. 2005, \mnras, 362, 799, \dodoi{10.1111/j.1365-2966.2005.09270.x}

\bibitem[{{Marchesini} {et~al.}(2023){Marchesini}, {Brammer}, {Morishita}, {Bergamini}, {Wang}, {Bradac}, {Roberts-Borsani}, {Strait}, {Treu}, {Fontana}, {Jones}, {Santini}, {Vulcani}, {Acebron}, {Calabr{\`o}}, {Castellano}, {Glazebrook}, {Grillo}, {Mercurio}, {Nanayakkara}, {Rosati}, {Tubthong}, \& {Vanzella}}]{DMarchesini2023}
{Marchesini}, D., {Brammer}, G., {Morishita}, T., {et~al.} 2023, \apjl, 942, L25, \dodoi{10.3847/2041-8213/acaaac}

\bibitem[{{Maseda} {et~al.}(2021){Maseda}, {van der Wel}, {Franx}, {Bell}, {Bezanson}, {Muzzin}, {Sobral}, {D'Eugenio}, {Gallazzi}, {de Graaff}, {Leja}, {Straatman}, {Whitaker}, {Williams}, \& {Wu}}]{MMaseda2021}
{Maseda}, M.~V., {van der Wel}, A., {Franx}, M., {et~al.} 2021, \apj, 923, 18, \dodoi{10.3847/1538-4357/ac2bfe}

\bibitem[{{McCracken} {et~al.}(2012){McCracken}, {Milvang-Jensen}, {Dunlop}, {Franx}, {Fynbo}, {Le F{\`e}vre}, {Holt}, {Caputi}, {Goranova}, {Buitrago}, {Emerson}, {Freudling}, {Hudelot}, {L{\'o}pez-Sanjuan}, {Magnard}, {Mellier}, {M{\o}ller}, {Nilsson}, {Sutherland}, {Tasca}, \& {Zabl}}]{HMcCracken2012}
{McCracken}, H.~J., {Milvang-Jensen}, B., {Dunlop}, J., {et~al.} 2012, \aap, 544, A156, \dodoi{10.1051/0004-6361/201219507}

\bibitem[{{McLeod} {et~al.}(2021){McLeod}, {McLure}, {Dunlop}, {Cullen}, {Carnall}, \& {Duncan}}]{DMcLeod2021}
{McLeod}, D.~J., {McLure}, R.~J., {Dunlop}, J.~S., {et~al.} 2021, \mnras, 503, 4413, \dodoi{10.1093/mnras/stab731}

\bibitem[{{Mendel} {et~al.}(2020){Mendel}, {Beifiori}, {Saglia}, {Bender}, {Brammer}, {Chan}, {F{\"o}rster Schreiber}, {Fossati}, {Galametz}, {Momcheva}, {Nelson}, {Wilman}, \& {Wuyts}}]{JMendel2020}
{Mendel}, J.~T., {Beifiori}, A., {Saglia}, R.~P., {et~al.} 2020, \apj, 899, 87, \dodoi{10.3847/1538-4357/ab9ffc}

\bibitem[{{Miller} {et~al.}(2023){Miller}, {van Dokkum}, \& {Mowla}}]{TMiller2023}
{Miller}, T.~B., {van Dokkum}, P., \& {Mowla}, L. 2023, \apj, 945, 155, \dodoi{10.3847/1538-4357/acbc74}

\bibitem[{{Momcheva} {et~al.}(2017){Momcheva}, {van Dokkum}, {van der Wel}, {Brammer}, {MacKenty}, {Nelson}, {Leja}, {Muzzin}, \& {Franx}}]{IMomcheva2017}
{Momcheva}, I.~G., {van Dokkum}, P.~G., {van der Wel}, A., {et~al.} 2017, \pasp, 129, 015004, \dodoi{10.1088/1538-3873/129/971/015004}

\bibitem[{{Mowla} {et~al.}(2018){Mowla}, {van Dokkum}, {Brammer}, {Momcheva}, {van der Wel}, {Whitaker}, {Nelson}, {Bezanson}, {Muzzin}, {Franx}, {MacKenty}, {Leja}, {Kriek}, \& {Marchesini}}]{LMowla2018}
{Mowla}, L., {van Dokkum}, P., {Brammer}, G., {et~al.} 2018, ArXiv e-prints.
\newblock \doarXiv{1808.04379}

\bibitem[{{Muzzin} {et~al.}(2013{\natexlab{a}}){Muzzin}, {Marchesini}, {Stefanon}, {Franx}, {McCracken}, {Milvang-Jensen}, {Dunlop}, {Fynbo}, {Brammer}, {Labb{\'e}}, \& {van Dokkum}}]{AMuzzin2013b}
{Muzzin}, A., {Marchesini}, D., {Stefanon}, M., {et~al.} 2013{\natexlab{a}}, \apj, 777, 18, \dodoi{10.1088/0004-637X/777/1/18}

\bibitem[{{Muzzin} {et~al.}(2013{\natexlab{b}}){Muzzin}, {Marchesini}, {Stefanon}, {Franx}, {Milvang-Jensen}, {Dunlop}, {Fynbo}, {Brammer}, {Labb{\'e}}, \& {van Dokkum}}]{AMuzzin2013a}
---. 2013{\natexlab{b}}, \apjs, 206, 8, \dodoi{10.1088/0067-0049/206/1/8}

\bibitem[{{Naab} {et~al.}(2009){Naab}, {Johansson}, \& {Ostriker}}]{TNaab2009}
{Naab}, T., {Johansson}, P.~H., \& {Ostriker}, J.~P. 2009, \apjl, 699, L178, \dodoi{10.1088/0004-637X/699/2/L178}

\bibitem[{{Naab} {et~al.}(2014){Naab}, {Oser}, {Emsellem}, {Cappellari}, {Krajnovi{\'c}}, {McDermid}, {Alatalo}, {Bayet}, {Blitz}, {Bois}, {Bournaud}, {Bureau}, {Crocker}, {Davies}, {Davis}, {de Zeeuw}, {Duc}, {Hirschmann}, {Johansson}, {Khochfar}, {Kuntschner}, {Morganti}, {Oosterloo}, {Sarzi}, {Scott}, {Serra}, {van de Ven}, {Weijmans}, \& {Young}}]{TNaab2014}
{Naab}, T., {Oser}, L., {Emsellem}, E., {et~al.} 2014, \mnras, 444, 3357, \dodoi{10.1093/mnras/stt1919}

\bibitem[{{Nanayakkara} {et~al.}(2024){Nanayakkara}, {Glazebrook}, {Jacobs}, {Kawinwanichakij}, {Schreiber}, {Brammer}, {Esdaile}, {Kacprzak}, {Labbe}, {Lagos}, {Marchesini}, {Marsan}, {Oesch}, {Papovich}, {Remus}, \& {Tran}}]{TNanayakkara2024}
{Nanayakkara}, T., {Glazebrook}, K., {Jacobs}, C., {et~al.} 2024, Scientific Reports, 14, 3724, \dodoi{10.1038/s41598-024-52585-4}

\bibitem[{{Newman} {et~al.}(2018){Newman}, {Belli}, {Ellis}, \& {Patel}}]{ANewman2018b}
{Newman}, A.~B., {Belli}, S., {Ellis}, R.~S., \& {Patel}, S.~G. 2018, \apj, 862, 126, \dodoi{10.3847/1538-4357/aacd4f}

\bibitem[{{Newman} {et~al.}(2012){Newman}, {Ellis}, {Bundy}, \& {Treu}}]{ANewman2012}
{Newman}, A.~B., {Ellis}, R.~S., {Bundy}, K., \& {Treu}, T. 2012, \apj, 746, 162, \dodoi{10.1088/0004-637X/746/2/162}

\bibitem[{{Nidever} {et~al.}(2024){Nidever}, {Gilbert}, {Tollerud}, {Siders}, {Escala}, {Prieto}, {Smith}, {Cunha}, {Debattista}, {Ting}, \& {Kirby}}]{DNidever2024}
{Nidever}, D.~L., {Gilbert}, K., {Tollerud}, E., {et~al.} 2024, in Early Disk-Galaxy Formation from JWST to the Milky Way, ed. F.~{Tabatabaei}, B.~{Barbuy}, \& Y.-S. {Ting}, Vol. 377, 115--122, \dodoi{10.1017/S1743921323002016}

\bibitem[{{Noll} {et~al.}(2009){Noll}, {Pierini}, {Cimatti}, {Daddi}, {Kurk}, {Bolzonella}, {Cassata}, {Halliday}, {Mignoli}, {Pozzetti}, {Renzini}, {Berta}, {Dickinson}, {Franceschini}, {Rodighiero}, {Rosati}, \& {Zamorani}}]{SNoll2009}
{Noll}, S., {Pierini}, D., {Cimatti}, A., {et~al.} 2009, \aap, 499, 69, \dodoi{10.1051/0004-6361/200811526}

\bibitem[{{Oke} \& {Gunn}(1983)}]{JOke1983}
{Oke}, J.~B., \& {Gunn}, J.~E. 1983, \apj, 266, 713, \dodoi{10.1086/160817}

\bibitem[{{Park} {et~al.}(2023){Park}, {Belli}, {Conroy}, {Tacchella}, {Leja}, {Cutler}, {Johnson}, {Nelson}, \& {Emami}}]{MPark2023}
{Park}, M., {Belli}, S., {Conroy}, C., {et~al.} 2023, \apj, 953, 119, \dodoi{10.3847/1538-4357/acd54a}

\bibitem[{{Park} {et~al.}(2024){Park}, {Belli}, {Conroy}, {Johnson}, {Davies}, {Leja}, {Tacchella}, {Mendel}, {Benton}, {Bugiani}, {Emami}, {Khoram}, {Li}, {Maheson}, {Mathews}, {Naidu}, {Nelson}, {Terrazas}, \& {Weinberger}}]{MPark2024}
---. 2024, arXiv e-prints, arXiv:2404.17945, \dodoi{10.48550/arXiv.2404.17945}

\bibitem[{{Poggianti} {et~al.}(2013){Poggianti}, {Calvi}, {Bindoni}, {D'Onofrio}, {Moretti}, {Valentinuzzi}, {Fasano}, {Fritz}, {De Lucia}, {Vulcani}, {Bettoni}, {Gullieuszik}, \& {Omizzolo}}]{BPoggianti2013}
{Poggianti}, B.~M., {Calvi}, R., {Bindoni}, D., {et~al.} 2013, \apj, 762, 77, \dodoi{10.1088/0004-637X/762/2/77}

\bibitem[{Schlawin {et~al.}(2020)Schlawin, Leisenring, Misselt, Greene, McElwain, Beatty, \& Rieke}]{ESchlawin_2020}
Schlawin, E., Leisenring, J., Misselt, K., {et~al.} 2020, The Astronomical Journal, 160, 231, \dodoi{10.3847/1538-3881/abb811}

\bibitem[{{Schreiber} {et~al.}(2018){Schreiber}, {Glazebrook}, {Nanayakkara}, {Kacprzak}, {Labb{\'e}}, {Oesch}, {Yuan}, {Tran}, {Papovich}, {Spitler}, \& {Straatman}}]{CSchreiber2018}
{Schreiber}, C., {Glazebrook}, K., {Nanayakkara}, T., {et~al.} 2018, \aap, 618, A85, \dodoi{10.1051/0004-6361/201833070}

\bibitem[{{Scoville} {et~al.}(2007){Scoville}, {Aussel}, {Brusa}, {Capak}, {Carollo}, {Elvis}, {Giavalisco}, {Guzzo}, {Hasinger}, {Impey}, {Kneib}, {LeFevre}, {Lilly}, {Mobasher}, {Renzini}, {Rich}, {Sanders}, {Schinnerer}, {Schminovich}, {Shopbell}, {Taniguchi}, \& {Tyson}}]{NScoville2007}
{Scoville}, N., {Aussel}, H., {Brusa}, M., {et~al.} 2007, \apjs, 172, 1, \dodoi{10.1086/516585}

\bibitem[{{Setton} {et~al.}(2024){Setton}, {Khullar}, {Miller}, {Bezanson}, {Greene}, {Suess}, {Whitaker}, {Antwi-Danso}, {Atek}, {Brammer}, {Cutler}, {Dayal}, {Feldmann}, {Furtak}, {Fujimoto}, {Glazebrook}, {Goulding}, {Kokorev}, {Labbe}, {Leja}, {Ma}, {Marchesini}, {Nanayakkara}, {Pan}, {Price}, {Siegel}, {Shipley}, {Weaver}, {van Dokkum}, {Wang}, \& {Williams}}]{DSetton2024}
{Setton}, D.~J., {Khullar}, G., {Miller}, T.~B., {et~al.} 2024, arXiv e-prints, arXiv:2402.05664, \dodoi{10.48550/arXiv.2402.05664}

\bibitem[{{Speagle}(2020)}]{JSpeagle2020}
{Speagle}, J.~S. 2020, \mnras, 493, 3132, \dodoi{10.1093/mnras/staa278}

\bibitem[{{Straatman} {et~al.}(2014){Straatman}, {Labb{\'e}}, {Spitler}, {Allen}, {Altieri}, {Brammer}, {Dickinson}, {van Dokkum}, {Inami}, {Glazebrook}, {Kacprzak}, {Kawinwanichakij}, {Kelson}, {McCarthy}, {Mehrtens}, {Monson}, {Murphy}, {Papovich}, {Persson}, {Quadri}, {Rees}, {Tomczak}, {Tran}, \& {Tilvi}}]{CStraatman2014}
{Straatman}, C.~M.~S., {Labb{\'e}}, I., {Spitler}, L.~R., {et~al.} 2014, \apjl, 783, L14, \dodoi{10.1088/2041-8205/783/1/L14}

\bibitem[{{Suess} {et~al.}(2019){Suess}, {Kriek}, {Price}, \& {Barro}}]{KSuess2019a}
{Suess}, K.~A., {Kriek}, M., {Price}, S.~H., \& {Barro}, G. 2019, \apj, 877, 103, \dodoi{10.3847/1538-4357/ab1bda}

\bibitem[{{Suess} {et~al.}(2020){Suess}, {Kriek}, {Price}, \& {Barro}}]{KSuess2020}
---. 2020, \apjl, 899, L26, \dodoi{10.3847/2041-8213/abacc9}

\bibitem[{{Suess} {et~al.}(2021){Suess}, {Kriek}, {Price}, \& {Barro}}]{KSuess2021}
---. 2021, \apj, 915, 87, \dodoi{10.3847/1538-4357/abf1e4}

\bibitem[{{Suess} {et~al.}(2022{\natexlab{a}}){Suess}, {Kriek}, {Bezanson}, {Greene}, {Setton}, {Spilker}, {Feldmann}, {Goulding}, {Johnson}, {Leja}, {Narayanan}, {Hall-Hooper}, {Hunt}, {Lower}, \& {Verrico}}]{KSuess2022a}
{Suess}, K.~A., {Kriek}, M., {Bezanson}, R., {et~al.} 2022{\natexlab{a}}, \apj, 926, 89, \dodoi{10.3847/1538-4357/ac404a}

\bibitem[{{Suess} {et~al.}(2022{\natexlab{b}}){Suess}, {Leja}, {Johnson}, {Bezanson}, {Greene}, {Kriek}, {Lower}, {Narayanan}, {Setton}, \& {Spilker}}]{KSuess2022b}
{Suess}, K.~A., {Leja}, J., {Johnson}, B.~D., {et~al.} 2022{\natexlab{b}}, \apj, 935, 146, \dodoi{10.3847/1538-4357/ac82b0}

\bibitem[{{Suess} {et~al.}(2023){Suess}, {Williams}, {Robertson}, {Ji}, {Johnson}, {Nelson}, {Alberts}, {Hainline}, {D'Eugenio}, {{\"U}bler}, {Rieke}, {Rieke}, {Bunker}, {Carniani}, {Charlot}, {Eisenstein}, {Maiolino}, {Stark}, {Tacchella}, \& {Willott}}]{KSuess2023}
{Suess}, K.~A., {Williams}, C.~C., {Robertson}, B., {et~al.} 2023, \apjl, 956, L42, \dodoi{10.3847/2041-8213/acf5e6}

\bibitem[{{Tanaka} {et~al.}(2019){Tanaka}, {Valentino}, {Toft}, {Onodera}, {Shimakawa}, {Ceverino}, {Faisst}, {Gallazzi}, {G{\'o}mez-Guijarro}, {Kubo}, {Magdis}, {Steinhardt}, {Stockmann}, {Yabe}, \& {Zabl}}]{MTanaka2019}
{Tanaka}, M., {Valentino}, F., {Toft}, S., {et~al.} 2019, \apjl, 885, L34, \dodoi{10.3847/2041-8213/ab4ff3}

\bibitem[{{Toft} {et~al.}(2017){Toft}, {Zabl}, {Richard}, {Gallazzi}, {Zibetti}, {Prescott}, {Grillo}, {Man}, {Lee}, {G{\'o}mez-Guijarro}, {Stockmann}, {Magdis}, \& {Steinhardt}}]{SToft2017}
{Toft}, S., {Zabl}, J., {Richard}, J., {et~al.} 2017, \nat, 546, 510, \dodoi{10.1038/nature22388}

\bibitem[{{Treu} {et~al.}(2010){Treu}, {Auger}, {Koopmans}, {Gavazzi}, {Marshall}, \& {Bolton}}]{TTreu2010}
{Treu}, T., {Auger}, M.~W., {Koopmans}, L. V.~E., {et~al.} 2010, \apj, 709, 1195, \dodoi{10.1088/0004-637X/709/2/1195}

\bibitem[{{Valentino} {et~al.}(2020){Valentino}, {Tanaka}, {Davidzon}, {Toft}, {G{\'o}mez-Guijarro}, {Stockmann}, {Onodera}, {Brammer}, {Ceverino}, {Faisst}, {Gallazzi}, {Hayward}, {Ilbert}, {Kubo}, {Magdis}, {Selsing}, {Shimakawa}, {Sparre}, {Steinhardt}, {Yabe}, \& {Zabl}}]{FValentino2020}
{Valentino}, F., {Tanaka}, M., {Davidzon}, I., {et~al.} 2020, \apj, 889, 93, \dodoi{10.3847/1538-4357/ab64dc}

\bibitem[{{Valentino} {et~al.}(2023){Valentino}, {Brammer}, {Gould}, {Kokorev}, {Fujimoto}, {Jespersen}, {Vijayan}, {Weaver}, {Ito}, {Tanaka}, {Ilbert}, {Magdis}, {Whitaker}, {Faisst}, {Gallazzi}, {Gillman}, {Gim{\'e}nez-Arteaga}, {G{\'o}mez-Guijarro}, {Kubo}, {Heintz}, {Hirschmann}, {Oesch}, {Onodera}, {Rizzo}, {Lee}, {Strait}, \& {Toft}}]{FValentino2023}
{Valentino}, F., {Brammer}, G., {Gould}, K. M.~L., {et~al.} 2023, \apj, 947, 20, \dodoi{10.3847/1538-4357/acbefa}

\bibitem[{{van der Wel} {et~al.}(2011){van der Wel}, {Rix}, {Wuyts}, {McGrath}, {Koekemoer}, {Bell}, {Holden}, {Robaina}, \& {McIntosh}}]{AvanderWel2011}
{van der Wel}, A., {Rix}, H.-W., {Wuyts}, S., {et~al.} 2011, \apj, 730, 38, \dodoi{10.1088/0004-637X/730/1/38}

\bibitem[{{van der Wel} {et~al.}(2014){van der Wel}, {Franx}, {van Dokkum}, {Skelton}, {Momcheva}, {Whitaker}, {Brammer}, {Bell}, {Rix}, {Wuyts}, {Ferguson}, {Holden}, {Barro}, {Koekemoer}, {Chang}, {McGrath}, {H{\"a}ussler}, {Dekel}, {Behroozi}, {Fumagalli}, {Leja}, {Lundgren}, {Maseda}, {Nelson}, {Wake}, {Patel}, {Labb{\'e}}, {Faber}, {Grogin}, \& {Kocevski}}]{AvanderWel2014}
{van der Wel}, A., {Franx}, M., {van Dokkum}, P.~G., {et~al.} 2014, \apj, 788, 28, \dodoi{10.1088/0004-637X/788/1/28}

\bibitem[{{van Dokkum} \& {Franx}(2001)}]{PvanDokkum2001a}
{van Dokkum}, P.~G., \& {Franx}, M. 2001, \apj, 553, 90, \dodoi{10.1086/320645}

\bibitem[{{van Dokkum} {et~al.}(2008){van Dokkum}, {Franx}, {Kriek}, {Holden}, {Illingworth}, {Magee}, {Bouwens}, {Marchesini}, {Quadri}, {Rudnick}, {Taylor}, \& {Toft}}]{PvanDokkum2008}
{van Dokkum}, P.~G., {Franx}, M., {Kriek}, M., {et~al.} 2008, \apjl, 677, L5, \dodoi{10.1086/587874}

\bibitem[{{van Dokkum} {et~al.}(2010){van Dokkum}, {Whitaker}, {Brammer}, {Franx}, {Kriek}, {Labb{\'e}}, {Marchesini}, {Quadri}, {Bezanson}, {Illingworth}, {Muzzin}, {Rudnick}, {Tal}, \& {Wake}}]{PvanDokkum2010}
{van Dokkum}, P.~G., {Whitaker}, K.~E., {Brammer}, G., {et~al.} 2010, \apj, 709, 1018, \dodoi{10.1088/0004-637X/709/2/1018}

\bibitem[{{Weller} {et~al.}(2024){Weller}, {Pacucci}, {Ni}, {Hernquist}, \& {Park}}]{EWeller2024}
{Weller}, E.~J., {Pacucci}, F., {Ni}, Y., {Hernquist}, L., \& {Park}, M. 2024, arXiv e-prints, arXiv:2406.02664, \dodoi{10.48550/arXiv.2406.02664}

\bibitem[{{Whitaker} {et~al.}(2012){Whitaker}, {Kriek}, {van Dokkum}, {Bezanson}, {Brammer}, {Franx}, \& {Labb{\'e}}}]{KWhitaker2012}
{Whitaker}, K.~E., {Kriek}, M., {van Dokkum}, P.~G., {et~al.} 2012, \apj, 745, 179, \dodoi{10.1088/0004-637X/745/2/179}

\bibitem[{{Wild} {et~al.}(2020){Wild}, {Taj Aldeen}, {Carnall}, {Maltby}, {Almaini}, {Werle}, {Wilkinson}, {Rowlands}, {Bolzonella}, {Castellano}, {Gargiulo}, {McLure}, {Pentericci}, \& {Pozzetti}}]{VWild2020}
{Wild}, V., {Taj Aldeen}, L., {Carnall}, A., {et~al.} 2020, \mnras, 494, 529, \dodoi{10.1093/mnras/staa674}

\bibitem[{{Wu} {et~al.}(2018){Wu}, {van der Wel}, {Bezanson}, {Gallazzi}, {Pacifici}, {Straatman}, {Bari{\v{s}}i{\'c}}, {Bell}, {Chauke}, {van Houdt}, {Franx}, {Muzzin}, {Sobral}, \& {Wild}}]{PFWu2018}
{Wu}, P.-F., {van der Wel}, A., {Bezanson}, R., {et~al.} 2018, \apj, 868, 37, \dodoi{10.3847/1538-4357/aae822}

\bibitem[{{Yan} \& {Blanton}(2012)}]{RYan2012}
{Yan}, R., \& {Blanton}, M.~R. 2012, \apj, 747, 61, \dodoi{10.1088/0004-637X/747/1/61}

\bibitem[{{Yan} {et~al.}(2006){Yan}, {Newman}, {Faber}, {Konidaris}, {Koo}, \& {Davis}}]{RYan2006}
{Yan}, R., {Newman}, J.~A., {Faber}, S.~M., {et~al.} 2006, \apj, 648, 281, \dodoi{10.1086/505629}

\bibitem[{{Yano} {et~al.}(2016){Yano}, {Kriek}, {van der Wel}, \& {Whitaker}}]{MYano2016}
{Yano}, M., {Kriek}, M., {van der Wel}, A., \& {Whitaker}, K.~E. 2016, \apjl, 817, L21, \dodoi{10.3847/2041-8205/817/2/L21}

\bibitem[{{Zhuang} {et~al.}(2023){Zhuang}, {Leethochawalit}, {Kirby}, {Nightingale}, {Steidel}, {Glazebrook}, {Barone}, {Skobe}, {Sweet}, {Nanayakkara}, {Allen}, {Vasan}, {Jones}, {Kacprzak}, {Tran}, \& {Jacobs}}]{ZZhuang2023}
{Zhuang}, Z., {Leethochawalit}, N., {Kirby}, E.~N., {et~al.} 2023, \apj, 948, 132, \dodoi{10.3847/1538-4357/acc79b}

\bibitem[{{Zibetti} {et~al.}(2013){Zibetti}, {Gallazzi}, {Charlot}, {Pierini}, \& {Pasquali}}]{SZibetti2013}
{Zibetti}, S., {Gallazzi}, A., {Charlot}, S., {Pierini}, D., \& {Pasquali}, A. 2013, \mnras, 428, 1479, \dodoi{10.1093/mnras/sts126}

\end{thebibliography}
\bibliographystyle{aasjournal}

\appendix
\restartappendixnumbering
\section{NIRSpec Resolution}\label{ap:LSF}
The instrumental resolution for NIRSpec is provided for a uniformly-illuminated slit in the JWST User Documentation (JDox)\footnote{\url{https://jwst-docs.stsci.edu/jwst-near-infrared-spectrograph/nirspecinstrumentation/nirspec-dispersers-and-filters}}. However, since our quiescent sources are relatively compact, they will not uniformly illuminate the MSA slits, which will improve the actual instrumental resolution of our observed spectra \citep[e.g.][]{AdeGraaff2023,DNidever2024, TNanayakkara2024}.

For each galaxy we derive the wavelength dependent resolution in a rectified, combined spectrum, taking into account the source morphology and slit placement.  Our method builds upon the work presented in \citet{AdeGraaff2023}, who model the LSF for an individual unrectified frame. Following \citet{AdeGraaff2023}, we use \texttt{MSAFIT} to create a 3D model cube of the source profile as a function of wavelength ($I(x,y,\lambda)$). We define this source profile as a S\'ersic profile with the morphological parameters from \citet{SCutler2022}, and sample the wavelength grid in intervals of $\Delta\lambda = 0.1~\mu$m, spanning the entire wavelength range of our filter/disperser combination. At each wavelength point we insert an emission line with an intrinsic velocity dispersion of zero, and a total normalized intensity of 1. We set the location of the source in the slit as the source location given by the MSA reference file that is generated during the observations, and use the shutters in which the source was observed. The model cube is then convolved with the point-spread function (PSF) and modeled onto the detector plane by \texttt{MSAFIT}.

The resulting model corresponds to a spectrum from a single, unrectified frame. Rectifying and combining the individual frames will further broaden the spectra. To obtain an estimate of the instrumental resolution in the final rectified, combined frames we thus have to apply the same reduction steps to these modeled individual frames as we do for the real observed data. To this end, we input the model spectra (one for each nodding position in our observation set-up) into the same modified pipeline we used to reduce our data, starting from Stage 2 of the JWST Calibration Pipeline (see Section \ref{sec:data_red}). After processing the model spectra through Stages 2 and 3 of the reduction pipeline we obtain a rectified, combined 2D spectrum with the same format as the observed 2D spectrum. From this model 2D spectrum we extract the 1D spectrum, and fit the inserted emission lines with single Gaussians. The FWHM of these Gaussians give us the resolution as a function of wavelength.

\begin{figure}
    \centering
    \includegraphics[width = 1.0\textwidth]{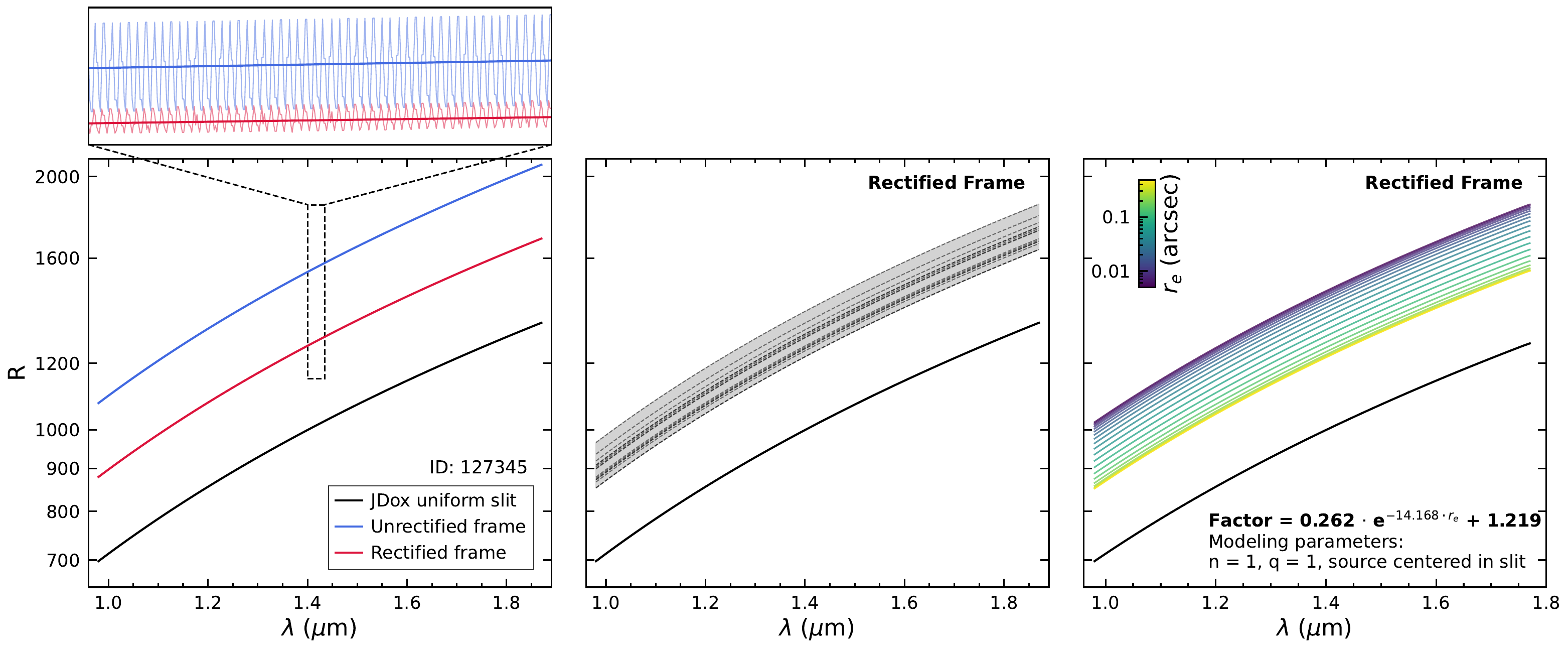}
    \caption{Left: The modeled resolution curve of NIRSpec for the F100LP/G140M filter/grating for source 127345 in the rectified frame (red line) and unrectified frame (blue line). The black solid line shows the dispersion curve for a uniformly illuminated slit from JWST User Documentation (JDox). The inlay at the top shows sinusoidal pattern in the measured FWHMs due to wavelength undersampling. Middle: The range of resolutions for the rectified frame covered by the quiescent galaxies in our sample (grey area), with the mean curves of individual galaxies shown as dashed grey lines. The black solid line shows the dispersion curve for a uniformly illuminated slit from JDox. For the galaxies in our sample the resolution is a factor $1.22-1.38$ higher than the reported resolution from JDox. Right: The resolutions in the rectified frame for sources with sizes ranging from $r_e = 0.005$" (point-source) to $r_e = 0.48$" (uniformly illuminated slit), for a source that is centered in the slit, has a S\'ersic index of 1, and $q=1$. The black solid line shows the dispersion curve for a uniformly illuminated slit from JDox. To obtain the LSF for a source with a given $r_e$, we multiply the JDox resolution curve by the factor calculated using the formula in the bottom right of the plot.}
    \label{fig:LSF}
\end{figure}
In the left panel of Figure \ref{fig:LSF} we show the modeled LSF for source 127345 in the rectified frame in red, as well as the LSF calculated in \texttt{MSAFIT} for an individual unrectified frame in blue. We also show the LSF for an uniformly illuminated slit from JDox in black. As expected, the resolution of the modeled source is better than that of a uniformly illuminated slit, and after rectifying and combining the frames the resolution decreases slightly. In the right panel of Figure \ref{fig:LSF} we show the resolution curves for the rectified spectra of all galaxies in our sample. The improvement in resolution for the galaxies in our sample is a factor $1.22-1.38$.

We caution that when deriving the LSF for an individual galaxy it is important to sample the inserted emission lines at sufficiently small wavelength intervals. We find that our measured LSF is not a smooth curve, for both the rectified and unrectified frame. The pattern we measure does not reflect a true discontinuity in the LSF, but originates from the fact that the wavelength grid is undersampled for our filter/disperser combination, causing the inserted emission lines to not all be dispersed over the same pixel area. For a constant line intensity this means that the measured width of the emission line in the spectrum will vary depending on what part of the pixels it is dispersed over. We note that this undersampling effect is especially clear in our modeling, as we model lines without intrinsic broadening; for broadened emission lines the wavelength undersampling effect is less strong. To quantify how strong the variation in resolution due to undersampling is, we insert emission lines in an oversampled wavelength grid. We do this by creating 100 model cubes, each with emission lines at $\Delta \lambda = 100$~\AA, but we shift the position of the emission lines by 1~\AA\ for each model cube. We then measure the FWHMs of the emission lines for each of these models to obtain the LSF sampled at $\Delta \lambda = 1$~\AA. We show the results of this test in the zoomed in plot in Figure \ref{fig:LSF}. This test shows that in the rectified frame, the variation from the median resolution due to undersampling is $R\sim60$, while in the unrectified frame the variation is $R\sim235$. We obtain the final LSF for each galaxy from the average of the oversampled FWHMs from this test.

To understand how morphology and slit position affect the LSF, we additionally perform two tests. First, we assess the difference in resolution from source size by varying the half-light radius ($r_e$) from 0.005" (point source) to 0.48" (fully illuminated slit). We center the source in the slit, use a S\'ersic index of 1, and an axis ratio of 1. From this test we find a difference in resolution of $\sim18\%$, with a point source having the highest resolution and a uniformly illuminated slit having the lowest resolution. We note that the estimated resolution of the uniformly illuminated slit is still significantly higher (a factor $\sim1.2$) than the resolution reported by JDox. We fit an exponential relation to the multiplication factors we find for our range of $r_e$ values, and find that the multiplication factor depends on $r_e$ as
\begin{equation}
    \text{Factor} = 0.262 \cdot e^{-14.168 \cdot r_e} + 1.219.
\end{equation}

Secondly, we test the effect of the source position in the slit by calculating the resolution of a source with a S\'ersic index of 1, an axis ratio of 1, and an effective radius of 0.05\farcs, for different positions in the slit. We first test the effect of the placement along the dispersion direction, and find that the maximum difference in resolution from moving it along the slit in this direction is 11\%. We also note that in some cases the resolution of a source that is slightly off-center is somewhat higher than a completely centered source. This is likely due to undersampling effects. When we vary the source placement along the spatial direction, we find that the improvement in resolution from a source at the edge of the slit compared to a centered source is 3\%.

While these tests are insightful to see the approximate effect of source morphology and slit placement on the resolution, we emphasize that the full source needs to be modeled in the \texttt{MSAFIT} software to obtain a reasonable estimate for the true resolution of the source. We also note that the modeling using \texttt{MSAFIT} is still subject to further calibration (for details, see \citet{AdeGraaff2023}), but is sufficiently accurate for the purposes of this work.

\section{Quiescent galaxy spectra}\label{ap:all_spec}
\begin{figure*}[b]
\begin{center}
    \centering
    \includegraphics[width = 0.91\textwidth]{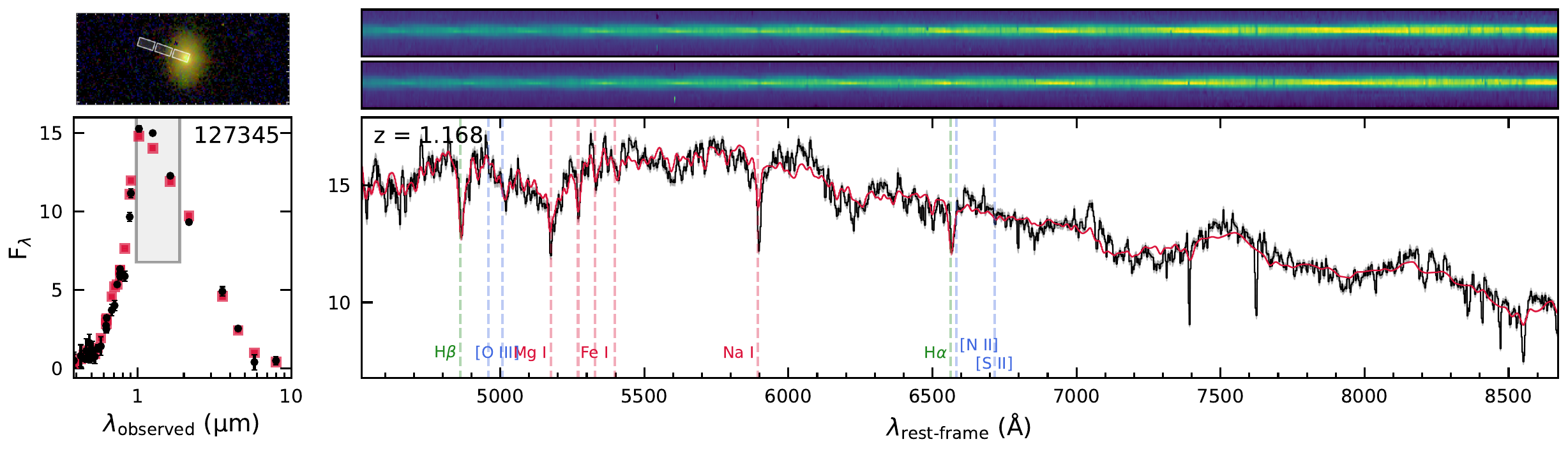}
    \includegraphics[width = 0.91\textwidth]{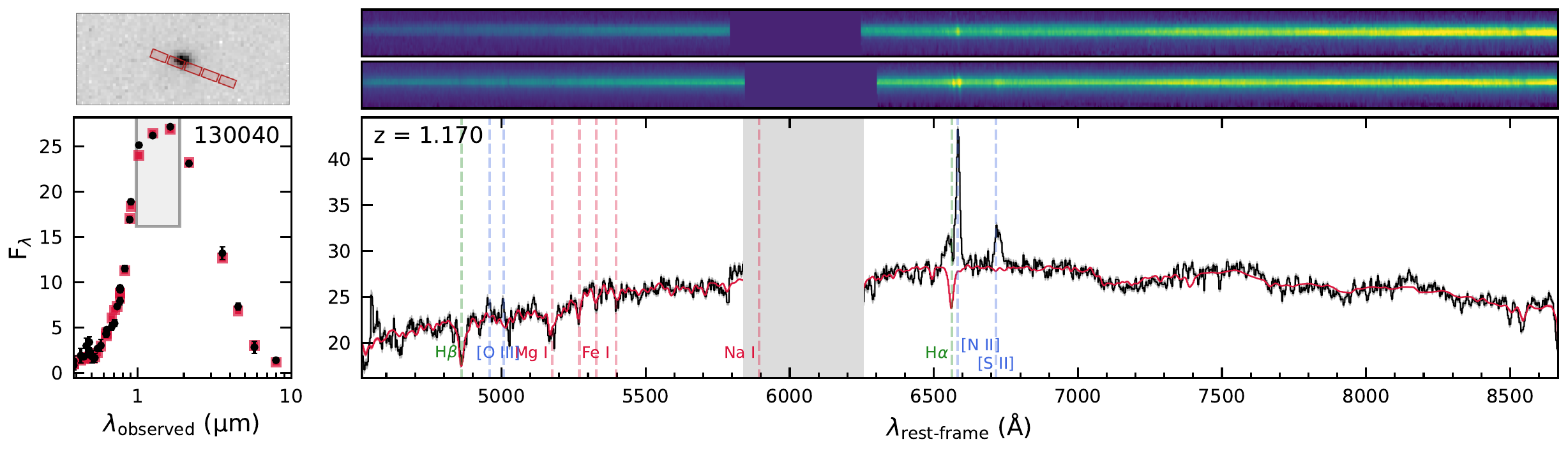}
    \includegraphics[width = 0.91\textwidth]{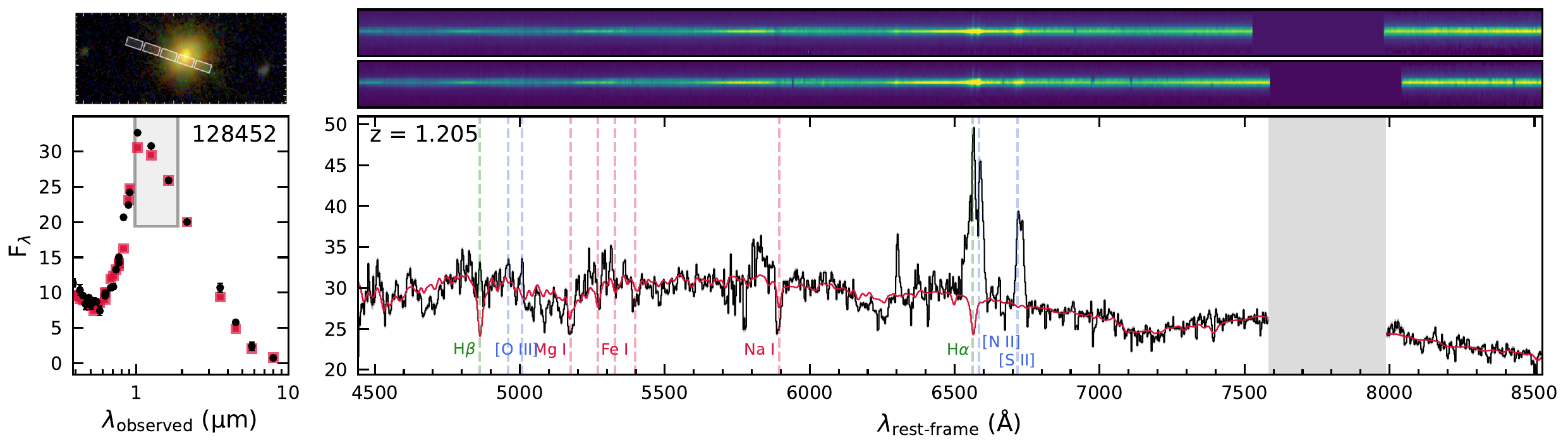}
    \includegraphics[width = 0.91\textwidth]{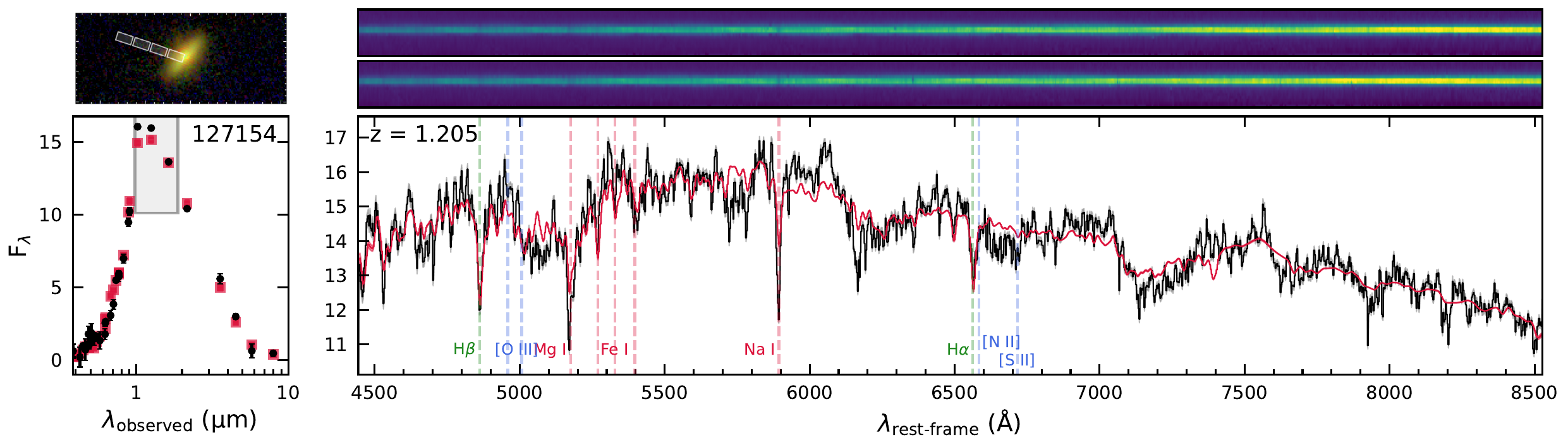}
    \caption{Same as Figure \ref{fig:example_q_spec}.}
\end{center}
\end{figure*}

\restartappendixnumbering
\begin{figure*}
\begin{center}
    \centering
    \includegraphics[width = 1\textwidth]{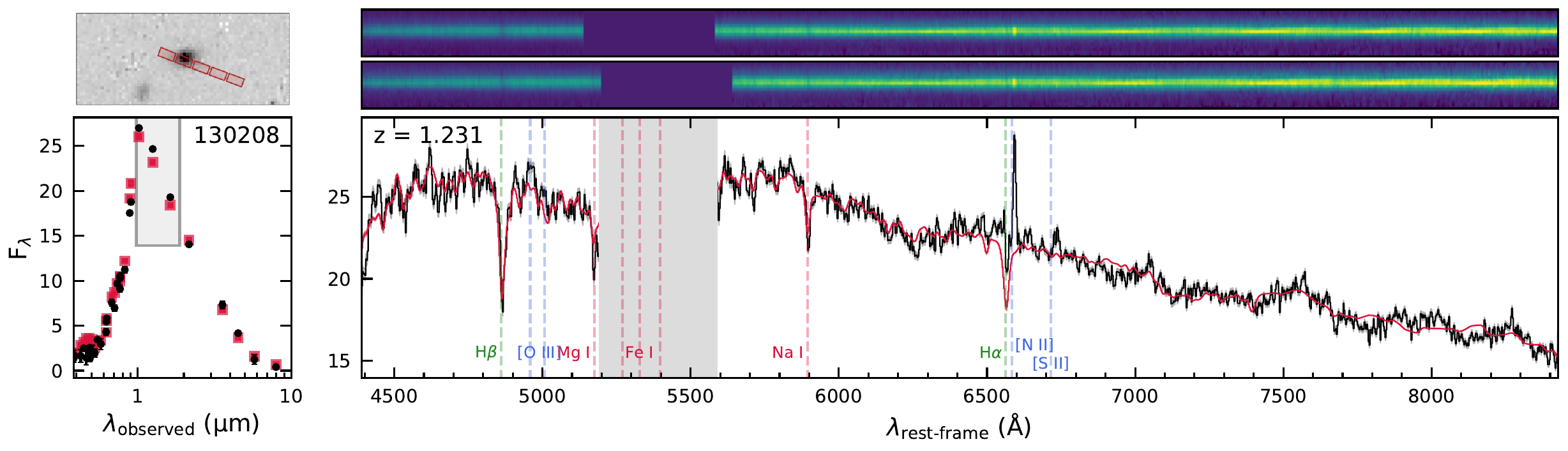}
    \includegraphics[width = 1\textwidth]{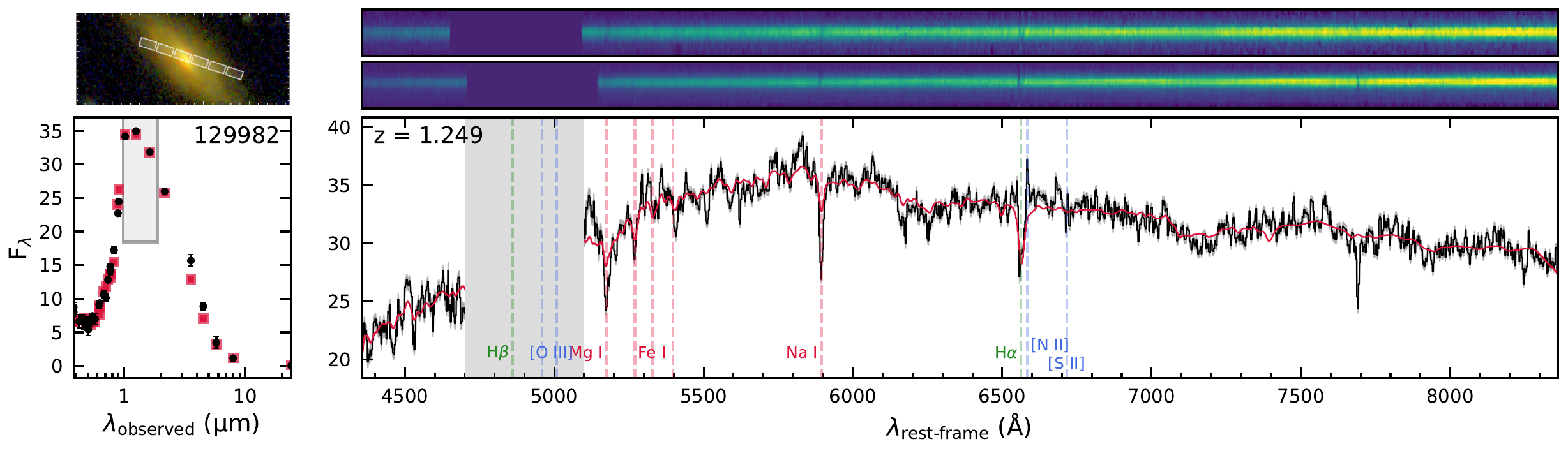}
    \includegraphics[width = 1\textwidth]{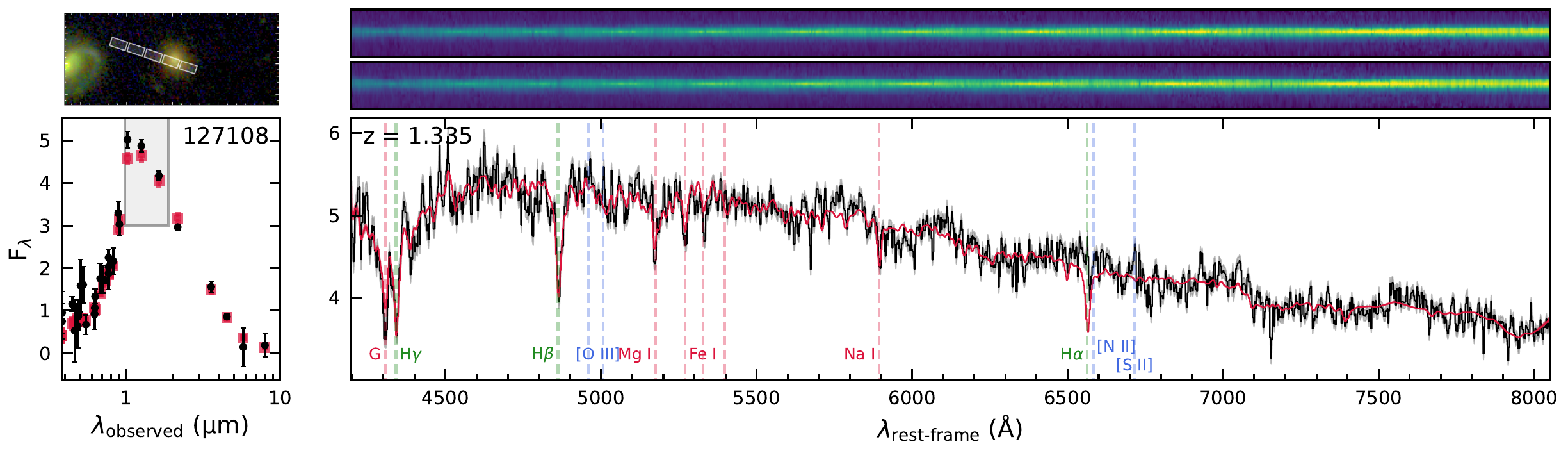}
    \includegraphics[width = 1\textwidth]{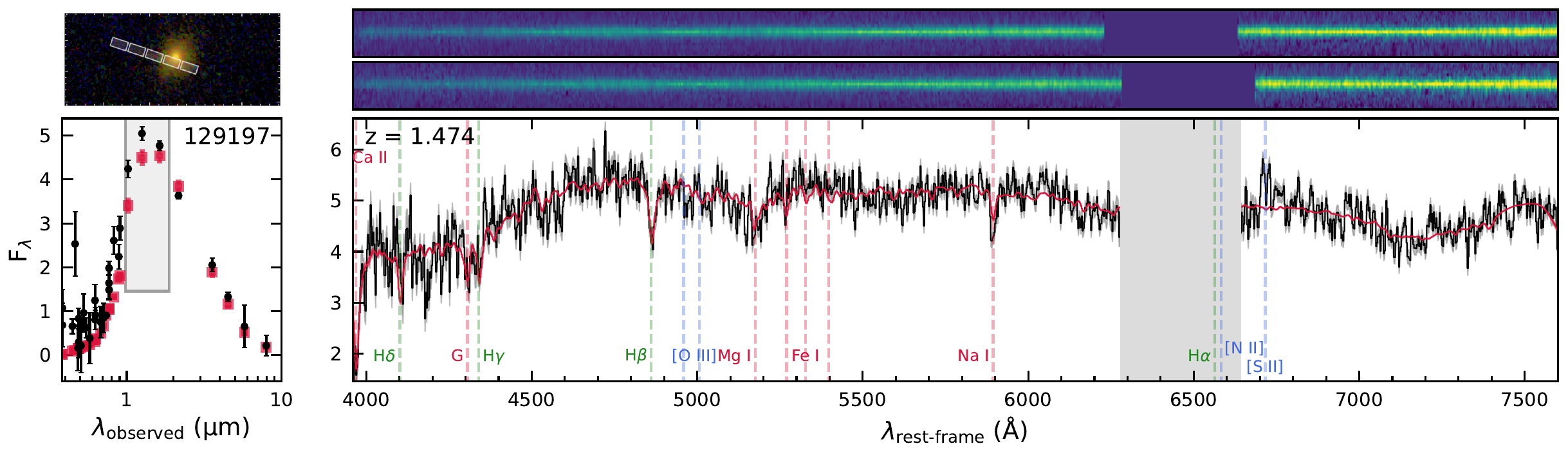}
    
    \caption{Continued.}
    \label{fig:all_spec}
\end{center}
\end{figure*}

\restartappendixnumbering
\begin{figure*}
\begin{center}
    \centering
    \includegraphics[width = 1\textwidth]{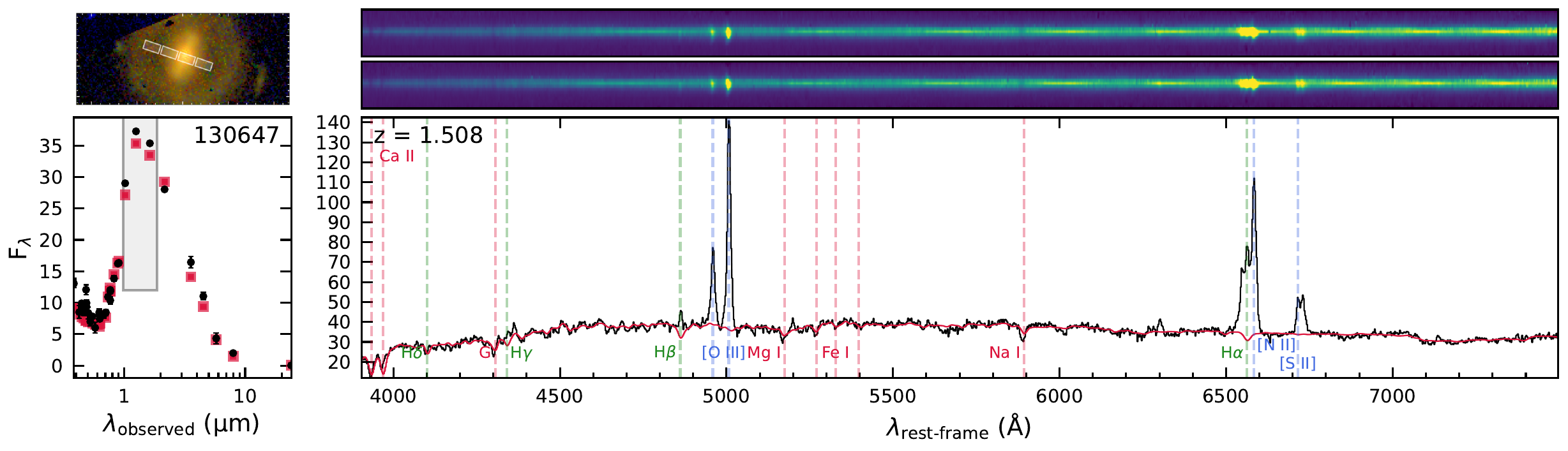}
    \includegraphics[width = 1\textwidth]{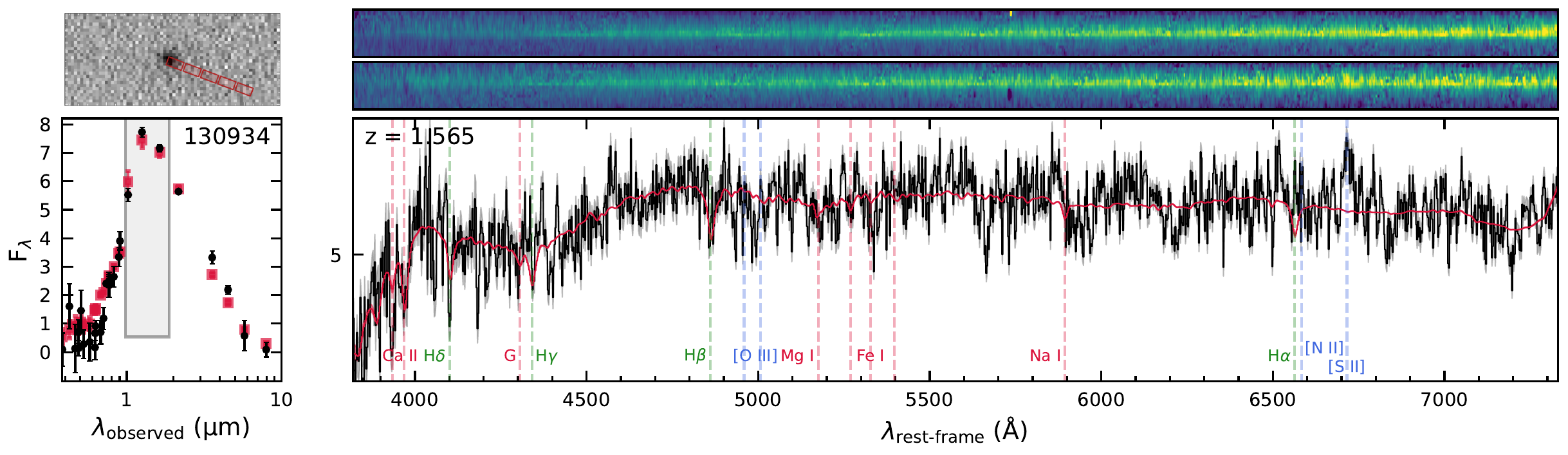}
    \includegraphics[width = 1\textwidth]{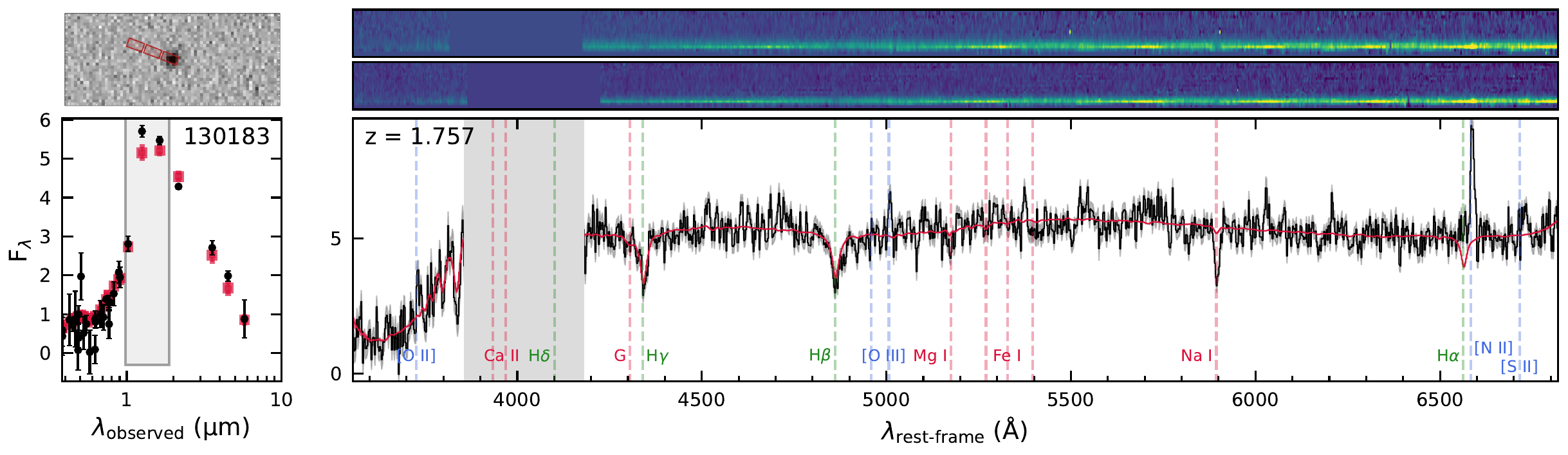}
    \includegraphics[width = 1\textwidth]{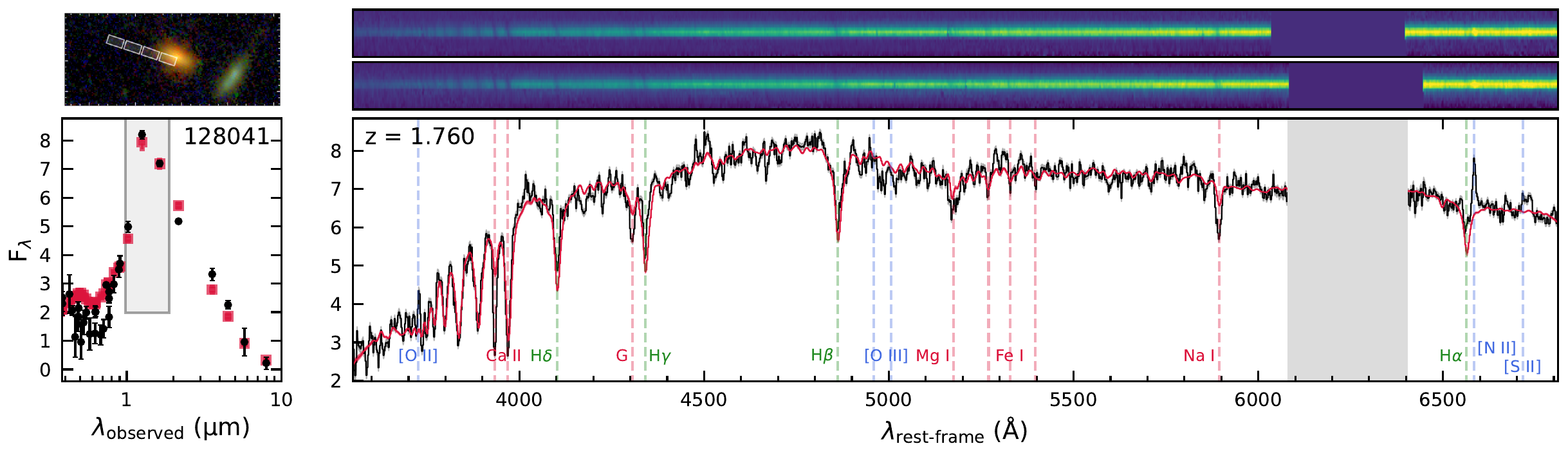}
    
    \caption{Continued.}
\end{center}
\end{figure*}

\restartappendixnumbering
\begin{figure*}
\begin{center}
    \centering
    \includegraphics[width = 1\textwidth]{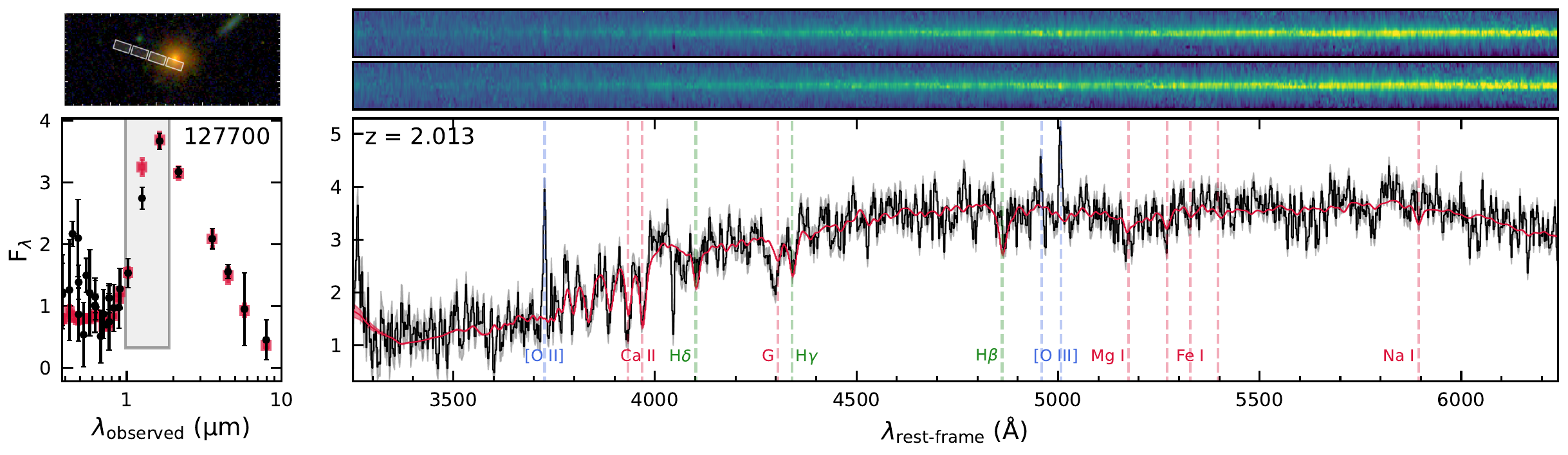}
    \includegraphics[width = 1\textwidth]{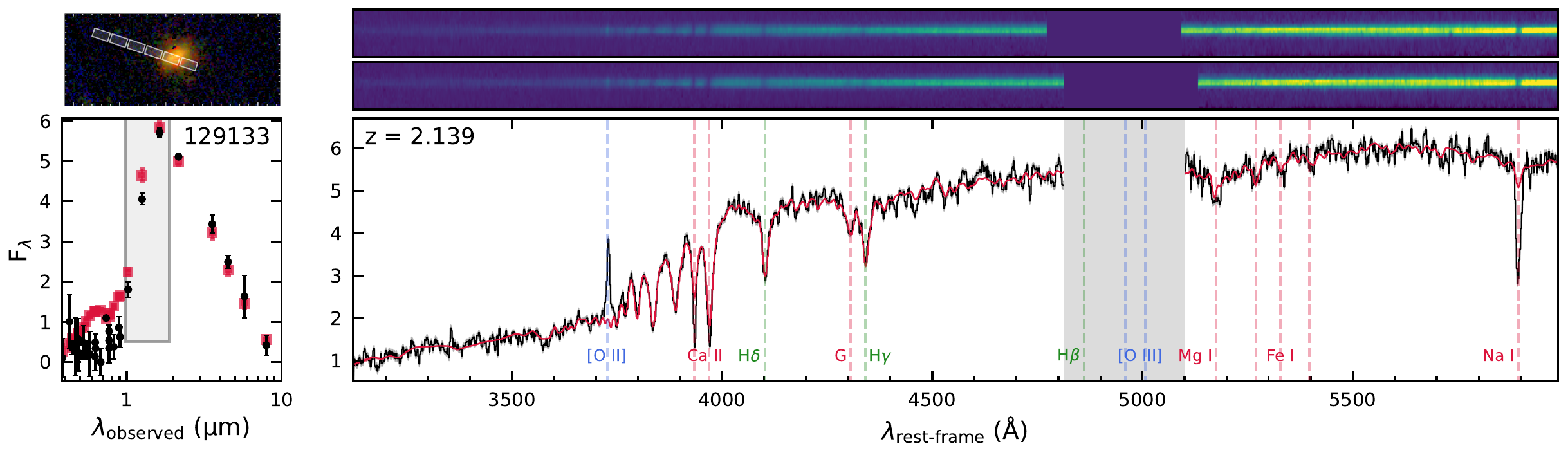}
    \includegraphics[width = 1\textwidth]{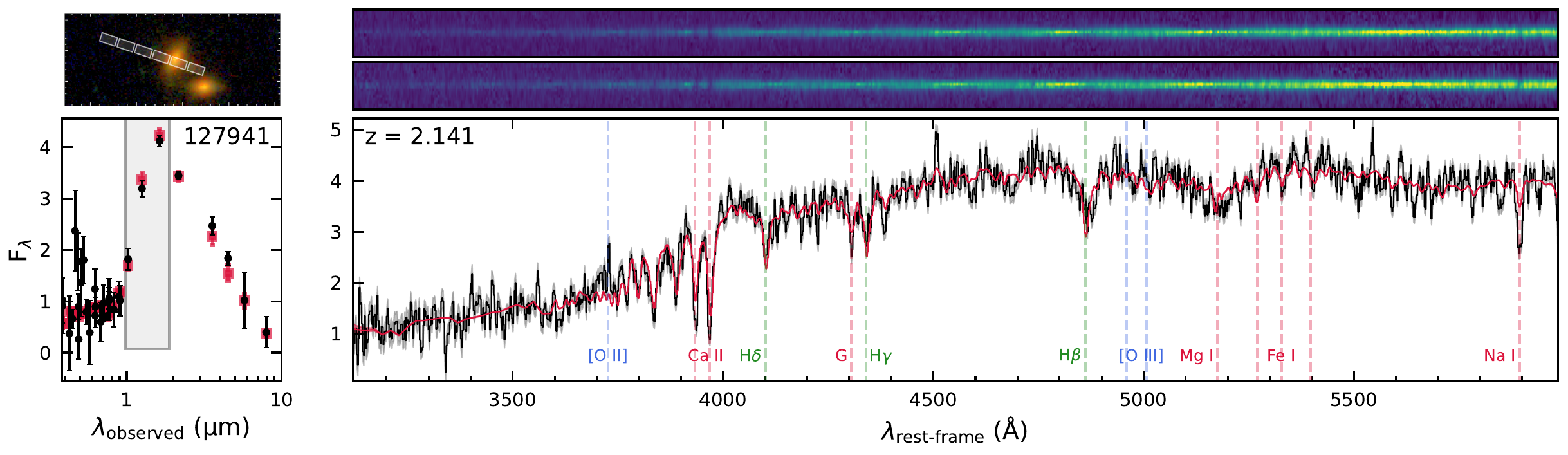}
    \includegraphics[width = 1\textwidth]{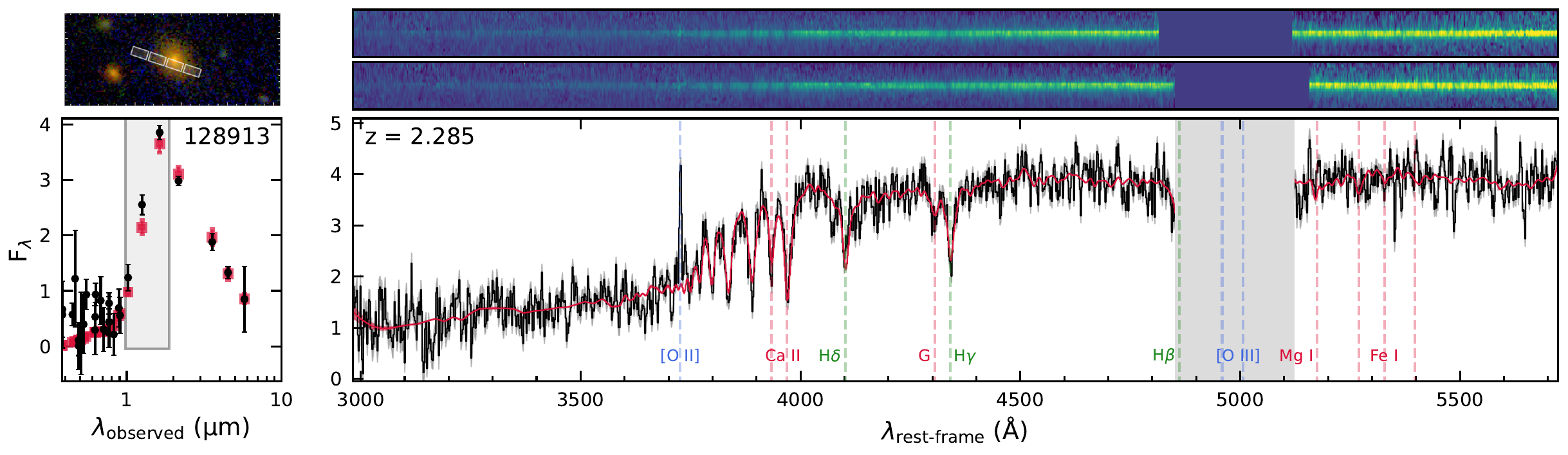}
    \caption{continued.}
\end{center}
\end{figure*}

\restartappendixnumbering
\begin{figure*}
\begin{center}
    \centering
    \includegraphics[width = 1\textwidth]{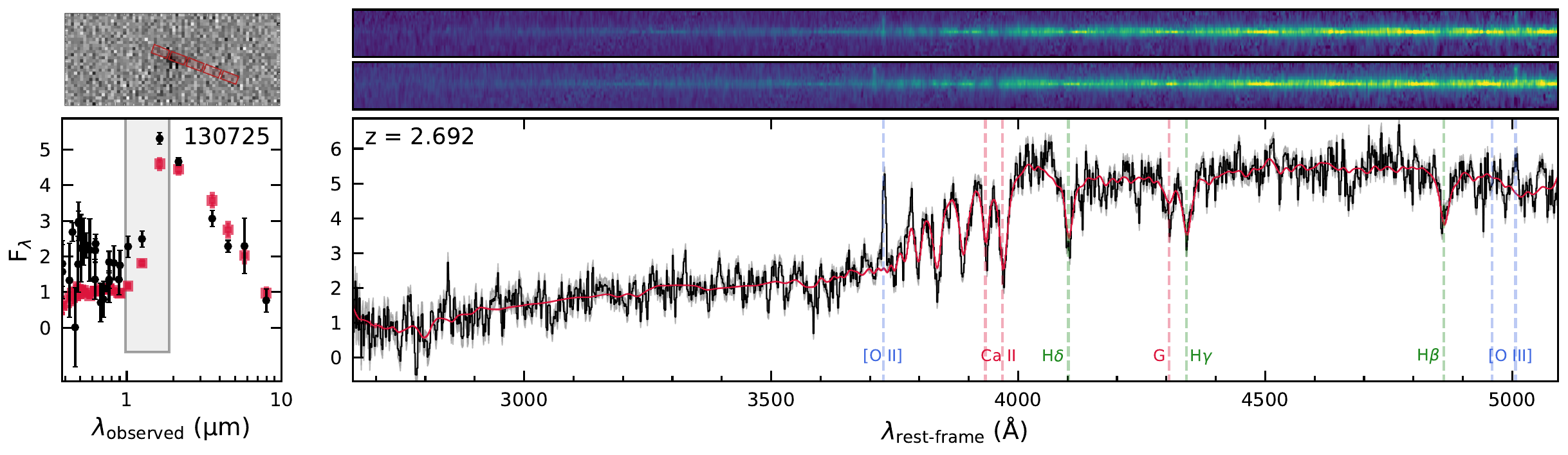}
    \includegraphics[width = 1\textwidth]{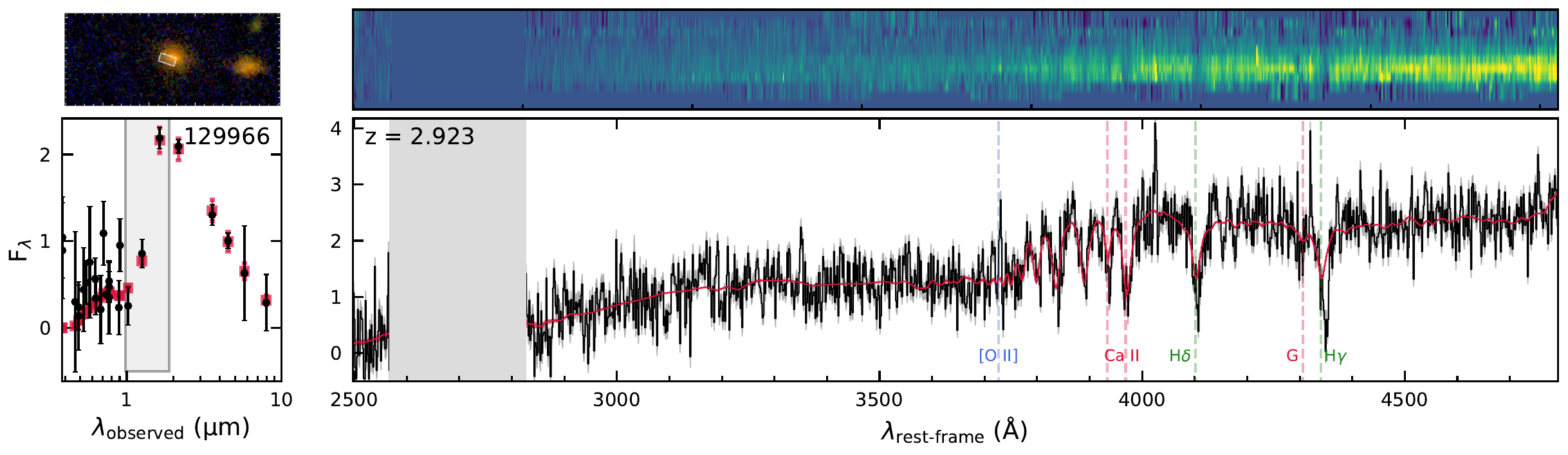}
    \caption{continued.}
\end{center}
\end{figure*}

\section{SFHs}\label{ap:all_SFHs}
\restartappendixnumbering
\begin{figure*}[!h]
\begin{center}
    \centering
    \includegraphics[width = 1.\textwidth]{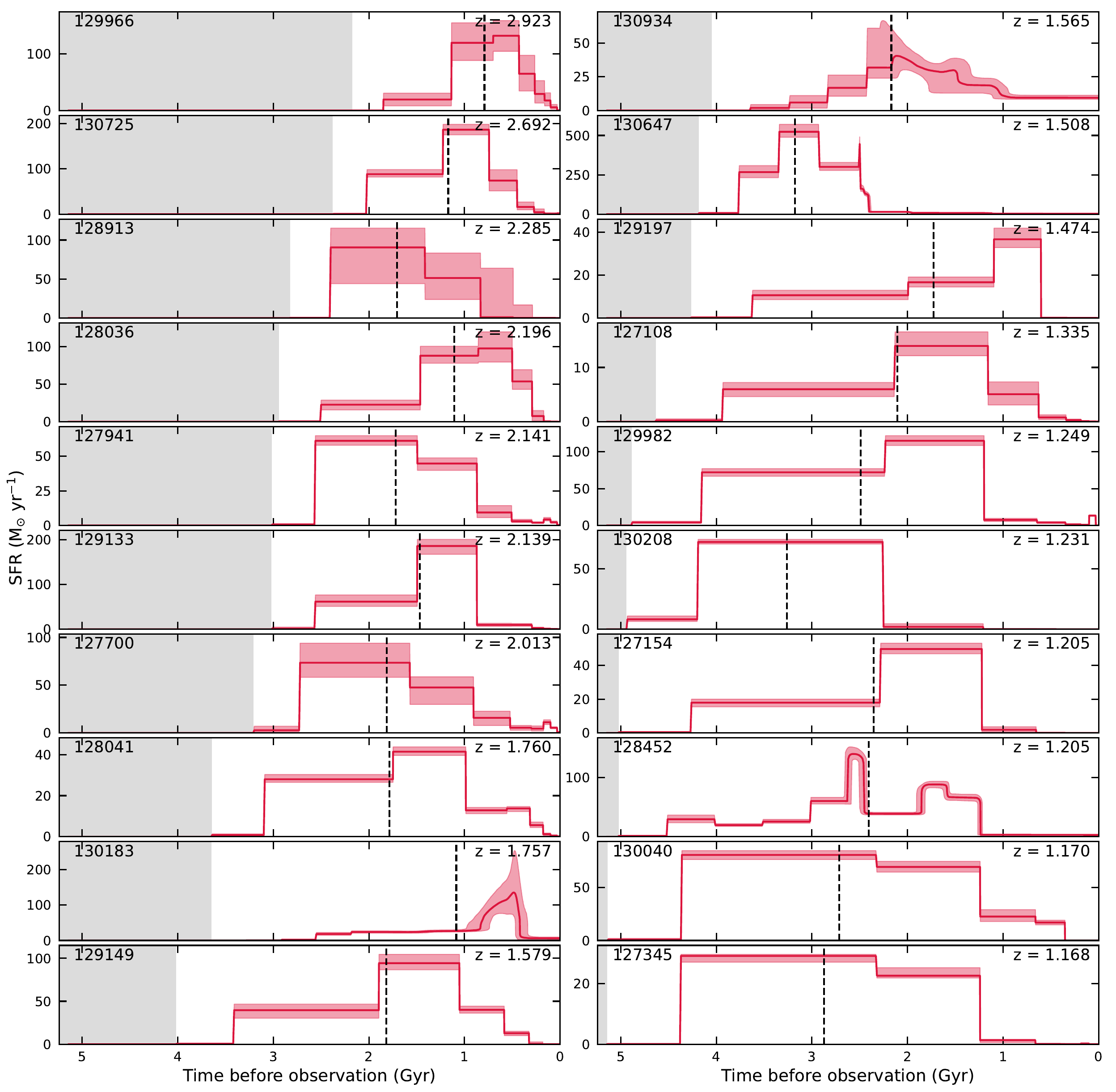}
    \caption{Best-fit SFHs for all galaxies in the quiescent sample. The SFHs are ordered by decreasing redshift, as indicated in the top right of each panel. The shaded areas represent the 16th-84th percentile confidence intervals of the SFHs. The black, dashed vertical lines indicate the time where 50\% of the galaxy's stellar mass was formed ($t_{50}$). The gray shaded area in each panel corresponds to ages that are older than the age of the universe for the observed redshift.}
    \label{fig:all_sfhs}
\end{center}
\end{figure*}

\section{Filler galaxies}\label{ap:SF_gals}
\begin{table*}[b]
\centering
\caption{Overview of confirmed star-forming filler galaxy sample}
\label{tab:filler_sample}
\begin{tabular}{l l c c c c c r}
\hline \hline
\multicolumn{1}{c}{ID} & \multicolumn{2}{c}{\underline{Coordinates}} & \multicolumn{1}{c}{$H$}  & \multicolumn{1}{c}{$z_{\rm spec}$} &  \multicolumn{2}{c}{\underline{Rest-frame colors}} & \multicolumn{1}{c}{$t_{\text{exp}}$} \\
& \multicolumn{1}{c}{R.A.} & 
\multicolumn{1}{c}{Decl.} & & & $U-V$ & $V-J$ & \multicolumn{1}{c}{(hr)} \\
\hline 
126887 & 10:01:58.27 & 2:25:44.3 & 25.1 & 1.579 & 0.63 & -0.42 & 4.9 \\ 
127350 & 10:02:01.03 & 2:26:00.2 & 22.3 & 2.030 & 1.36 & 1.25 & 6.6 \\ 
127491 & 10:01:57.73 & 2:26:13.9 & 21.9 & 1.169 & 0.63 & 0.26 & 13.1 \\ 
127646 & 10:02:02.00 & 2:26:28.2 & 25.0 & 2.832 & 0.45 & -0.51 & 16.4 \\ 
127680 & 10:01:59.64 & 2:26:25.1 & 22.7 & 1.738 & 0.94 & 1.09 & 16.4 \\ 
127894 & 10:02:04.90 & 2:26:41.2 & 25.1 & 3.518 & 0.64 & 0.09 & 11.5 \\ 
127900 & 10:01:57.88 & 2:26:39.4 & 22.6 & 1.202 & 0.67 & 0.42 & 13.1 \\ 
128019 & 10:02:00.72 & 2:26:44.3 & 22.5 & 2.007 & 1.15 & 1.28 & 16.4 \\ 
128029 & 10:02:05.44 & 2:26:44.4 & 22.8 & 2.206 & 0.56 & 0.38 & 16.4 \\ 
128047 & 10:02:01.12 & 2:26:42.6 & 21.4 & 2.197 & 1.46 & 1.48 & 6.6 \\ 
128080 & 10:02:00.32 & 2:26:48.2 & 23.0 & 2.791 & 0.59 & 0.72 & 8.2 \\ 
128188 & 10:02:01.66 & 2:26:52.4 & 21.9 & 1.664 & 1.45 & 1.70 & 16.4 \\ 
128344 & 10:01:59.05 & 2:27:02.7 & 22.3 & 1.370 & 1.03 & 0.78 & 16.4 \\ 
128345 & 10:02:05.39 & 2:27:05.0 & 22.6 & 2.008 & 1.03 & 0.89 & 3.3 \\ 
128422 & 10:01:59.92 & 2:27:05.7 & 22.0 & 1.372 & 1.19 & 0.87 & 8.2 \\ 
128444 & 10:02:05.94 & 2:27:09.0 & 22.7 & 2.786 & 0.62 & 0.54 & 9.8 \\ 
128561 & 10:02:09.67 & 2:27:17.8 & 23.3 & 2.925 & 0.73 & 1.02 & 16.4 \\ 
128827 & 10:02:04.39 & 2:27:35.3 & 23.7 & 3.148 & 0.80 & 0.78 & 6.6 \\ 
128948 & 10:02:08.83 & 2:27:43.7 & 24.8 & 2.005 & 0.51 & -0.01 & 9.8 \\ 
129015 & 10:01:57.29 & 2:27:47.4 & 24.7 & 2.309 & 0.51 & -0.19 & 4.9 \\ 
129024 & 10:02:08.43 & 2:27:45.5 & 22.9 & 1.280 & 0.51 & 0.17 & 16.4 \\ 
129161 & 10:02:07.06 & 2:27:49.5 & 21.9 & 1.437 & 1.18 & 0.72 & 16.4 \\ 
129264 & 10:01:58.60 & 2:28:01.0 & 26.0 & 1.653 & 0.24 & -0.96 & 4.9 \\ 
129315 & 10:01:55.81 & 2:28:03.1 & 24.7 & 2.944 & 0.29 & -0.23 & 8.2 \\ 
129363 & 10:02:07.84 & 2:28:03.7 & 22.9 & 1.336 & 1.15 & 1.36 & 6.6 \\ 
129508 & 10:01:55.56 & 2:28:12.6 & 23.1 & 2.920 & 0.92 & 1.34 & 8.2 \\ 
129654 & 10:02:03.19 & 2:28:22.2 & 23.1 & 2.506 & 1.23 & 1.03 & 4.9 \\ 
129663 & 10:02:02.95 & 2:28:25.3 & 24.2 & 3.719 & 0.65 & 0.79 & 8.2 \\ 
129689 & 10:01:55.20 & 2:28:26.5 & 25.6 & 2.087 & 0.58 & -1.06 & 4.9 \\ 
129695 & 10:01:55.81 & 2:28:21.0 & 22.7 & 2.688 & 1.18 & 1.31 & 8.2 \\ 
129776 & 10:01:55.76 & 2:28:31.9 & 24.0 & 2.972 & 0.43 & 0.73 & 9.8 \\ 
129829 & 10:01:54.33 & 2:28:32.0 & 22.6 & 1.580 & 0.59 & 0.39 & 6.6 \\ 
129992 & 10:01:56.00 & 2:28:35.7 & 21.5 & 1.443 & 1.06 & 0.75 & 16.4 \\ 
130128 & 10:02:07.20 & 2:28:51.5 & 25.8 & 2.780 & 0.39 & -1.55 & 8.2 \\ 
130180 & 10:02:04.82 & 2:28:54.6 & 25.6 & 3.947 & 0.17 & -0.05 & 9.8 \\ 
130293 & 10:01:55.28 & 2:28:58.0 & 22.9 & 1.515 & 0.57 & 0.42 & 16.4 \\ 
130320 & 10:02:04.20 & 2:28:59.5 & 22.7 & 1.547 & 1.01 & 0.54 & 3.3 \\ 
130390 & 10:02:06.12 & 2:29:06.5 & 26.2 & 3.353 & 0.55 & 0.13 & 16.4 \\ 
130547 & 10:01:58.17 & 2:29:15.9 & 24.7 & 2.400 & 0.52 & -0.10 & 9.8 \\ 
130782 & 10:02:05.29 & 2:29:26.1 & 22.1 & 1.514 & 0.85 & 0.68 & 3.3 \\ 
130874 & 10:02:04.54 & 2:29:28.9 & 21.8 & 1.290 & 0.99 & 0.68 & 9.8 \\ 
130907 & 10:02:04.55 & 2:29:31.8 & 21.4 & 1.517 & 1.05 & 1.12 & 16.4 \\ 
130998 & 10:02:03.05 & 2:29:43.3 & 25.4 & 1.403 & 1.07 & 0.50 & 9.8 \\ 
131179 & 10:02:03.86 & 2:29:49.5 & 22.6 & 2.145 & 1.41 & 1.81 & 6.6 \\ 
131143 & 10:02:03.06 & 2:29:46.8 & 22.9 & 2.914 & 0.63 & 0.22 & 9.8 \\ 
131221 & 10:02:04.55 & 2:29:54.1 & 24.3 & 2.691 & 1.63 & 1.38 & 16.4\\ 
\hline\hline
\end{tabular}
\end{table*}

\end{document}